\documentclass[12pt]{iopart}

\usepackage{iopams}
\usepackage{enumerate}

\usepackage{amsthm}
\newtheorem{thm}{Theorem}

\newcommand{\be}{\begin{equation}}
\newcommand{\ee}{\end{equation}}
\newcommand{\ba}{\begin{eqnarray}}
\newcommand{\ea}{\end{eqnarray}}

\begin{document}

\title[Conformally reducible 1+3 spacetimes]
{Conformally reducible 1+3 spacetimes}

\author{Jaume Carot$^1$, Aidan J Keane$^2$ and Brian O J Tupper$^3$}
\address{$^1$ Departament de F\'{i}sica, Universitat de les Illes Balears,
Cra. Valldemossa km 7.5, E-07122 Palma de Mallorca, Spain\\
$^2$ 87 Carlton Place, Glasgow G5 9TD, Scotland, UK\\
$^3$ Department of Mathematics and Statistics, University of
New Brunswick, Fredericton, New Brunswick, E3B 5A3, Canada
}

\eads{\mailto{jcarot@uib.es, aidan@countingthoughts.com and bt32@rogers.com}}

\begin{abstract}
Spacetimes which are conformally related to reducible 1+3 spacetimes are considered.
We classify these spacetimes according to the conformal algebra of the underlying
reducible spacetime, giving in each case canonical expressions for the metric and conformal
Killing vectors, and provide physically meaningful examples.
\end{abstract}

\pacs{02.40.Ky, 04.20.Jb}

\section{Introduction}
In a previous article \cite{CarotTupper}, conformally reducible 2+2 spacetimes,
i.e., spacetimes conformal to reducible (decomposable) 2+2 spacetimes were classified
according to their conformal symmetries. In this article we extend our study to
conformally reducible 1+3 spacetimes. Reducible 1+3 spacetimes have been the
subject of previous studies.
Coley and Tupper \cite{ColeyTupper} found the general form for reducible 1+3 spacetimes
which admit proper conformal Killing vectors.
Tsamparlis, Nikolopoulos and Apostolopoulos \cite{Tsamparlis} related the conformal Killing
vectors of reducible 1+3 spacetimes to the conformal Killing vectors of the underlying 3-space
and showed that the latter can be obtained from a combination of a gradient conformal Killing
vector and a Killing vector or homothetic Killing vector. Capocci and Hall \cite{CapocciHall}
studied the conformal symmetries in terms of the holonomy group classification.

A spacetime $(M,g)$ is said to be a {\em conformally reducible 1+3 spacetime}
if, for every point $p\in M$, there exists a coordinate chart
$\{x^a\}$ such that the line element takes the form
\be
ds^2 = \exp (2 \mu (x^a))
\left(d\sigma_0^2 +d\sigma^2\right), \label{2.1}
\ee where
\be
d\sigma_0^2 = \epsilon_0 d\eta^2, \;\; \epsilon_0 = \pm 1
\label{2.2}\ee \be d\sigma^2 = h_{AB}(x^C) dx^A dx^B \;\;\;\;
A,B, \cdots =1,2,3  \label{2.3}
\ee
with $d\sigma_0^2$ of signature plus or minus one and $d\sigma^2$ of signature
+1 or +3 respectively. That is, $(M,g)$ is conformally related to a reducible 1+3
spacetime, say $(M,\hat{g})$ whose associated line element shall be written,
from now on, as:
\[
d\Sigma^2 = d\sigma_0^2 + d\sigma^2
\]
or equivalently, and in the above  coordinate system:
\be
d\Sigma^2 = \epsilon_0 d\eta^2 + h_{AB}(x^C) dx^A dx^B, \;\; \epsilon_0 = \pm 1,
\;\; A,B, \cdots =1,2,3 \label{2.4}
\ee
Henceforth, the 3-space with metric
\be
d \sigma^2 = h_{AB}(x^C) dx^A dx^B
\ee
will be referred to as $(V,h)$.

Now, the spacetime $(M,\hat g)$ is (locally) reducible 1+3 if and
only if it admits a global, non-null, nowhere zero covariantly
constant vector field $\vec \eta$, and one can then distinguish
between 1+3-spacelike (whenever $\vec \eta$ is timelike, hence
$(V,h)$ is spacelike and $\epsilon_0 =-1$) or 1+3-timelike
($\vec \eta$ spacelike, $(V,h)$ Lorentz and $\epsilon_0 =+1$). A
characterization by means of the spacetime holonomy group is also
possible; thus in the first case (1+3-spacelike) the holonomy type
is $R_{13}$, whereas in the second case it is $R_{10}$. If other
non-null covariantly constant vector fields exist, the spacetime
reduces still further, and the holonomy types are different,
but the spacetimes can be described  in a similar fashion (see
\cite{HallKay} for details).

Before proceeding we recall that (see for example \cite{KSMH2}) a vector field $\vec X$ is said to be a
\emph{conformal Killing vector} (CKV) iff $\mathcal{L}_{\vec X} g
=2\phi g$ where $\phi$ is some function of the coordinates
(\emph{conformal scalar}),
$g$ is the metric tensor, and $\mathcal{L}_{\vec X}$ stands for the
Lie derivative operator with
respect to the vector field $\vec X$. The above equation can also
be written in an arbitrary coordinate chart as
\be
X_{a;b} = \phi g_{ab} + F_{ab} \label{ckv10}
\ee
where $X_a$ and $g_{ab}$ are the covariant components of $\vec X$
and the metric in the chosen chart and $F_{ab} =-F_{ba}$ is the
conformal bivector. When $\phi \ne \mathrm{constant}$ the CKV is said
to be {\it proper}, and if  $\phi_{;ab}=0$ the CKV is a {\it special} CKV (SCKV).
When $\phi$ is a constant, $\vec X$ is a {\it homothetic vector} (HV) and
$F_{ab}$ is called the {\it homothetic bivector}. When $\phi = 0$,
$\vec X$ is a {\it Killing vector} (KV) and $F_{ab}$ is the
{\it Killing bivector}. If $F_{ab} = 0$, i.e., $X_a = X_{,a}$, the CKV
is a gradient CKV (GCKV). Similarly, gradient HV and gradient KV are
referred to as GHV and GKV respectively. The set of CKVs admitted by a spacetime
$(M,g)$ form, under the usual Lie bracket operation, a Lie algebra of vector
fields which we shall designate as $\mathcal{C}_r(M,g)$, $r$ being its dimension.
Also the SCKV, HV and KV admitted by $(M,g)$ form Lie algebras designated as
$\mathcal{S}_r(M,g)$, $\mathcal{H}_r(M,g)$ and $\mathcal{G}_r(M,g)$, respectively.
Similar notations are used to designate the corresponding Lie algebras for the
spacetime $(M,\hat g)$ and the 3-space $(V,h)$. It follows that in any given
spacetime (or 3-space) $\mathcal{C}_r \supseteq \mathcal{S}_m \supseteq \mathcal{H}_s
\supseteq \mathcal{G}_n$, with $r\geq m \geq s \geq n$.
We refer the reader to \cite{HallSteele} for further details on CKV
and their Lie algebra.

The invariant characterisation of $(M,{\hat g})$ in terms of the existence of a non-null
covariantly constant vector field provides an invariant characterisation of $(M,g)$
which was given in \cite{Ramos} by the following theorem:

\begin{thm}The necessary and sufficient condition for $(M,g)$ to
be conformally related to a reducible 1+3 spacetime
$(M,\hat g)$ is that it admits a non-null, nowhere vanishing global conformal
Killing vector (CKV) $\vec X$ which is hypersurface
orthogonal.\label{theorem1}\end{thm}

We note that warped spacetimes of class A, as defined in \cite{CarotdaCosta}
and \cite{Ramos}, are particular instances of conformally reducible 1+3
spacetimes.

If $(M,g)$ is a conformally reducible 1+3 spacetime its conformal algebra
will be the same as that of the underlying reducible 1+3 spacetime
$(M,\hat g)$ thus providing a classification for conformally reducible
1+3 spacetimes. An investigation of the conformal algebra of $(M,\hat g)$
was carried out in \cite{Ramos}. We will not repeat the details of that
investigation here, but we will repeat the resulting theorem
(Theorem 10 of \cite{Ramos}) which gives results that are crucial to the
present investigation. However, the theorem presented here differs from that
in \cite{Ramos} in that in parts (2) and (3) we have added comments
pertaining to the conformally flat spaces and in part (4) we have a more
general result than the corresponding part in \cite{Ramos}.

\begin{thm} \label{thm:four}
 Let $(M,\hat g)$ a reducible 1+3 spacetime;
the following results hold regarding its conformal Lie algebra:
\begin{enumerate}
\item If $(V,h)$ admits no GCKV then the only CKV that $(M,\hat
g)$ admits are HV and KV.
\item $(M,\hat g)$ can admit a proper CKV $\vec Y$ if and only if $(V,h)$
admits a GCKV, which can be either a GKV $\vec \xi$, or a GHV
$\vec \gamma$, or a (proper) GCKV $\vec\zeta$.
However, a non-special proper CKV exists only if $(M,\hat g)$ is a conformally flat
or conformally reducible 2+2 spacetime (see case 4 below).
\item Two or more GCKV are admitted by $(V,h)$ if and only if $(V,h)$ is
of constant curvature (hence conformally flat) and it is flat if
one of the  GCKV admitted is a GHV. The resulting reducible 1+3
spacetime $(M, \hat g)$ is conformally flat also. The converse also
holds: given a conformally flat reducible 1+3 spacetime, the
three-dimensional subspace $(V, h)$ is necessarily of constant
curvature.
\item If only one GCKV is admitted by $(V,h)$, then $(M,\hat g)$ may admit a proper CKV or SCKV
which is unique up to the addition of HV and the following
situations may arise:
\begin{enumerate}
\item If it is a GKV $\vec \xi$ then, it must be null
else $(M,\hat g)$ degenerates into a 1+1+2 reducible spacetime \cite{Ramos}. Thus $\vec \xi$ is a covariantly
constant null KV and $(M,\hat g)$ is a pp-wave spacetime specialised to a 1+3 spacetime with metric
(see section 35.1 of \cite{KSMH2})
\begin{equation}\label{wth0}
  ds^2 = d\eta^2 -2dudv - 2H(u,x) du^2 + dx^2
\end{equation}
and the null KV $\vec \xi = \partial_v$.
Note that this metric is of isometry class 1(i) in \cite{keanetupper} where it is shown that no member
of this class can admit a non-special proper CKV. However, such a spacetime can admit a proper SCKV,
$\vec Y$, iff the metric function $H(u,x)$ and two other functions $f(u)$, $g(u)$, can be found
satisfying the differential equation
\newline
\ba
\fl (\rho u^2 + \alpha u + \beta) H(u,x)_{,u} + [\case{1}{2} x (2 \rho u + \alpha + k) + f(u)] H(u,x)_{,x}
\nonumber\\
+ x f(u)_{,uu} + g(u)_{,u} + (2 \rho u + \alpha - k) H(u,x)) = 0
\label{eq:ppwavecase1icondition}
\ea
where $\rho$, $\alpha$, $\beta$, $k$ are constants and $\rho \ne 0$ for a proper SCKV to exist
(see \cite{keanetupper} for details). If $\rho \ne 0$, a translation along the $u$-axis and a rescaling
of the $u$, $v$ coordinates changes (\ref{eq:ppwavecase1icondition}) into the form
\ba
& & (u^2 + \sigma) H(u,x)_{,u} + [x(u + \case{1}{2} k) +f(u)] H(u,x)_{,x}
\nonumber\\
& & + x f(u)_{,uu} + g(u)_{,u} + (2u - k) H(u,x) = 0.
\label{eq:ppwavecase1icondition2}
\ea
where $\sigma$ is an arbitrary constant. The SCKV $\vec Y$ is given by
\ba
\vec Y & = & \eta (u + \case{1}{2}k) \partial_\eta + (u^2 + \sigma) \partial_u
\nonumber\\
& & + [kv + \case{1}{2} x^2 + \case{1}{2} \eta^2 + x f(u)_{,u} + g(u)] \partial_v
\nonumber\\
& & +[(u + \case{1}{2} k)x + f(u)] \partial_x
\label{wth01}
\ea
with $\phi = u + \case{1}{2} k$.
The corresponding metric and SCKV $\vec X$ in $(V, h)$ are
\ba
d\Sigma^2 & = & -2dudv - 2H(u,x) du^2 + dx^2.
\\
\vec X & = & (u^2 + \sigma) \partial_u +[(u + \case{1}{2} k)x + f(u)] \partial_x.
\nonumber\\
& & + [kv + \case{1}{2} x^2 + x f(u)_{,u} + g(u)] \partial_v .
\label{eq:thm4asckvonvh}
\ea
The SCKV $\vec Y$ can be written as
\[
\vec Y = \eta (u + \case{1}{2}k) \partial_\eta + \case{1}{2} \eta^2 \vec \xi + \vec X.
\]
If in (\ref{eq:ppwavecase1icondition}), $\rho = 0$, then no proper SCKV exist, only HV and, possibly, KV.

\item If it is a GHV then $(M,\hat g)$ admits a proper
SCKV $\vec Y$ and if it is a GCKV then $(M,\hat g)$ admits a
proper CKV $\vec Y$. In each case $\vec Y$ is unique up to
the addition of HV. The metric in each of these cases is given by
\ba
ds^2 = \epsilon_0 d\eta^2+\epsilon_1 du^2 +
M^2(u)\Omega^2(v,x) \left[\epsilon_2 dv^2 + dx^2 \right]
\nonumber\\
\epsilon_\alpha = \pm 1 \: \: (\alpha=0,1,2) ,
\: \:
\epsilon_0\epsilon_1\epsilon_2 = -1 ,
\: \:
\epsilon_0+\epsilon_1+\epsilon_2 = +1 ,
\label{wth3}
\ea
where $M(u) = u$ in the GHV case while in the GCKV case, depending
on the sign of $\epsilon_\alpha$, $M(u)$ is given by one of the four functions
\begin{equation}\label{M(w)}
\sin ku, \;\; \sinh ku, \;\; \cosh ku, \;\; e^{ku} \;\; (k \ne 0)
\end{equation}
which coincide with the expressions found in \cite{ColeyTupper}, apart from the
expression $M(u) = e^{ku}$ which was omitted in \cite{ColeyTupper}. Note that the
metric (\ref{wth3}) is a conformally reducible 2+2 spacetime and so has been dealt
with in \cite{CarotTupper}.
\end{enumerate}
\end{enumerate}
\label{thconformal1}\end{thm}

In case (3) in the above theorem, namely; whenever  $(M, \hat g)$ is
conformally flat, $ \mathcal{C}(M, \hat{g})$ is 15-dimensional and
is the same as that of Minkowski spacetime. This case is also
contained in the study carried out in \cite{CarotTupper} and the
reader is referred there, where expressions for the metric and
generators of the conformal algebra are given in different
coordinate gauges.

Thus the cases of interest in Theorem \ref{thm:four} are case (1) in
which $(M, \hat{g})$ admits only HV and KV, as is always so for a
reducible 2+2 spacetime, and case (4a), the only case in which
$(M, \hat{g})$ admits a proper SCKV without degenerating into a
conformally flat or conformally reducible 2+2 spacetime. This latter
case, i.e.: (4a), occurs iff $(V, h)$ admits a null GKV and a
SCKV, in which case $(M, {\hat g})$ is a pp-wave spacetime.

In section \ref{sec:fixedpoints} we briefly deal with with those
reducible 1+3 spacetimes admitting KVs and HVs with fixed points.
In the subsequent sections we will
tacitly restrict to the cases in which the KVs or HVs in the
reducible 1+3 spacetime have no fixed points (or at least one of the
KVs or HVs has no fixed points, so that one could choose coordinates adapted to it).
We investigate the possible solutions
that can arise in case (1) of Theorem \ref{thm:four}, i.e., when
$(V,h)$ admits KVs or HVs and we also give examples of the pp-wave
case (4a). In section \ref{sec:ih} we obtain the canonical metrics
admitting maximal $\mathcal{G}_r(V,h)$ and the corresponding
reducible 1+3 spacetimes. In section \ref{sec:homothetygroups}
we obtain the canonical metrics admitting maximal $\mathcal{H}_r(V,h)$
and the corresponding reducible 1+3 spacetimes. Finally, in section
\ref{sec:discussion} some examples of reducible and conformally reducible 1+3
spacetimes are presented.

\section{Fixed points of KV and HV in 1+3 spacetimes}
\label{sec:fixedpoints}

Here we will turn our attention briefly
to the cases in which the spacetime admits KVs and/or HVs which do
have fixed points. We shall not attempt to give detailed coordinate
expressions for the metric and the relevant vector field admitting a
fixed point, as this would be tedious and lengthy, in any case this
can be done  following the same methods as in \cite{CarotTupper}.
Instead we shall focus on the  general results that can be mostly
gathered from \cite{HallLowPulham}.

Given any vector field $\vec X$ on a manifold $M$, a point $p\in M$
is said to be a \emph{fixed point} of $\vec X$ iff $\vec X (p)= 0$.
This is equivalent to saying that the 1-parameter group of (local)
diffeomorphisms  $\{\varphi_t \}$ that $\vec X$ generates is such
that $\varphi_t (p) = p$ for all values of $t$ where this makes
sense.

We shall be interested in the case in which $\vec X$ is an affine
vector field (AVF), KVs and HVs being then special instances of
that, and therefore it satisfies
$$ X^a_{;\, bc} = R^{a}_{\; bcd} X^d$$ or equivalently, decomposing
$X_{a;b}$ into is symmetric and skew-symmetric parts $h_{ab}$ and
$F_{ab}$ respectively $$ X_{a;b} = h_{ab} + F_{ab}, \qquad
h_{ab;c}=0, \quad F_{ab;c}= R_{abcd} X^d$$ If $h_{ab} = k g_{ab}$
with $k =$ constant then $\vec X$ is a HV (KV if $k=0$).

If $\vec X (p)= 0$ then $$\varphi_{t*}:\, T_pM \rightarrow T_pM
\qquad \mathrm{and \;is \;such \; that} \qquad \varphi_{t*} =
\exp{tA}$$ where $A$ denotes a matrix whose elements are $A^a_{\, b}
= X^{a}_{\; ,b} (p)$. Using then normal coordinates $x^a$ in
neighbourhood of $p$ and since the exponential map $\chi$ satisfies
$\chi \circ \varphi_{t*} = \varphi_{t} \circ \chi$ it follows that
$X^a = A^a_{\, b} x^b$; that is: the components of $\vec X$ are
linear functions of the coordinates in a neighborhood of $p$. This
allows classifying and studying the properties of the set of fixed
points of $\vec X$ as well as finding coordinate expressions for
both the metric and $\vec X$ as it was done in  \cite{CarotTupper}.

We shall next consider the case of a 1+3 decomposable spacetime
$(M,g)$ admitting  an AVF $\vec X$  which has a fixed point. We
shall distinguish between the cases 1+3 spacelike and 1+3 timelike
decomposable, see  \cite{HallLowPulham} for details.

\subsection{1+3 spacelike decomposable.}

Suppose the line element of $(M,g)$ is given in coordinates $x^a =
\eta, x^B$, $B=1,2,3$ by
\begin{equation}\label{spacelikedecomp}
    ds^2 = -d\eta^2 + h_{AB}(x^C) dx^A dx^B
\end{equation}
Clearly, $\vec \eta = \partial_\eta$ is a nowhere vanishing covariantly
constant KV such that $\eta_{a} = \eta_{,a}$. Suppose that an AVF $\vec X$
exists which has fixed points and decompose it as   $\vec X= -
\kappa \vec \eta + \vec k$, where $k_a = (g_{ab} + \eta_a \eta_b) X^b$ is a
vector field tangent everywhere to $(V,h)$. It is then immediate to
see that
\begin{equation}\label{k1}
    k_{a;b} = \alpha (g_{ab} +  \eta_a \eta_b) + F_{ab},
\end{equation}
where $\alpha \in \mathbb{R}$ and $F_{ab}$ is the affine bivector (i.e.: $F_{ab} = X_{[a;b]}$),
hence $\vec k$ is a HV (KV if $\alpha =0$) of $(V,h)$.

Now, $\vec X(p)=0$ implies $\kappa (p) \vec \eta (p) = \vec k(p) =0$,
and it then follows that the only non-trivial possibility compatible
with $\vec X$ being either a KV or a HV is that $\vec X = \vec k$ is
a KV (this corresponds to $\alpha=\beta=\gamma=0$ and $F^a_{\, b}
(p) \neq 0$), the set of its fixed points is then a 2-dimensional
submanifold (since $F^a_{\, b} (p) \neq 0$ and is necessarily simple
as $\vec k$ is tangent to $(V,h)$)  through $p$ containing the
integral curve of $\vec \eta$ that passes through that point. Other
non-trivial possibilities exist but they all correspond to $\vec X$
being a proper AVF.

\subsection{1+3 timelike decomposable.}
Choose coordinates $x^a = \eta, x^B$ such that
\begin{equation}\label{timelikedecomp}
    ds^2 = d\eta^2 + h_{AB}(x^C) dx^A dx^B
\end{equation}
that is: $(V,h)$ is now Lorentz. Equation (\ref{k1}) and all of
the comments in the previous subsection still hold writing now
$\vec X = \kappa \vec \eta + \vec k$. In this case
there exist some more possibilities for an AVF $\vec X$ with a
fixed point, but only two of them give rise to KV or HV, the rest of
them corresponding to $\vec X$ being a proper AVF. The ones we are
interested in are:
\begin{enumerate}
  \item  $\alpha=\beta=\gamma=0$ and $\vec X = \vec
  k$ is then a KV everywhere tangent to $(V,h)$ whose set of  fixed points is
  a 2-dimensional submanifold through $p$ containing
  the integral curve of $\vec \eta$ that passes through that point.
  \item $\alpha \neq 0$, $\beta = 0$ and $F^a_{\; b} (p) \neq  0$
  (necessarily timelike). Then $\vec k \neq 0$ is a proper HV on
  $(V,h)$ vanishing at $p$ (which is then necessarily isolated in
  $(V,h)$). $\vec X$  is a HV and has an isolated zero at
  $p$ in the hypersurface $\eta= -\gamma/\alpha$.

\end{enumerate}

\section{Isometry Groups on $(V,h)$}
\label{sec:ih}
We wish to enumerate all relevant isometry groups on three-dimensional manifolds $(V,h)$ of either signature.
Isometry groups will be denoted as ${\mathcal G}_r$, $r$ being as usual the
dimension of the group (and that of the associated algebra). Note that a ${\mathcal G}_r$ on $(V,h)$ will
lead to a ${\mathcal G}_{r+1}$ on $(M, {\hat g})$ on account of the existence of the KV $\partial_\eta$.
However, in the case of a pp-wave with $(V,h)$ conformally flat, but not of constant curvature,
a ${\cal G}_r$ on $(V,h)$ leads to a ${\cal G}_{r+2}$ on $(M, {\hat g})$.
We denote by a slash (\lq$/\,$') the covariant derivative with respect to the
three-metric $h$.

The isometry groups ${\mathcal G}_r$ admitted by three-dimensional spaces $(V,h)$ of
either signature are as follows. We first consider the case with a ${\cal G}_1$ acting on null
orbits and then a ${\cal G}_1$ acting on non-null orbits.
Developing this, we consider ${\cal G}_2$ acting on non-null orbits, and ${\cal G}_2$ acting on null orbits
(there are three distinct cases, see section \ref{eq:G2ofisometries}).
When we come to consider the ${\cal G}_3$ algebras we must distinguish between those which
admit a ${\cal G}_2$ subalgebra and those which do not. Those ${\cal G}_3$ that do admit a ${\cal G}_2$
subalgebra can be derived from those metrics admitting an abelian ${\cal G}_2$ or non-abelian ${\cal G}_2$
structure, with either null or non-null orbits.
Finally, every ${\cal G}_4$ admits a ${\cal G}_3$ subalgebra and so every 3-space admitting a ${\cal G}_4$
structure can be derived from an appropriate 3-space metric admitting a ${\cal G}_3$.

The isometry structures on $(V,h)$ can be deduced from Petrov's work (chapter 5 of \cite{Petrov}) on
four-dimensional manifolds. However, this needs to be refined for the case of reducible 1+3 spacetimes.
${\cal G}_3$ on two-dimensional null orbits are also examined in detail in Barnes \cite{barnes79}.
The ${\cal G}_3$ acting transitively on $(V,h)$ correspond to the Bianchi types, see \cite{KSMH2}.
The group actions on three-dimensional null orbits are irrelevant for a reducible 1+3 spacetime.
There are no ${\cal G}_5$ groups with three-dimensional orbits (Fubini's Theorem, Theorem 57.1 of
\cite{eisenhart33}, Hall \cite{hall03}).
Of course, if a $(V,h)$ admits a ${\cal G}_6$ then it is necessarily a space of constant curvature and the
corresponding 1+3 spacetime will be conformally flat and need not be considered here.
The Lie algebras are shown in table \ref{tab:liealgebras}.
\begin{table}[h]
\caption{The Lie algebra structures taken from section 10 of \cite{Petrov}. The quantities $p$ ($p^2 < 4$), $q$ ($q \ne 0$, $q \ne 1$), $s$ and $c_i^\alpha$ are real constants, where $i = 1,2,3$, $\alpha = 1,2,3,4$. It can be seen by inspection that every 4-dimensional algebra admits a 3-dimensional subalgebra with basis $\vec X_1$, $\vec X_2$, $\vec X_3$.}
\footnotesize\rm
\begin{tabular*}{\textwidth}{@{}l*{15}{@{\extracolsep{0pt plus12pt}}l}}
\br
$r$ & $Algebra$ & $[\vec X_1, \vec X_2]$ & $[\vec X_1, \vec X_3]$ & $[\vec X_2, \vec X_3]$
& $[\vec X_1, \vec X_4]$ & $[\vec X_2, \vec X_4]$ & $[\vec X_3, \vec X_4]$\\
\mr
2 & $I$ & 0 & - & - & - & - & - \\
2 & $II$ & $\vec X_1$ & - & - & - & - & - \\
3 & $I$ & $0$ & $0$ & $0$ & - & - & - \\
3 & $II$ & $0$ & $0$ & $\vec X_1$ & - & - & - \\
3 & $III$ & $0$ & $\vec X_1$ & $0$ & - & - & - \\
3 & $IV$ & $0$ & $\vec X_1$ & $\vec X_1 + \vec X_2$ & - & - & - \\
3 & $V$ & $0$ & $\vec X_1$ & $\vec X_2$ & - & - & - \\
3 & $VI_q$ & $0$ & $\vec X_1$ & $q \vec X_2$ & - & - & - \\
3 & $VII_p$ & $0$ & $\vec X_2$ & $-\vec X_1 + p \vec X_2$ & - & - & - \\
3 & $VIII$ & $\vec X_1$ & $2\vec X_2$ & $\vec X_3$ & - & - & - \\
3 & $IX$ & $\vec X_3$ & $-\vec X_2$ & $\vec X_1$ & - & - & - \\
4 & $I_s$ & $0$ & $0$ & $\vec X_1$ & $s \vec X_1$ & $\vec X_2$ & $(s-1) \vec X_3$ \\
4 & $II$ & $0$ & $0$ & $\vec X_1$ & $2 \vec X_1$ & $\vec X_2$ & $\vec X_2 + \vec X_3$ \\
4 & $III_p$ & $0$ & $0$ & $\vec X_1$ & $p \vec X_1$ & $\vec X_3$ & $-\vec X_2 + p\vec X_3$ \\
4 & $IV$ & $0$ & $0$ & $\vec X_2$ & $\vec X_1$ & $0$ & $0$ \\
4 & $V$ & $0$ & $\vec X_1$ & $\vec X_2$ & $\vec X_2$ & $-\vec X_1$ & $0$ \\
4 & $VI$ & $0$ & $0$ & $0$ & $c^\alpha_1 \vec X_\alpha$ & $c^\alpha_2 \vec X_\alpha$ & $c^\alpha_3 \vec X_\alpha$ \\
4 & $VII$ & $\vec X_1$ & $2 \vec X_2$ & $\vec X_3$ & $0$ & $0$ & $0$ \\
4 & $VIII$ & $\vec X_3$ & $-\vec X_2$ & $\vec X_1$ & $0$ & $0$ & $0$ \\
\br
\end{tabular*}
\label{tab:liealgebras}
\end{table}
\begin{table}[h]
\caption{Canonical metric forms for $(M, {\hat g})$ from $(V, h)$ admitting ${\cal G}_1$. The term hypersurface orthogonal is abbreviated to $h.o.$. $\epsilon_0 = \pm 1$ and $\epsilon_1 = \pm 1$ are not both negative.}
\footnotesize\rm
\begin{tabular*}{\textwidth}{@{}l*{15}{@{\extracolsep{0pt plus12pt}}l}}
\br
Algebra & $(V, h)$ & Spacetime metric/${\cal G}_2$ basis & Condition\\
\mr

${\cal G}_1$ on $N_1$ & $(\ref{eq:g1nullorbitsfinalmetric})$ & $ds^2 = \epsilon_0 d\eta^2
+ P^{-2}(u,x) dx^2$ & $\epsilon_0 w^2 > 0$ \\
& & $\hspace{10 pt} - 2 du [w(u,x) dv + H(u,x) du]$ & \\
& & $\vec X_0 = \partial_\eta$, $\vec l = \partial_v$ & \\

${\cal G}_1$ on $S_1$ or $T_1$ & $(\ref{metricnonnullHV2a})$ & $ds^2 = \epsilon_0 d\eta^2
+ H(u,w) (\epsilon_1 du^2 + \epsilon_2 dw^2)$ &
$\epsilon_0 \epsilon_1 H(u,w) $ \\
($\vec X$ is not $h.o.$) & & $ \hspace{10 pt} + 2K(u,w) dv dw + L(u,w) dv^2$ & $\hspace{10 pt} [K^2(u,w) - \epsilon_2 L(u,w)H(u,w)] > 0$ \\
& & $\vec X_0 = \partial_\eta$, $\vec X = \partial_v$ & \\

${\cal G}_1$ on $S_1$ or $T_1$ & $(\ref{eq:confflat2space})$ & $ds^2 = \epsilon_0 d\eta^2
+ A(u,w) du^2$ &
$\epsilon_0 A(u,w)B(u,w)C(u,w) < 0$ \\
($\vec X$ is $h.o.$) & & $\hspace{10 pt}+ B(u,w) dv^2 + C(u,w)dw^2$ & \\
& & $\vec X_0 = \partial_\eta$, $\vec X = \partial_v$ & \\
\br
\end{tabular*}
\label{tab:g1}
\end{table}
Any conformally reducible 2+2 spacetimes obtained in this analysis can be discarded since they
are treated in \cite{CarotTupper}.

We will treat separately the cases of null orbits and non-null orbits.
The results are summarized in table \ref{tab:g1}.

\subsection{$(V,h)$ admits a  ${\mathcal G}_1$ on null orbits.}\label{nullG1}
Let $\vec l$ be a null KV; i.e.: $l^Al_A =0$ and put $l_{A;B} =
F_{AB}$ its associated Killing bivector. Choose two other vector
fields to complete a null triad, say $\{\vec l,\vec n,\vec x\}$
such that: $l^Al_A = n^An_A = 0$, $l^An_A=-1$, $x^Ax_A=1$ and the
remaining products are zero.

From $(l^Al_A)_{;B} =0$ it follows $l^AF_{AB} =0$ and therefore
\be F_{AB} = \alpha \left(l_A x_B -x_A l_B\right) \Rightarrow
l_{[A}l_{B;C]} =0\ee
that is: $\vec l$ is a null KV which is hypersurface orthogonal
and geodesic (the latter follows from the above form for the
Killing bivector).

It then follows that one may choose coordinates $v,u,x$ so
that
\ba \vec l = \partial_v, \qquad l_A dx^A = -w(u,x) du
\label{eq:g1nullkv}\\
d\sigma^2 = P^{-2} dx^2 -2 du\left(w dv -m dx + H du\right)
\ea
where the metric functions $P,w,m,H$ depend on  $u,x$. (See
\cite{KSMH2}, p 380).

Notice that one can still perform the following coordinate changes that preserve the form of $\vec l$ and $l_A dx^A $,
namely: $v=v', u= u'$ and $x=f(u',x')$; by performing one such coordinate change, the line element reads (dropping primes for convenience):
\ba
d\sigma^2 & = & P^{-2}f_{,x}^2 dx^2 +2 f_{,x}\left(m+f_{,u}P^{-2}\right)du dx
\nonumber\\
& & + \left(P^{-2}f_{,u}^2 +2m f_{,u}-2H\right)du^2 -2w du dv
\ea
and it follows that $f$ can be chosen so that $m+f_{,u}P^{-2}=0$.
Rewriting $2H + m^2P^2$ as $2H$ the metric becomes
\be
d\sigma^2 = P^{-2}(u,x) dx^2 -2 du \left[ w(u,x) dv  + H(u,x) du \right]
\label{eq:g1nullorbitsfinalmetric}
\ee
and the null KV is given by equation (\ref{eq:g1nullkv}).

The 3-space $(V, h)$ with metric (\ref{eq:g1nullorbitsfinalmetric}) admits only
the KV $\vec l = \partial_v$ for general functions $P(u,x)$, $w(u,x)$ and $H(u,x)$
and the corresponding spacetime $(M, {\hat g})$ admits a ${\cal G}_2$ only. However,
if $\vec l$ is covariantly constant then $w(u,x)$ is a constant that can be
rescaled to unity and $(M, {\hat g})$ is a 1+3 pp-wave spacetime with metric given by
equation (\ref{wth0}). In this case, although $(V, h)$ admits only a ${\cal G}_1$,
the corresponding $(M, {\hat g})$ admits a ${\cal G}_3$, the additional KVs being
\[
\vec X_2 = \partial_\eta, \qquad \vec X_3 = \eta \partial_v + u \partial_\eta.
\]
If, in addition, functions $f(u)$, $g(u)$ can be found such that $H(u,x)$ satisfies
equation (\ref{eq:ppwavecase1icondition}) with $\rho \ne 0$, i.e.,
equation (\ref{eq:ppwavecase1icondition2}), then $(V, h)$ admits a ${\cal S}_2$ with the SCKV
given by equation (\ref{eq:thm4asckvonvh}) and $(M, {\hat g})$ admits a
${\cal S}_4 \supset {\cal G}_3$. However, if $\rho = 0$ then $(V, h)$ admits a
${\cal H}_2$ and $(M, {\hat g})$ admits a ${\cal H}_4 \supset {\cal G}_3$.
Thus the condition that $\vec l$ is covariantly constant implies that the dimension
of the conformal algebra of $(M, {\hat g})$ is two more than that of $(V,h)$, as
mentioned earlier.

\subsection{$(V,h)$ admits a  ${\mathcal G}_1$ on non-null orbits.}
\label{sec:g1onnonnullorbits}
Let $\vec X$ be a non-null KV, i.e.,
\begin{equation}\label{HVnonnull1}
    X_{A/B} =  F_{AB},
\end{equation}
where $F_{AB} = -F_{BA}$ is the Killing bivector. There are two possibilities:
\begin{enumerate}

  \item $\vec X$ is not hypersurface orthogonal (h.o.).
  Assuming one is not at a fixed point,
  one can choose an adapted coordinate system, say $u,v,w$ such that
  $\vec X = \partial_v$, and the metric reads then of the form
  \begin{equation}
    d\sigma^2 = h_{AB}(u,w) dx^A dx^B.
  \end{equation}
  The coordinate change $v' \mapsto v + f(u,w)$ allows one to set
  $h'_{vw} = 0$ without altering the form of the KV $\vec X$ and a
  transformation of the form $u = g(u',w')$, $w = h(u',w')$ can be
  used to render the 2-metric in the $u$, $w$ plane in an explicitly
  conformally flat form. Thus, dropping the primes, the metric of
  $(V,h)$ can be written as
  \begin{equation}\label{metricnonnullHV2a}
    d\sigma^2 =  H(u,w) (\epsilon_1 du^2 + \epsilon_2 dw^2)
  + 2K(u,w) dv dw + L(u,w) dv^2
  \end{equation}

  \item $\vec X$ is h.o., i.e. $X_{[A}X_{B/C]} = 0$. Then, using the metric
  (\ref{metricnonnullHV2a}), we find that
  \[
    6 X_{[A}X_{B/C]} =  (L K_{,u} - K L_{,u}) \epsilon_{ABC}
  \]
  with $\epsilon_{uvw} = +1$. Thus $(L K_{,u} - K L_{,u}) = 0$ which implies that
  $K(u,w) = f(w) L(u,w)$ for some function $f(w)$.
  Substituting this into metric (\ref{metricnonnullHV2a}), making the transformation
  $v = v' - \int f(w) dw$, and dropping the primes, the metric of $(V,h)$  takes the
  diagonal form
  \begin{equation}\label{eq:confflat2space}
    d\sigma^2 = A(u,w) du^2 + B(u,w) dv^2 + C(u,w)dw^2.
  \end{equation}
\end{enumerate}

\subsection{$(V,h)$ admits a group ${\mathcal G}_2$ of isometries.}
\label{eq:G2ofisometries}

A group ${\mathcal G}_2$ can act on two-dimensional spacelike ($S_2$), timelike ($T_2$) or null ($N_2$) orbits,
and in each case there exist two inequivalent structures; namely
\be
{\mathcal G}_2I: \quad [\vec X,\vec Y] = 0
\qquad {\mathcal G}_2II:\quad [\vec X,\vec Y] = k \vec X,
\label{eq:G2brackets}
\ee
where $\vec X,\vec Y$ are two independent KVs generating the group and $k$ is a constant that can be set equal
to unity without loss of generality.

There are four sub-cases $(a) - (d)$ according to the nature of the ${\cal G}_2$ orbits.
Type $(a)$ has non-null orbits $S_2$ or $T_2$, types $(b) - (d)$ have null orbits.
Type $(b)$ contains a subgroup ${\cal G}_1$ generated by a null KV, type $(c)$ has no null
subgroup ${\cal G}_1$, with a null vector orthogonal to the orbits of the ${\cal G}_2$,
and type $(d)$ has no null subgroup ${\cal G}_1$, with a null vector in the orbits of the ${\cal G}_2$.
Some of the $(V,h)$ metrics found in this section correspond to pp-wave $(M, {\hat g})$
spacetimes and, in general, such $(V,h)$ admit a ${\cal G}_r$ with $r>2$. However, they are
included here because they arise naturally in our search for $(V,h)$ admitting a ${\cal G}_2$.

This section is organised in the following way: The abelian and non-abelian cases are dealt with
separately, considering each sub-case $(a) - (d)$ as follows:
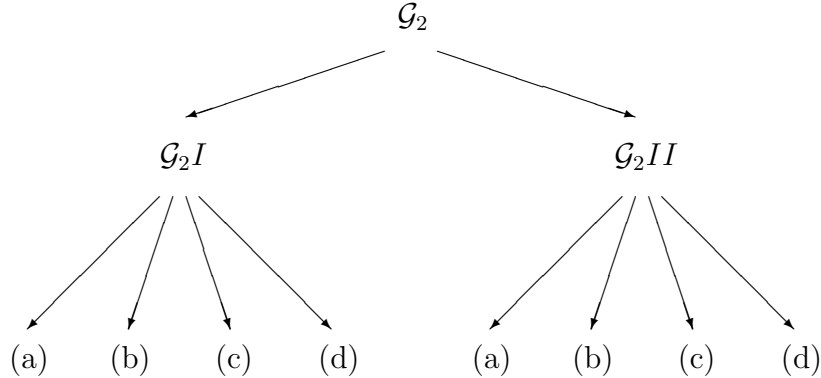
\begin{center}{
\begin{picture}(120,145)
\put(70,140){${\cal G}_2$}

\put(65,130){\vector(-3,-1){75}}
\put(-20,87){${\cal G}_2 I$}
\put(85,130){\vector(3,-1){75}}
\put(152,87){${\cal G}_2 II$}

\put(-20,75){\vector(-1,-1){50}}
\put(-15,75){\vector(-1,-3){16.75}}
\put(-10,75){\vector(1,-3){16.75}}
\put(-5,75){\vector(1,-1){50}}
\put(-77,10){(a)}
\put(-39,10){(b)}
\put(1,10){(c)}
\put(40,10){(d)}

\put(155,75){\vector(-1,-1){50}}
\put(160,75){\vector(-1,-3){16.75}}
\put(165,75){\vector(1,-3){16.75}}
\put(170,75){\vector(1,-1){50}}
\put(98,10){(a)}
\put(136,10){(b)}
\put(176,10){(c)}
\put(215,10){(d)}

\end{picture}
}
\end{center}

\subsubsection{The abelian case, ${\mathcal G}_2 I$}\label{G2abelian}
In the following three sub-cases (a), (b) and (c) we may choose coordinates such that $\vec X = \partial_v$,
$\vec Y = \partial_w$, so that the metric of $(V, h)$ is of the form
\ba
d \sigma^2 & = & h_{uu} (u) du^2 + 2 h_{uv} (u) du dv + 2 h_{uw} (u) du dw
\nonumber\\
& & + h_{vv} (u) dv^2 + 2 h_{vw} (u) dv dw + h_{ww} (u) dw^2 \, .
\label{eq:generalAbelian}
\ea

\subsubsection*{(a) $(V,h)$ admits a group ${\mathcal G}_2 I$ acting on non-null orbits $S_2$ or $T_2$.}\label{G2onV2}
We can use a coordinate transformation of the form
\be
v \mapsto v + f(u), \qquad w \mapsto w + g(u)
\label{eq:coordTransformationsCase1}
\ee
to eliminate the $h_{uv}$ and $h_{uw}$ terms from the metric (\ref{eq:generalAbelian}) provided
that $h_{vv} h_{ww} - (h_{vw})^2 \ne 0$. Rescaling the $u$-coordinate the metric takes the form
\be
d\sigma^2 = \epsilon_1 du^2 + A(u) dv^2 + 2B(u) dv dw + C(u) dw^2 \, .
\label{eq:G2Abeliancasea}
\ee
If one of the KVs is h.o. then, by an argument similar to that in section
\ref{sec:g1onnonnullorbits}, the metric of $(V,h)$ can be diagonalized, i.e.,
\be
d \sigma^2 = \epsilon_1 du^2 + A(u) dv^2 + C(u) dw^2.
\label{eq:g2abelianadiagonal}
\ee
Note that, in equation (\ref{eq:G2Abeliancasea}), if $A(u) = 0$ (equivalently $C(u) = 0$),
then $\vec X_1 = \partial_v$ (equivalently $\vec X_2 = \partial_w$) is a
non-covariantly constant null vector and $\vec X_1$, $\vec X_2$ are KVs
(see section \ref{sec:discussion}). If, in addition, $B(u)$ in equation
(\ref{eq:G2Abeliancasea}) is a constant then $\vec X_1 = \partial_v$ is covariantly
constant and $(V,h)$ is a 3-dimensional pp-wave spacetime admitting a ${\cal G}_2 I$
acting on non-null orbits, and so does not contradict the statement in (b) below.
The coordinate change $u \mapsto x$, $w \mapsto -u$ transforms the metric
of $(V,h)$ into
\be
d \sigma^2 = -2 H(x) du^2 - 2 du dv + dx^2.
\label{eq:g2abeliana3dimppwave}
\ee
The corresponding spacetime $(M, {\hat g})$ is an isometry class 8 ($\epsilon = 0$)
pp-wave solution \cite{keanetupper} admitting a ${\cal G}_4$ with basis
\be
\vec X_1 = \partial_v, \qquad \vec X_2 = \partial_u, \qquad
\vec X_3 = \partial_\eta, \qquad \vec X_4 = \eta \partial_v + u \partial_x.
\label{eq:g2abeliana4dimppwavebasis}
\ee
In the special case when $H(x) = x^n$, $n = constant$, there is an additional HV given by
\be
\vec H = \case{1}{2} (2 - n) u \partial_u + \case{1}{2} (2 + n) v \partial_v
+ x \partial_x + \eta \partial_\eta
\label{eq:g2abeliana4dimppwavehv}
\ee
with $\psi = 1$ and $(M, {\hat g})$ admits a ${\cal H}_5 \supset {\cal G}_4$.
If $n = -2$ there is a further symmetry, namely a SCKV of the form
\be
\vec S = u^2 \partial_u + \case{1}{2} (x^2 + \eta^2) \partial_v
+ u x \partial_x + u \eta \partial_\eta
\label{eq:g2abeliana4dimppwavesckv}
\ee
so that $(M, {\hat g})$ admits a ${\cal S}_6 \supset {\cal H}_5 \supset {\cal G}_4$.
If $n = 2$, $(V,h)$ is the special case (iii) of the following subsection (b)
and admits a ${\cal H}_7 \supset {\cal G}_6$. In this case there exists a group
${\cal G}_2 I$ on null orbits, but $\vec X_2 = \partial_u$ does not lie on a null orbit.

\subsubsection*{(b) $(V,h)$ admits a group ${\mathcal G}_2 I$ acting on null orbits $N_2$ containing a subgroup ${\cal G}_1$ generated by a null KV}
\label{eq:G2IonnullorbitsN2G1generatedbynullKV}

This is clearly a special case of the one discussed in section \ref{nullG1}, but it is nevertheless an interesting
case to be analyzed. Let the null KV be $\vec l$ and suppose that $\vec X$ is another KV which generates
the group ${\mathcal G}_2$ along with $\vec l$. It is then easy to show that $\vec X$ must be spacelike and
orthogonal to $\vec l$ for if any of these two conditions failed, a second  null vector, independent of $\vec l$,
would exist at every point on the orbits thus contradicting the hypothesis that they are null.

Now, from $l^AX_A =0$ and Killing's equation for both $X_A$ and $l_A$ it follows
$$[\vec l,\vec X]^A = -2 l^A_{\, ;B} X^B$$
Recalling now section \ref{nullG1}, we may write $l_{A;B} = \alpha (l_A x_B - x_A l_B)$ where $x^Ax_A =1$
and $x^A l_A =0$. It is now easy to show\footnote{e.g.: choose a null triad $l_A,n_A,x_A$ such
that $-l_A n^A =x^A x_A =1$ and the rest of the products zero. Now $X_A = a l_A + cx_A$, and a
null rotation around $l_A$ can be used to set $X_A \propto x_A$.} that $\vec x$ can be chosen
parallel to the KV $\vec X$, say $\vec X = \lambda \vec x$. The above commutator gives then:
$$[\vec l,\vec X]^A = -2 \alpha \lambda l^A$$
and the two inequivalent Lie algebra structures arise then depending on the value of $ -2 \alpha \lambda$.
If it is non-zero, then it must be constant and we have a ${\cal G}_2 II$ structure which we discuss later.
If it is zero, which implies $\alpha =0$, then $\vec l$ is covariantly constant, in which case $(M, {\hat g})$
is a pp-wave spacetime with metric of the form (\ref{wth0}). However, rather than use this metric, it is more
convenient to note that the group is abelian (${\cal G}_2 I$) and we can set up coordinates
such that $\vec l = \partial_v$ and $\vec X = \partial_x$, then $l_A l^A = l_B X^B =0$ together
with the fact that they are KVs, leads immediately to the metric of $(V, h)$ in the form
\be
d\sigma^2 = P^{-2}(u) dx^2 - 2 \omega(u) du dv + 2 m(u) du dx - 2H(u) du^2.
\ee
Rescaling $u$ so that $\omega(u) = 1$ and changing $x \mapsto x - \int m(u) P^2(u) du$, replacing
$2H + m(u) P^2(u)$ by $2H(u)$ the metric becomes
\be
d\sigma^2 = P^{-2}(u) dx^2 - 2dudv - 2H(u) du^2,
\label{eq:g2Ibmetric2}
\ee
a form that will be useful later. The KVs $\vec l = \partial_v$ and $\vec X = \partial_x$ are
unchanged and the metric admits a third KV, namely $\vec Y = x\partial_v + u \partial_x$ and a
HV, $\vec Z = (2v + 2 \int H(u) du)\partial_v + x \partial_x$, and thus admits a
${\cal H}_4 \supset {\cal G}_3$ on orbits $N_2$. In fact, defining new coordinates by
\[
u \mapsto u, \qquad
x \mapsto P(u) x, \qquad
v \mapsto v - \int H(u) du + \case{1}{2} P^{-1}(u) P(u)_{,u} x^2
\]
the metric becomes
\be
d\sigma^2 = -A(u) x^2 du^2 - 2 du dv + dx^2,
\label{eq:g2Ibmetric3}
\ee
where $A(u) = P(u)^{-1} P(u)_{,uu} - 2 P^{-2}(u) (P(u)_{,u})^2$.
The corresponding reducible 1+3 spacetime $(M, \hat{g})$, i.e.,
\be
d \sigma^2 = -A(u) x^2 du^2 - 2 du dv + dx^2 + dy^2
\label{eq:isometryclass10pp-wave}
\ee
is a special case of an isometry class 10 pp-wave spacetime \cite{keanetupper}.
For arbitrary $A(u)$, $(M, \hat{g})$ admits a ${\cal H}_6 \supset {\cal G}_5$. The five KVs are
\ba
\vec X_1 & = & \partial_v,
\qquad
\vec X_2 = f(u)_{,u} x \partial_v + f(u) \partial_x,
\qquad
\vec X_3 = g(u)_{,u} x \partial_v + g(u) \partial_x,
\nonumber\\
\vec X_4 & = & y \partial_v + u \partial_y,
\qquad
\vec X_5 = \partial_y
\label{eq:isometryclass10pp-wavebasis}
\ea
where $f(u)$, $g(u)$ are independent solutions of the differential equation
\[
C(u)_{,uu} + A(u) C(u) = 0,
\]
and the HV is
\[
\vec H = 2 v \partial_v + x \partial_x + y \partial_y.
\]
For certain special forms of $A(u)$ there exists an additional symmetry.
From table 4 of reference \cite{keanetupper} we find that for a 1+3
pp-wave spacetime the only possible choices for $A(u)$ leading to
an additional symmetry are:

$(i)$ $A(u) = (u^2 + \delta)^{-2}$ which results in a SCKV given in
$(V,h)$ by
\[
\vec S = (u^2 + \delta) \partial_u + \case{1}{2} x^2 \partial_v + u x \partial_x
\]
and in $(M, {\hat g})$ by
\[
\vec S = (u^2 + \delta) \partial_u + \case{1}{2} (x^2 + \eta^2) \partial_v
+ u x \partial_x + u \eta \partial_\eta
\]
so that $(M, {\hat g})$ admits a ${\cal S}_7 \supset {\cal H}_6 \supset {\cal G}_5$.
This is an isometry class 10(ii) spacetime in the classification of \cite{keanetupper}.

$(ii)$ $A(u) = \alpha u^{-2}$, $\alpha = constant$, which results in a KV given
by $\vec Y = u \partial_u - v \partial_v$, so that $(V,h)$ admits a
${\cal H}_5 \supset {\cal G}_4$ and $(M, {\hat g})$, which is an isometry class
11 spacetime \cite{keanetupper}, admits a ${\cal H}_7 \supset {\cal G}_6$.

$(iii)$ $A(u) = \alpha$, constant, which results in the KV $\vec X = \partial_u$
so that $(V,h)$ admits a ${\cal H}_5 \supset {\cal G}_4$. $(M, {\hat g})$ admits
a ${\cal H}_7 \supset {\cal G}_6$ and is an isometry class 13 spacetime \cite{keanetupper}.

\subsubsection*{(c) $(V,h)$ admits a group ${\mathcal G}_2 I$ acting on null orbits $N_2$ with
no null subgroup ${\cal G}_1$, with a null vector orthogonal to the orbits of the ${\cal G}_2$.}
\label{sec:g2IonN2nonullsubgroup}

In this case the metric (\ref{eq:generalAbelian}) satisfies
\be
h_{vv} h_{ww} - (h_{vw})^2 = 0
\label{eq:degeneratecondition}
\ee
so that the $(v,w)$ space is degenerate. The coordinate transformations (\ref{eq:coordTransformationsCase1})
cannot be used to eliminate $h_{uv}$ and $h_{uw}$, but can be used to eliminate $h_{uu}$ and $h_{uv}$ to
obtain
\be
d \sigma^2 = 2 du dw + (A(u) dv + B(u) dw)^2 \, ,
\label{eq:G2Abeliancasec}
\ee
where $A(u) \ne 0$ for non-degeneracy.
We note that in this case there exists a null vector {\it orthogonal to the orbits of} the ${\cal G}_2$.
The null vector, $\partial_u$, is orthogonal to the null 2-space spanned by $v$ and $w$.

Note that if $B(u) = 0$ or if $B(u)$ and $A(u)$ are proportional to each other, a simple
coordinate change transforms the metric into precisely the form (\ref{eq:g2Ibmetric3}).
Thus $(M, {\hat g})$ is the special isometry class 10 pp-wave spacetime considered in detail in
case (b) above. In further consideration of the 3-space with metric (\ref{eq:G2Abeliancasec})
we will assume that $B(u)$ is neither zero or proportional to $A(u)$.

\subsubsection*{(d) $(V,h)$ admits a group ${\mathcal G}_2 I$ acting on null orbits $N_2$ with
no null subgroup ${\cal G}_1$, with a null vector in the orbits of the ${\cal G}_2$.}
If there exists a null vector in the orbits of the ${\cal G}_2$ then we must choose coordinates such
that $\vec X = \partial_v$ is one KV and require a second non-null KV
$\vec Z$ satisfying $[\vec X,\vec Z] = 0 $. We choose $\vec Z = u \partial_v + \partial_w$. Starting with
the general metric for $(V, h)$, applying the Killing equations for $\vec X$ and $\vec Z$, and making a
coordinate transformation of the form $v \mapsto v + f(u)$ leads to a metric for $(V, h)$ of the form
\ba
d \sigma^2 &= (A(u) w^2 - 2 B(u) w + C(u)) du^2 - 2(A(u)w - B(u)) du dv
\nonumber\\
&+ A(u) dv^2 + 2 D(u) dudw.
\label{eq:G2Abeliancased}
\ea
The corresponding reducible 1+3 spacetime $(M, \hat{g})$ is equivalent to metric (29.4) in \cite{Petrov}.

\begin{table}
\caption{The reducible 1+3 spacetimes $(M, {\hat g})$ with $(V, h)$ metrics admitting
an abelian ${\cal G}_2$. $\epsilon_0 = \pm 1$ and $\epsilon_1 = \pm 1$ are not both negative.
The type ${\cal G}_2 I (b)$ cannot occur, since imposition of this condition leads to
a ${\cal G}_3$ on $(V, h)$.}
\footnotesize\rm
\begin{tabular*}{\textwidth}{@{}l*{15}{@{\extracolsep{0pt plus12pt}}l}}
\br
Algebra & $(V, h)$ & Spacetime metric/${\cal G}_3$ basis & Condition\\
\mr

${\cal G}_2 I (a)$ & $(\ref{eq:G2Abeliancasea})$ & $ds^2 = \epsilon_0 d\eta^2
+ \epsilon_1 du^2 + A(u) dv^2 + 2B(u) dv dw + C(u) dw^2$ &
$\epsilon_0 \epsilon_1 (A(u) C(u) - B^2(u))$ \\
& & $\vec X_0 = \partial_\eta$, $\vec X = \partial_v$, $\vec Y = \partial_w$ & $\hspace{10 pt}< 0$ \\

${\cal G}_2 I (c)$ & $(\ref{eq:G2Abeliancasec})$ & $ds^2 = \epsilon_0 d\eta^2
+ 2 du dw + (A(u) dv + B(u) dw)^2$ & $\epsilon_0 A^2(u) > 0$ \\
& & $\vec X_0 = \partial_\eta$, $\vec X = \partial_v$, $\vec Y = \partial_w$ & \\

${\cal G}_2 I (d)$ & $(\ref{eq:G2Abeliancased})$ & $ds^2 = \epsilon_0 d\eta^2
+ (A(u) w^2 - 2 B(u) w + C(u)) du^2 + A(u) dv^2$ & $\epsilon_0 A(u) D^2(u) > 0$ \\
& & $\hspace{10 pt} - 2(A(u)w - B(u)) du dv + 2 D(u) dudw$ & \\
& & $\vec X_0 = \partial_\eta$, $\vec X = \partial_v$, $\vec Y = u \partial_v + \partial_w$ & \\

\br
\end{tabular*}
\label{tab:g2}
\end{table}
The results of this section are summarised in table \ref{tab:g2}.

\subsubsection{The non-abelian case, ${\cal G}_2 II$}\label{G2nonabelian}
We set $k = 1$ in (\ref{eq:G2brackets}), i.e., the group structure is
$[\vec X, \vec Y] = \vec X$. In the following three sub-cases (a), (b) and (c)
we choose coordinates $v$, $w$ adapted to the orbits such
that $\vec X = \partial_v$, $\vec Y = v \partial_v + \partial_w$.
It is then easy to see that the metric takes the form
\ba
d \sigma^2 & = h_{uu}(u) du^2 + 2 h_{uv}(u) e^{-w} du dv + 2 h_{uw}(u)dudw
\nonumber\\
& + h_{vv}(u) e^{-2w} dv^2 + 2 h_{vw} e^{-w} dvdw + h_{ww}(u) dw^2.
\label{eq:G2nonabeliangeneralmetric}
\ea

\subsubsection*{(a) ${\cal G}_2 II$ acting on non-null orbits $S_2$ or $T_2$.}
We can use a coordinate transformation of the form
\be
v \mapsto v + f(u) e^w, \qquad w \mapsto w + g(u)
\label{eq:G2IIcoordinatetransformation}
\ee
to eliminate the $h_{uv}$ and $h_{uw}$ terms from the metric (\ref{eq:G2nonabeliangeneralmetric})
provided that $h_{vv} h_{ww} - (h_{vw})^2 \ne 0$. Rescaling the $u$-coordinate the metric takes
the form
\be
d \sigma^2 = \epsilon_1 du^2 + A(u) e^{-2w} dv^2 + 2 B(u) e^{-w} dvdw + C(u) dw^2.
\label{eq:G2nonAbeliancasea}
\ee
If $\vec X$ is a hypersurface orthogonal KV; then again
$6 X_{[A} X_{B;C]} = (A B_{,u}- B A_{,u}) \epsilon_{ABC} = 0$ implies that $B(u)= aA(u)$ and the coordinate
change $v = v'-a e^{w'}$ and $w=w'$ allows setting $B(u)=0$. If $\vec Y$ is hypersurface orthogonal,
then
\ba
\fl 6 Y_{[A} Y_{B;C]} = e^{-w} [v^2 e^{-2w}(A B_{,u}- B A_{,u}) + v e^{-w}(A C_{,u}- C A_{,u})
\nonumber\\
+ (B C_{,u}- C B_{,u})]\epsilon_{ABC} = 0
\ea
implies $B(u)= aA(u)$ and $C(u)= bA(u)$ for some constants $a,b$. The above change of coordinates (along with
a constant re-scaling of the coordinate $w$) produces then the following line element:
\begin{equation}\label{G2IIb}
   d\sigma^2 = \epsilon_1 du^2 + A(u) \left[ e^{-2w} dv^2 + dw^2 \right].
\end{equation}
Note that this implies that $(M, {\hat g})$ is conformally related to a reducible 2+2 spacetime.
Thus we consider only the case in which $\vec X$ is not h.o., i.e., metric (\ref{eq:G2nonAbeliancasea}).

Note that if $A(u) = 0$ in equation (\ref{eq:G2nonAbeliancasea}), the KV $\vec X = \partial_v$
is a non-covariantly constant null vector. If, in addition, $B(u)$ is a constant, which we can
scale to unity, then $\vec X$ is covariantly constant and $(V,h)$ is a 3-dimensional pp-wave
space with metric $d \sigma^2 = du^2 + 2 e^{-w} dv dw + C(u) dw^2$. The transformations
\[
u \mapsto x, \qquad w \mapsto - \ln |u|, \qquad v \mapsto v
\]
change this metric into the form
\be
d \sigma^2 = dx^2 - 2 dudv - 2 K(x) u^{-2} du^2
\ee
and the corresponding $(M, {\hat g})$ is an isometry class 5 pp-wave spacetime \cite{keanetupper}.
For an arbitrary $K(x)$, $(v,h)$ admits only the two KVs
\[
\vec X_1 = \partial_v, \qquad \vec X_2 = u \partial_u - v \partial_v.
\]
There are two special cases of $(V,h)$ that admit an additional symmetry:

$(i)$ If $K(x) = \alpha e^{kx} + 2 k^{-1} x$, where $\alpha$, $k$ are constants, there
is a third KV, $\vec X_3$, given by
\[
\vec X_3 = \partial_u + 2k^{-2} (1 - kx) u^{-2} \partial_v + 2k^{-1} u^{-1} \partial_x
\]
and $(V,h)$ admits a ${\cal G}_3 V$.

$(ii)$ If $K(x) = \alpha \ln |x|$, where $\alpha$ is a constant, the additional symmetry
is a SCKV given by
\[
\vec S = u^2 \partial_u + (\case{1}{2} x^2 - \alpha \ln |u|)\partial_v + u x \partial_x
\]
with $\psi = u$, so that $(V,h)$ admits a ${\cal S}_3 \supset {\cal G}_2$. The corresponding
$(M, {\hat g})$ admits the SCKV
\[
\vec S = \eta u \partial_\eta + u^2 \partial_u
+ [\case{1}{2} (x^2 + \eta^2) - \alpha \ln |u|] \partial_v + u x \partial_x
\]
which corresponds to the expression (\ref{wth01}) with $k = 0$, $g(u) = - \alpha \ln|u|$, $f(u) = 0$.

\subsubsection*{(b) ${\cal G}_2 II$ acting on null orbits $N_2$ containing a null subgroup ${\cal G}_1$.}
We set up coordinates so that $\vec l = \partial_v$ and $\vec X = v \partial_v + \partial_x$,
then $l_A l^A = l_A X^A = 0$ together with the fact that they are KV implies
\be
d \sigma^2 = P^{-2}(u) dx^2 - 2 e^{-x} du dv - 2H(u) du^2.
\label{eq:G2nonAbeliancaseb}
\ee
Unlike the corresponding ${\cal G}_2 I$ case, $(V, h)$ does not admit, in general, any further KV.
In fact the 3-space (\ref{eq:G2nonAbeliancaseb}) is conformally related to the pp-wave 3-space
with metric (\ref{eq:g2Ibmetric2}) as can be seen by making the coordinate transformation
\[
u \mapsto u, \qquad v \mapsto v + \case{1}{2} P^{-1}(u) P(u)_{,u} x^2,
\qquad x \mapsto \case{1}{2} P(u)x
\]
which changes (\ref{eq:G2nonAbeliancaseb}) into
\be
d \sigma^2 = 4 P^{-2}(u) x^{-2} [dx^2 - 2 du dv - A(u) x^2 du^2]
\label{eq:g2IIbppwave3space}
\ee
where $A(u) = -2P^{-2}(u) (P(u)_{,u})^2 + P^{-1}(u) P(u)_{,uu} + \case{1}{2} H(u) P^2(u)$.
The conformal factor $4 P^{-2}(u) x^{-2}$ leaves $\vec X_1 = \partial_v$ as a KV and changes
the HV, $\vec H = 2v \partial_v + x \partial_x$ into a KV, but all other KVs of (\ref{eq:g2Ibmetric2})
are transformed into proper CKVs of (\ref{eq:g2IIbppwave3space}), which are not symmetries
of the corresponding $(M, {\hat g})$.
Note that if $P$ is a constant we can rescale the coordinate $x$ so that the metric takes the form
\be
d \sigma^2 = -2 e^{-2ax} dudv - 2H(u) du^2 + dx^2
\ee
where $a$ is a constant. In this case $(V, h)$ is a 3-space of constant curvature. Thus the corresponding
$(M, \hat{g})$ is conformally flat. The transformation
\[
e^{ax} \mapsto a((x')^2+ (y')^2)^{\case{1}{2}},
\qquad
y \mapsto \case{1}{a} tan^{-1} \case{y'}{x'}
\qquad u \mapsto u' \qquad v \mapsto av'
\]
changes the metric of $(M, {\hat g})$ into (dropping the primes)
\be
d \sigma^2 = (x^2+ y^2)^{-1} [-2 dudv -2H(u)(x^2+ y^2) du^2 + dx^2 + dy^2]
\label{eq:conformaltoconformallyflatppwave}
\ee
where the quantity in the square brackets is the standard metric for the
conformally flat pp-wave spacetime. However, the spacetime with metric
(\ref{eq:conformaltoconformallyflatppwave}) is not a pp-wave spacetime as it
does not admit a covariantly constant vector.

\subsubsection*{(c) ${\cal G}_2 II$ acting on null orbits $N_2$ with
no null subgroup ${\cal G}_1$, with a null vector orthogonal to the orbits of the ${\cal G}_2$.}
In this case the metric (\ref{eq:G2nonabeliangeneralmetric}) satisfies equation (\ref{eq:degeneratecondition})
so that, again, the $(v, w)$ space is degenerate. We use the coordinate transformations
(\ref{eq:G2IIcoordinatetransformation}) to eliminate $h_{uu}$ and $h_{uv}$ and, rescaling the
$u$-coordinate, we obtain
\be
d \sigma^2 = 2 du dw + [A(u) e^{-w} dv + B(u) dw]^2.
\label{eq:G2nonAbeliancasec}
\ee

We note that in this case there exists a null vector {\it orthogonal to the orbits of} the ${\cal G}_2$.
The null vector, $\partial_u$, is orthogonal to the null 2-space spanned by $v$ and $w$.

\subsubsection*{(d) ${\cal G}_2 II$ acting on null orbits $N_2$ with
no null subgroup ${\cal G}_1$, with a null vector in the orbits of the ${\cal G}_2$.}
In this case we could choose coordinates so that $\vec Y$ is one KV and find another non-null KV,
$\vec Z$, such that $[\vec Y, \vec Z] = \vec Y$. Instead we choose coordinates such that
$\vec X = \partial_v$ is the null vector and $\vec Y = \partial_w$, $\vec Z = \partial_v + w \partial_w$
are the two non-null KV satisfying $[\vec Y, \vec Z] = \vec Y$. The basic metric is the analog of metric
(\ref{eq:G2nonabeliangeneralmetric}) with $v$ and $w$ interchanged. From equation (\ref{eq:degeneratecondition})
and the fact that $\vec X$ is null, it follows that $h_{vv} = h_{vw} = 0$, and a coordinate transformation
$w \mapsto w$, $v \mapsto v + f(u)$, together with a rescaling of the $u$-coordinate leads to the metric
\be
d \sigma^2 = 2 dudv + 2 A(u) e^{-v} du dw + B(u) e^{-2v} dw^2.
\label{eq:G2nonAbeliancased}
\ee

The results of this section are summarised in table \ref{tab:g2nonabelian}.

\begin{table}
\caption{The reducible 1+3 spacetimes $(M, {\hat g})$ with $(V, h)$ metrics admitting
a non-abelian ${\cal G}_2$. $\epsilon_0 = \pm 1$ and $\epsilon_1 = \pm 1$ are not both negative.}
\footnotesize\rm
\begin{tabular*}{\textwidth}{@{}l*{15}{@{\extracolsep{0pt plus12pt}}l}}
\br
Algebra & $(V, h)$ & Spacetime metric/${\cal G}_3$ basis & Condition\\
\mr

${\cal G}_2 II (a)$ & $(\ref{eq:G2nonAbeliancasea})$ & $ds^2 = \epsilon_0 d\eta^2
+ \epsilon_1 du^2 + A(u) e^{-2w} dv^2 + 2 B(u) e^{-w} dvdw$ &
$\epsilon_0 \epsilon_1 (A(u) C(u) - B^2(u))$ \\
& & $\hspace{10 pt} + C(u) dw^2$ & $\hspace{10 pt}< 0$ \\
& & $\vec X_0 = \partial_\eta$, $\vec X = \partial_v$, $\vec Y = v \partial_v + \partial_w$ &  \\

${\cal G}_2 II (b)$ & $(\ref{eq:G2nonAbeliancaseb})$ & $ds^2 = \epsilon_0 d\eta^2
+ P^{-2}(u) dw^2 - 2 e^{-w} du dv - 2H(u) du^2$ &
$\epsilon_0 > 0$ \\
& & $\vec X_0 = \partial_\eta$, $\vec X = \partial_v$, $\vec Y = v \partial_v + \partial_w$ & \\

${\cal G}_2 II (c)$ & $(\ref{eq:G2nonAbeliancasec})$ & $ds^2 = \epsilon_0 d\eta^2
+ 2 du dw + (A(u) e^{-w} dv + B(u) dw)^2$ &
$\epsilon_0 A^2(u) > 0$ \\
& & $\vec X_0 = \partial_\eta$, $\vec X = \partial_v$, $\vec Y = v \partial_v + \partial_w$ & \\

${\cal G}_2 II (d)$ & $(\ref{eq:G2nonAbeliancased})$ & $ds^2 = \epsilon_0 d\eta^2
+ 2 dudv + 2 A(u) e^{-v} du dw + B(u) e^{-2v} dw^2$ & $\epsilon_0 B(u) > 0$ \\
& & $\vec X_0 = \partial_\eta$, $\vec X = \partial_w$, $\vec Y = \partial_v + w \partial_w$ & \\

\br
\end{tabular*}
\label{tab:g2nonabelian}
\end{table}

\subsection{$(V,h)$ admits a group ${\mathcal G}_3$ of isometries acting on two-dimensional orbits.}
\label{G3onN2}

In this case the orbits are of constant curvature and they can be either null or non-null. If they are null,
then it can be shown that a null KV exists necessarily, and therefore they are special cases of that discussed
in section \ref{nullG1}, see also \cite{Petrov} and \cite{barnes79}.
The line element can be written either in the form (\ref{eq:g2Ibmetric2}) with KVs
\[
\vec l = \partial_v, \qquad \vec X =\partial_x, \qquad \vec Y = x\partial_v + u \partial_x
\]
or in the transformed form (\ref{eq:g2Ibmetric3}).

If the orbits are non-null, then their normal is geodesic and invariant under the isometry group.
Furthermore, since the orbits are of constant curvature, we can
immediately write down the forms of the line element, (see for instance \cite{KSMH2}, p227):
\begin{equation}\label{G3onV2}
d\sigma^2 =\epsilon_1 du^2 + Y^2(u) \left[dx^2 +\epsilon_2 \Sigma^2(x,k) dy^2\right]
\end{equation}
where $\epsilon_1 = \pm 1$, $\epsilon_2 = \pm 1$, and
$\Sigma(x,k)= \sin x, \, x \, \mathrm{or}\, \sinh x$ for $k=1,0,-1$ respectively.
However, the spacetimes $(M, {\hat g})$ corresponding to the 3-space metric (\ref{G3onV2})
are conformally reducible 2+2 spacetimes and can be discarded.

\subsection{${\cal G}_3$ of isometries acting transitively on $(V,h)$.}\label{G3transitive}
We consider separately ${\cal G}_3 \supset {\cal G}_2$ and ${\cal G}_3 \not\supset {\cal G}_2$.
Each 3-space metric admitting a ${\cal G}_3 \supset {\cal G}_2$ can be obtained from the
appropriate 3-space metric admitting a ${\cal G}_2$,
in which case, using the coordinate system $\{u, v, w \}$ from section \ref{eq:G2ofisometries},
we shall take the general form for the third KV to be
\[
\vec X_3 = X^u(u, v, w) \partial_u + X^v(u, v, w) \partial_v + X^w(u, v, w) \partial_w .
\]
There is only one ${\cal G}_3$ type which does not admit a ${\cal G}_2$ subalgebra,
that being type ${\cal G}_3 IX$.
The following diagram illustrates the analysis of the ${\cal G}_3$ structures.
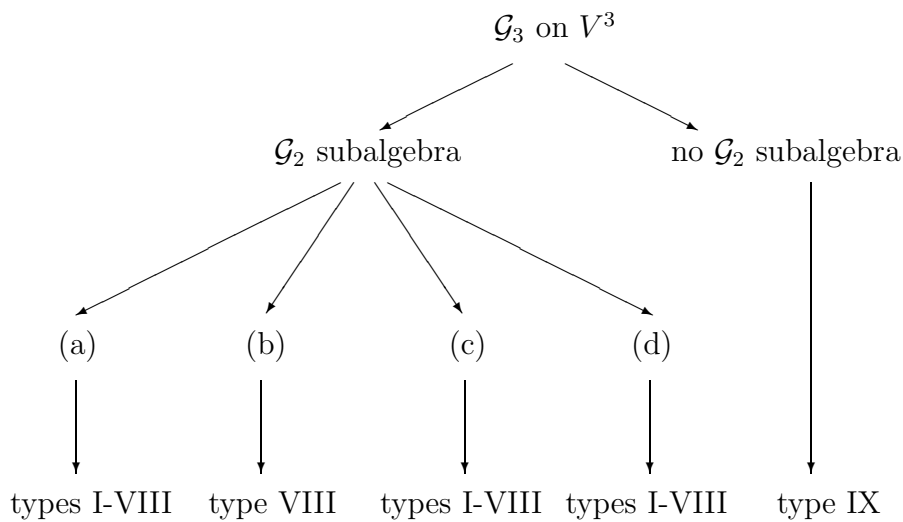
\begin{center}{
\begin{picture}(120,200)
\put(58,185){${\cal G}_3$ on $V^3$}

\put(65,175){\vector(-2,-1){50}}
\put(-25,138){${\cal G}_2$ subalgebra}
\put(85,175){\vector(2,-1){50}}
\put(125,138){no ${\cal G}_2$ subalgebra}

\put(0,130){\vector(-2,-1){100}}
\put(5,130){\vector(-2,-3){33}}
\put(13,130){\vector(2,-3){33}}
\put(18,130){\vector(2,-1){100}}
\put(-107,65){(a)}
\put(-36,65){(b)}
\put(41,65){(c)}
\put(110,65){(d)}

\put(-100,55){\vector(0,-1){35}}
\put(-30,55){\vector(0,-1){35}}
\put(47,55){\vector(0,-1){35}}
\put(117,55){\vector(0,-1){35}}
\put(-125,5){types I-VIII}
\put(-50,5){type VIII}
\put(15,5){types I-VIII}
\put(85,5){types I-VIII}

\put(178,130){\vector(0,-1){110}}
\put(165,5){type IX}

\end{picture}
}
\end{center}
The case ${\cal G}_3 \supset {\cal G}_2 I$ (b) need not be considered here because, as was shown in
section \ref{G2abelian}, imposing the ${\cal G}_2 I$ (b) condition automatically admits a
${\cal G}_3$ on $N_2$.
The only ${\cal G}_3 \supset {\cal G}_2$ (b) structure which needs to be considered here is
the ${\cal G}_3 VIII \supset {\cal G}_2 II$ (b).

\subsubsection{${\cal G}_3$ (from ${\cal G}_2$ case (a))}
Within the ${\cal G}_3 \supset {\cal G}_2$ only the type ${\cal G}_3 VIII$ has a non-abelian ${\cal G}_2$.
The 3-space admitting an abelian ${\cal G}_2$ is given by equation (\ref{eq:G2Abeliancasea})
with basis
\[
\vec X_1 = \partial_v, \qquad \vec X_2 = \partial_w.
\]

\subsubsection*{${\cal G}_3 I$}
This is flat 3-space.

\subsubsection*{${\cal G}_3 II$}
The 3-space is
\be
d \sigma^2 = \epsilon_1 du^2 + (dv-udw)^2 + k dw^2
\label{eq:metricg3IIa}
\ee
where $k \ne 0$ is a constant. The ${\cal G}_3$ basis is
\[
\vec X_1 = \partial_v, \qquad \vec X_2 = \partial_w, \qquad \vec X_3 = \partial_u + w \partial_v
\]

\subsubsection*{${\cal G}_3 III$}
The 3-space is
\be
d \sigma^2 = \epsilon_1 a^2 du^2 + \alpha e^{-2u} dv^2 + 2\beta e^{-u} dvdw + \gamma dw^2
\label{eq:metricg3IIIa}
\ee
where $a \ne 0$, $\alpha$, $\beta$ and $\gamma$ are constants such that $\alpha \gamma - \beta^2 \ne 0$.
If $\alpha = \gamma = 0$ then $(M, {\hat g})$ is conformally flat.
The ${\cal G}_3$ basis is
\[
\vec X_1 = \partial_v, \qquad \vec X_2 = \partial_w, \qquad  \vec X_3 = \partial_u + v \partial_v
\]

\subsubsection*{${\cal G}_3 IV$}
The 3-space is
\be
d \sigma^2 = \epsilon_1 a^2 du^2 + e^{-2 u} [\alpha dv^2 - 2(\alpha u - \beta) dvdw
+ (\alpha u^2 - 2\beta u + \gamma) dw^2]
\label{eq:metricg3IVa}
\ee
where $a \ne 0$, $\alpha$, $\beta$ and $\gamma$ are constants such that $\alpha \gamma - \beta^2 \ne 0$.
The ${\cal G}_3$ basis is
\[
\vec X_1 = \partial_v, \qquad \vec X_2 = \partial_w,
\qquad \vec X_3 = \partial_u + (v+w) \partial_v + w \partial_w
\]

\subsubsection*{${\cal G}_3 V$}
The 3-space is
\be
d \sigma^2 = \epsilon_1 a^2 du^2 + e^{-2 u}[\alpha dv^2 + 2 \beta dv dw + \gamma dw^2]
\label{eq:metricg3Va}
\ee
where $a \ne 0$, $\alpha$, $\beta$ and $\gamma$ are constants such that $\alpha \gamma - \beta^2 \ne 0$.
If $\alpha = \gamma = 0$ then $(M, {\hat g})$ is conformally flat.
The ${\cal G}_3$ basis is
\[
\vec X_1 = \partial_v, \qquad \vec X_2 = \partial_w,
\qquad \vec X_3 = \partial_u + v \partial_v + w \partial_w
\]

\subsubsection*{${\cal G}_3 VI_q$}
The 3-space is
\be
d \sigma^2 = \epsilon_1 a^2 du^2 + \alpha e^{-2 u} dv^2 + 2 \beta e^{-(1+q)u} dvdw + \gamma e^{-2qu} dw^2
\label{eq:metricg3VIa}
\ee
where $a$ is a non-zero constant and $\alpha$, $\beta$ and $\gamma$ are constants such that
$\alpha \gamma - \beta^2 \ne 0$. The ${\cal G}_3$ basis is
\[
\vec X_1 = \partial_v, \qquad \vec X_2 = \partial_w,
\qquad  \vec X_3 = \partial_u + v \partial_v + q w \partial_w
\]

\subsubsection*{${\cal G}_3 VII_p$}
The 3-space is given by
\ba
d \sigma^2 & = & \epsilon_1 a^2 du^2 + e^{-pu} (D(u) dv^2 + 2E(u) dv dw + F(u) dw^2)
\label{eq:metricg3VIIa}
\\
D(u) & = & \case{1}{2} [2 + ((p^2 - 2) c_1 + sp c_2) \cos(su) + ((p^2 - 2) c_2 - sp c_1) \sin (su)]
\label{eq:metricg3VIIacondition1}\\
E(u) & = & \case{1}{2} [ p + (s c_2 + p c_1) \cos(su) - (s c_1 - p c_2) \sin (su)]
\label{eq:metricg3VIIacondition2}\\
F(u) & = & 1 + c_1 \cos(su) + c_2 \sin (su)
\label{eq:metricg3VIIacondition3}
\ea
where $a$ is a non-zero constant, $c_1$ and $c_2$ are arbitrary constants and $p$ and $s$
are constants such that $p^2 < 4$ and $s = \sqrt{4-p^2}$, and the metric is non-degenerate.
The ${\cal G}_3$ basis is
\[
\vec X_1 = \partial_v, \qquad \vec X_2 = \partial_w, \qquad
\vec X_3 = \partial_u - w \partial_v + (pw + v) \partial_w
\]

\subsubsection*{${\cal G}_3 VIII$}
In this case we use the metric (\ref{eq:G2nonAbeliancasea}) of the 3-space admitting a
non-abelian ${\cal G}_2$. We find that
\ba
& & c (A(u) C(u)_{,u} + C(u) A(u)_{,u} - 2 B(u) B(u)_{,u}) = 0 \label{eq:g3viiicondition1}\\
& & A(u) f(u)_{,u} + B(u) g(u)_{,u} = 0 \label{eq:g3viiicondition2}\\
& & B(u) f(u)_{,u} + C(u) g(u)_{,u} = - \epsilon_1 c \label{eq:g3viiicondition3}
\ea
where $c$ is a constant which is necessarily non-zero else $(M, {\hat g})$ is conformal to a
2+2 reducible spacetime, and the functions $f(u)$ and $g(u)$ are given by
\ba
f(u) & = & - \case{1}{2} [c B(u)_{,u} + 2 C(u)] A^{-1}(u) \label{eq:g3viiicondition4}\\
g(u) & = & \case{1}{2} [c A(u)_{,u} + 4 B(u)] A^{-1}(u) \label{eq:g3viiicondition5}
\ea
Note that $A(u) C(u) - B^2(u) \ne 0$, and the non-degeneracy condition requires $A(u) \ne 0$
(since $A(u) = 0$ implies $B(u) = 0$).
Condition (\ref{eq:g3viiicondition1}) integrates to $A(u) C(u) - B^2(u) = k = constant$, $k \ne 0$ and
equations (\ref{eq:g3viiicondition2}) and (\ref{eq:g3viiicondition3}) give
\[
f(u)_{,u} = \epsilon_1 c k^{-1} B(u), \qquad g(u)_{,u} = \epsilon_1 c k^{-1} A(u)
\]
Differentiating (\ref{eq:g3viiicondition4}) and (\ref{eq:g3viiicondition5}) and equating the expressions
for $f(u)_{,u}$ and $g(u)_{,u}$ to those above gives
\ba
A(u)_{,u} & = & \pm A(u) [\beta - 4 \epsilon_1 k^{-1} A(u) - 16 k c^{-2} A^{-2}(u)
- 8 \alpha c^{-2} A^{-1}(u)]^{1 / 2}
\label{eq:metricg3VIIIafunctiona}\\
B(u) & = & - c^{-1} A(u) \biggl[\int (4 k A^{-2}(u) + \alpha A^{-1}(u)) du + \beta \biggr]
\label{eq:metricg3VIIIafunctionb}\\
C(u) & = & (k + B^2(u)) A^{-1}(u)
\label{eq:metricg3VIIIafunctionc}
\ea
where $\alpha$ and $\beta$ are arbitrary constants.

The ${\cal G}_3$ basis is
\ba
\vec X_1 & = & \partial_v, \qquad \vec X_2 = v \partial_v + \partial_w,
\nonumber\\
\vec X_3 & = & c e^w \partial_u + [v^2 + e^{2w} f(u)] \partial_v + [2v + e^w g(u)] \partial_w
\ea
Petrov presents this metric in an alternative coordinate system where
the KV $\vec X_3$ takes on a simpler form.

The canonical metric forms admitting transitive ${\cal G}_3 (a)$ algebras on $(V,h)$ are listed
in table \ref{tab:transitiveg3a}.
\begin{table}
\caption{Reducible 1+3 spacetimes admitting transitive ${\cal G}_3$ algebras on $(V,h)$,
with orbits of type (a). $\epsilon_0 = \pm 1$ and $\epsilon_1 = \pm 1$ are not both negative.
Those metrics which are conformally reducible 2+2 spacetimes are omitted from the table.}
\footnotesize\rm
\begin{tabular*}{\textwidth}{@{}l*{15}{@{\extracolsep{0pt plus12pt}}l}}
\br
Algebra & $(V, h)$ &  Spacetime metric/${\cal G}_4$ basis & Condition \\
\mr

${\cal G}_3 II (a)$ & $(\ref{eq:metricg3IIa})$ & $ds^2 = \epsilon_0 d\eta^2 + \epsilon_1 du^2 + (dv-udw)^2 + k dw^2$ &
$\epsilon_0 \epsilon_1 k < 0$ \\
& & $\vec X_0 = \partial_\eta$, $\vec X_1 = \partial_v$, $\vec X_2 = \partial_w$,
$\vec X_3 = \partial_u + w \partial_v$ & \\

${\cal G}_3 III (a)$ & $(\ref{eq:metricg3IIIa})$ &
$ds^2 = \epsilon_0 d\eta^2 + \epsilon_1 a^2 du^2 + \alpha e^{-2u} dv^2
+ 2\beta e^{-u} dvdw + \gamma dw^2$ &
$\epsilon_0 \epsilon_1 a^2 (\alpha \gamma - \beta^2) < 0$ \\
& & $\vec X_0 = \partial_\eta$, $\vec X_1 = \partial_v$, $\vec X_2 = \partial_w$,
$\vec X_3 = \partial_u + v \partial_v$ & \\

${\cal G}_3 IV (a)$  & $(\ref{eq:metricg3IVa})$ &
$ds^2 = \epsilon_0 d\eta^2 + \epsilon_1 a^2 du^2
$ & $\epsilon_0 \epsilon_1 a^2 (\alpha \gamma - \beta^2) < 0$ \\
& & $\hspace{10 pt} + e^{-2 u} [\alpha dv^2 - 2(\alpha u - \beta) dvdw + (\alpha u^2 - 2\beta u + \gamma) dw^2]$ & \\
& & $\vec X_0 = \partial_\eta$, $\vec X_1 = \partial_v$, $\vec X_2 = \partial_w$,
$\vec X_3 = \partial_u + (v+w) \partial_v + w \partial_w$ &\\

${\cal G}_3 V (a)$  & $(\ref{eq:metricg3Va})$ &
$ds^2 = \epsilon_0 d\eta^2 + \epsilon_1 a^2 du^2 + e^{-2 u}[\alpha dv^2 + 2 \beta dv dw + \gamma dw^2]$ &
$\epsilon_0 \epsilon_1 a^2 (\alpha \gamma - \beta^2) < 0$ \\
& & $\vec X_0 = \partial_\eta$, $\vec X_1 = \partial_v$, $\vec X_2 = \partial_w$,
$\vec X_3 = \partial_u + v \partial_v + w \partial_w$ & \\

${\cal G}_3 VI_q (a)$ & $(\ref{eq:metricg3VIa})$ &
$ds^2 = \epsilon_0 d\eta^2 + \epsilon_1 a^2 du^2 + \alpha e^{-2 u} dv^2
+ 2 \beta e^{-(1+q)u} dvdw $
& $\epsilon_0 \epsilon_1 a^2 (\alpha \gamma - \beta^2) < 0$\\
& & $\hspace{10 pt} + \gamma e^{-2qu} dw^2$ & $q \ne 0$, $q \ne 1$ \\
& & $\vec X_0 = \partial_\eta$, $\vec X_1 = \partial_v$, $\vec X_2 = \partial_w$,
$\vec X_3 = \vec X_3 = \partial_u + v \partial_v + q w \partial_w$ & \\

${\cal G}_3 VII_p (a)$ & $(\ref{eq:metricg3VIIa})$ &
$ds^2 = \epsilon_0 d\eta^2 + \epsilon_1 a^2 du^2$ &
$\epsilon_0 \epsilon_1 a^2 $ \\
& & $\hspace{10 pt} + e^{-pu} (D(u) dv^2 + 2E(u) dv dw + F(u) dw^2)$ & $(D(u) F(u) - E(u)^2)$ \\
& & Functions $D(u)$, $E(u)$, $F(u)$ given by
(\ref{eq:metricg3VIIacondition1}) - (\ref{eq:metricg3VIIacondition3}) & $\hspace{10 pt}< 0$ \\
& & $\vec X_0 = \partial_\eta$, $\vec X_1 = \partial_v$, $\vec X_2 = \partial_w$,
$\vec X_3 = \partial_u - w \partial_v + (pw + v) \partial_w$ & $p^2 < 4$
\\

${\cal G}_3 VIII (a)$ & $(\ref{eq:G2nonAbeliancasea})$ &
$ds^2 = \epsilon_0 d\eta^2 + \epsilon_1 du^2 + e^{-2w} A(u) dv^2$ &
$\epsilon_0 \epsilon_1 (A(u) C(u) - B(u)^2)$ \\
& & $\hspace{10 pt} + 2 B(u) e^{-w} dv dw + C(u) dw^2$ & $\hspace{10 pt}< 0$ \\
& & Functions $A(u)$, $B(u)$, $C(u)$ given by (\ref{eq:metricg3VIIIafunctiona})
- (\ref{eq:metricg3VIIIafunctionc})& $c \ne 0$ \\
& & $\vec X_0 = \partial_\eta$, $\vec X_1 = \partial_v$, $\vec X_2 = v\partial_v + \partial_w$, & \\
& & $\vec X_3 = c e^w \partial_u + [v^2 + e^{2w} f(u)] \partial_v + [2v + e^w g(u)] \partial_w$ & \\

\br
\end{tabular*}
\label{tab:transitiveg3a}
\end{table}

\subsubsection{${\cal G}_3$ (from ${\cal G}_2$ case (b))}
\label{sec:g3fromg2typeb}
There is only one possibility to consider here, ${\cal G}_3 VIII$, that is, the only algebra type
admitting a ${\cal G}_2 II$ subalgebra. As noted earlier the case (b) ${\cal G}_2 I$ automatically
admits a ${\cal G}_3$ on $N_2$, which is considered in section \ref{G3onN2}.
However, it is straightforward to show that type ${\cal G}_3 VIII (b)$ cannot occur since the
imposition of the algebra type on the ${\cal G}_2 II (b)$ metric leads to a contradiction.

\subsubsection{${\cal G}_3$ (from ${\cal G}_2$ case (c))}
\label{sec:G3fromG2c}
The 3-space with abelian ${\cal G}_2$ is given by equation (\ref{eq:G2Abeliancasec}) with basis
\[
\vec X_1 = \partial_v, \qquad \vec X_2 = \partial_w.
\]
Note that the metric function $A(u) \ne 0$ for non-degeneracy.

\subsubsection*{${\cal G}_3 I$}
This is flat 3-space and so $(M, {\hat g})$ is Minkowski spacetime.

\subsubsection*{${\cal G}_3 II$}
Imposing the third KV condition leads to ${\cal G}_4 I$ automatically (see metric
(\ref{eq:automaticG4I})), that is, there is no maximal ${\cal G}_3 II$.

\subsubsection*{${\cal G}_3 III$}
The 3-space is
\be
d \sigma^2 = 2 du dw + (\alpha e^{-u} dv+ \beta dw)^2
\label{eq:metricg3IIIc}
\ee
where $\alpha \ne 0$ and $\beta$ are constants.
\[
\vec X_1 = \partial_v, \qquad \vec X_2 = \partial_w \qquad \vec X_3 = \partial_u + v \partial_v
\]

\subsubsection*{${\cal G}_3 IV$}
The 3-space is
\be
d \sigma^2 = 2 du dw + k^2 u^2 [dv + \ln |u| dw]^2
\label{eq:metricg3IVc}
\ee
where $k \ne 0$ is a constant, and
\[
\vec X_1 = \partial_v, \qquad \vec X_2 = \partial_w \qquad
\vec X_3 = - u \partial_u + (v + w) \partial_v + w \partial_w
\]

\subsubsection*{${\cal G}_3 V$}
The 3-space is
\be
d \sigma^2 = 2 du dw + u^2 (\beta dv + \gamma dw)^2
\ee
where $\beta \ne 0$ and $\gamma$ are constants, and
\[
\vec X_1 = \partial_v, \qquad \vec X_2 = \partial_w \qquad
\vec X_3 = - u \partial_u + v \partial_v + w \partial_w
\]
This is a flat 3-space and so $(M, {\hat g})$ is Minkowski spacetime.

\subsubsection*{${\cal G}_3 VI_q$}
The 3-space is
\be
d \sigma^2 = 2 du dw + [k|u|^{1/q} dv +l |u| dw]^2
\label{eq:metricg3VIc}
\ee
where $k \ne 0$ and $l$ are constants. The ${\cal G}_3$
basis is
\[
\vec X_1 = \partial_v, \qquad \vec X_2 = \partial_w \qquad
\vec X_3 = - q u \partial_u + v \partial_v + q w \partial_w.
\]

\subsubsection*{${\cal G}_3 VII_p$}
The 3-space is
\be
d \sigma^2 = 2 du dw + (A(u) dv + B(u) dw)^2
\label{eq:metricg3VIIc}
\ee
where the following conditions are satisfied (where $X^u = a(u)$)
\ba
a(u) A(u)_{,u} + B(u) = 0
\nonumber\\
a(u) B(u)_{,u} - A(u) + p B(u) = 0
\nonumber\\
A^2(u) f(u)_{,u} + 1 = 0
\nonumber\\
a(u)_{,u} + A(u) B(u) f(u)_{,u} + p = 0
\label{eq:G3VIIIconditions}
\ea
and
\[
\vec X_1 = \partial_v, \qquad \vec X_2 = \partial_w, \qquad
\vec X_3 = a(u) \partial_u + (-w + f(u)) \partial_v + (pw + v) \partial_w
\]
The conditions (\ref{eq:G3VIIIconditions}) lead to $f(u)_{,u} = - A^{-2}(u)$,
$a(u)_{,u} = B(u) A^{-1}(u) - p$ and
\be
a(u) a(u)_{,uu} - (a(u)_{,u})^2 = p a(u)_{,u} + 1.
\label{eq:G3VIIIconditionsDefora}
\ee
Once a solution for $a(u)$ is determined then $A(u)$, $B(u)$ and $f(u)$ can be found using
\be
[a(u) A(u)]_{,u} = - p A(u),
\qquad
B(u) = - a(u) A(u)_{,u},
\qquad
f(u)_{,u} = - A^{-2}(u).
\label{eq:G3VIIcconditions}
\ee
For 3-dimensional orbits we require $a(u) \ne 0$.
Note that $a(u) = ku$, where $k$ is a constant, is a solution of equation
(\ref{eq:G3VIIIconditionsDefora}) but no real metric exists because equation
(\ref{eq:G3VIIIconditionsDefora}) and the condition $p^2 < 4$ lead to non-real
values for $k$. The general solution of (\ref{eq:G3VIIIconditionsDefora}) is unknown.

\subsubsection*{${\cal G}_3 VIII$}
Using the metric (\ref{eq:G2nonAbeliancasec}) of the 3-space with non-abelian ${\cal G}_2$ we find
that the imposition of this type leads to a contradiction in the Killing equations, and so is not possible.

The canonical metric forms admitting transitive ${\cal G}_3 (c)$ algebras on $(V,h)$ are listed
in table \ref{tab:transitiveg3c}.
\begin{table}
\caption{Reducible 1+3 spacetimes admitting transitive ${\cal G}_3$ algebras on $(V,h)$,
with orbits of type (c). $\epsilon_0 = + 1$. Those metrics which are conformally reducible 2+2 spacetimes are omitted from the table.}
\footnotesize\rm
\begin{tabular*}{\textwidth}{@{}l*{15}{@{\extracolsep{0pt plus12pt}}l}}
\br
Algebra & $(V, h)$ &  Spacetime metric/${\cal G}_4$ basis & Condition \\
\mr

${\cal G}_3 III (c)$ & $(\ref{eq:metricg3IIIc})$ &
$ds^2 = \epsilon_0 d\eta^2 + 2 du dw + (\alpha e^{-u} dv+ \beta dw)^2$ & $\epsilon_0 \alpha^2 > 0$ \\
& & $\vec X_0 = \partial_\eta$, $\vec X_1 = \partial_v$, $\vec X_2 = \partial_w$,
$\vec X_3 = \partial_u + v \partial_v$ & \\

${\cal G}_3 IV (c)$  & $(\ref{eq:metricg3IVc})$ &
$ds^2 = \epsilon_0 d\eta^2 + 2 du dw + k^2 u^2 [dv + \ln |u| dw]^2 $ & $\epsilon_0 k^2 u^2 > 0$ \\
& & $\vec X_0 = \partial_\eta$, $\vec X_1 = \partial_v$, $\vec X_2 = \partial_w$,
$\vec X_3 = - u \partial_u + (v + w) \partial_v + w \partial_w$ &\\

${\cal G}_3 VI_q (c)$ & $(\ref{eq:metricg3VIc})$ &
$ds^2 = \epsilon_0 d\eta^2 + 2 du dw + [k|u|^{1/q} dv +l |u| dw]^2$ & $\epsilon_0 k^2 |u|^{2/q} > 0$ \\
& & $\vec X_0 = \partial_\eta$, $\vec X_1 = \partial_v$, $\vec X_2 = \partial_w$,
$\vec X_3 = - q u \partial_u + v \partial_v + q w \partial_w$ & $q \ne 0$, $q \ne 1$ \\
& & & $u \ne 0$ \\

${\cal G}_3 VII_p (c)$ & $(\ref{eq:metricg3VIIc})$ &
$ds^2 = \epsilon_0 d\eta^2 + 2 du dw + (A(u) dv + B(u) dw)^2$ & $\epsilon_0 A^2(u) > 0$ \\
& & $\vec X_0 = \partial_\eta$, $\vec X_1 = \partial_v$, $\vec X_2 = \partial_w$, & $p^2 < 4$ \\
& & $\vec X_3 = a(u) \partial_u + (-w + f(u)) \partial_v + (pw + v) \partial_w$ & $a(u) \ne 0$\\
& & $a(u)$ given by (\ref{eq:G3VIIIconditionsDefora}),
$A(u)$ and $B(u)$ given by (\ref{eq:G3VIIcconditions}) &
\\

\br
\end{tabular*}
\label{tab:transitiveg3c}
\end{table}

\subsubsection{${\cal G}_3$ (from ${\cal G}_2$ case (d))}
The 3-space with abelian ${\cal G}_2$ is given by the metric (\ref{eq:G2Abeliancased}) with basis
$\vec X_1 = \partial_v$, $\vec X_2 = u\partial_v + \partial_w$.
Note that the determinant of this metric is $-A(u) D^2(u)$ so that $A(u) \ne 0$ and
$D(u) \ne 0$ for non-degeneracy.

\subsubsection*{${\cal G}_3 I$}
This case cannot occur since the third KV is forced to be a linear combination (with constant coefficients)
of $\vec X_1$ and $\vec X_2$.

\subsubsection*{${\cal G}_3 II$}
Imposing the ${\cal G}_3 II$ conditions leads to a ${\cal G}_4 I (s = 0)$ solution,
that is, no maximal ${\cal G}_3 II$ solution is possible. Details are given
in section \ref{sec:G4fromG3fromG2a}.

\subsubsection*{${\cal G}_3 III$}
The conditions on the functions are $A(u) = \alpha u^{-2}$, $D(u) = \delta u^{-1}$
where $\alpha$, $\delta$ are non-zero constants, with $B(u)$, $C(u)$ arbitrary, i.e.,
the metric of $(V,h)$ is
\ba
d \sigma^2 & = & [\alpha u^{-2} w^2 - 2 B(u) w + C(u)] du^2 -2[\alpha u^{-2}w - B(u)] du dv
\nonumber\\
& & + \alpha u^{-2} dv^2 + 2 \delta u^{-1} dudw.
\label{eq:G3IIId3spacemetric}
\ea
The basis for the ${\cal G}_3$ is
\[
\vec X_1 = \partial_v, \qquad \vec X_2 = u\partial_v + \partial_w,
\qquad
\vec X_3 = u \partial_u + [v + b(u)] \partial_v + c(u) \partial_w.
\]
where the functions $b(u)$, $c(u)$ are given by
\ba
b(u)_{,u} & = & \frac{1}{2 \alpha \delta}[-2 \delta u^3 B(u)_{,u} - 4 \delta u^2 B(u)
+ u^4 B^2(u) - \alpha u^2 C(u)]
\nonumber\\
c(u) & = & \case{1}{2 \alpha \delta} [u^4 B^2(u) - \alpha u^2 C(u)] .
\label{eq:G3IIIdconditions}
\ea
Note that $u \ne 0$ for 3-dimensional orbits.
Note the special case in which $B(u) = C(u) = 0$. Then $c(u) = 0$ and $b(u)$ is constant.
The KV $\vec X_3 = u \partial_u + v \partial_v$ and the metric is
\be
d \sigma^2 = \alpha u^{-2}(wdu - dv)^2 + 2 \delta u^{-1} du dw.
\label{eq:eq:G3IIIdconditionsii}
\ee

\subsubsection*{${\cal G}_3 IV$}
The metric functions are $A(u) = \alpha e^{2u}$, $D(u) = \delta e^u$
where $\alpha$ and $\delta$ are non-zero constants, with $B(u)$, $C(u)$ arbitrary,
i.e., the metric of $(V,h)$ is
\ba
d \sigma^2 & = & [\alpha e^{2u} w^2 - 2 B(u) w + C(u)] du^2 - 2[\alpha e^{2u}w - B(u)] du dv
\nonumber\\
& & + \alpha e^{2u} dv^2 + 2 \delta e^u dudw.
\label{eq:G3IVd3spacemetric}
\ea
The basis for the ${\cal G}_3$ is
\[
\vec X_1 = \partial_v, \qquad \vec X_2 = u\partial_v + \partial_w,
\qquad
\vec X_3 = -\partial_u + (v + b(u)) \partial_v + (w + c(u)) \partial_w .
\]
where the fuunctions $b(u)$, $c(u)$ are given by
\ba
b(u)_{,u} & = & c(u) + \alpha^{-1} e^{-u} [B(u) e^{-u}]_{,u} ,
\nonumber\\
c(u)_{,u} & = & \frac{1}{2 \alpha \delta} e^{-u}[\alpha C(u) - B^2(u) e^{-2u}]_{,u} .
\label{eq:G3IVdconditions}
\ea
In the special case $B(u) = C(u) = 0$ we have $c(u) = c$ is constant and $b(u) = cu + \beta$,
where $\beta$ is a constant. We can take $\vec X_3 = -\partial_u + v \partial_v + w \partial_w$
and the metric of $(V,h)$ is
\be
d \sigma^2 = \alpha e^{2u} (w du - dv)^2 + 2 \delta e^u dudw.
\ee

\subsubsection*{${\cal G}_3 V$}
This case cannot occur since $D(u)=0$ which contradicts the non-degeneracy condition.

\subsubsection*{${\cal G}_3 VI_q$}
The functions are $A(u) = \alpha u^{2p}$, $D(u) = \delta u^p$ where $\alpha$ and $\beta$ are
non-zero constants, $p = 1/(q - 1)$ with $B(u)$, $C(u)$ arbitrary. i.e., the metric of $(V,h)$ is
\ba
d \sigma^2 & = & [\alpha u^{2p} w^2 - 2 B(u) w + C(u)] du^2
- 2[\alpha u^{2p}w - B(u)] du dv
\nonumber\\
 & & + \alpha u^{2p} dv^2 + 2 \delta u^p dudw.
\label{eq:G3VId3spacemetric}
\ea
The basis for the ${\cal G}_3$ is
\ba
\vec X_1 & = & \partial_v, \qquad \vec X_2 = u\partial_v + \partial_w,
\nonumber\\
\vec X_3 & = & (1 - q) u \partial_u + (v + b(u)) \partial_v + (qw + c(u)) \partial_w ,
\nonumber
\ea
where the functions $b(u)$, $c(u)$ are given by
\ba
b(u)_{,u} & = & c(u) + \alpha^{-1} u^{-2p} [(q - 1) u B(u)_{,u} + (q - 2) B(u)],
\nonumber\\
c(u)_{,u} & = & \frac{(q - 1)}{2 \alpha \delta} u^{-pq}
[ \alpha u^2 C(u) - B^2(u) u^{2p(q - 2)} ]_{,u}.
\label{eq:G3VIdconditions}
\ea
In the special case $B(u) = C(u) = 0$ we have $c(u) = c$ is constant and $b(u) = cu + \beta$
where $\beta$ is a constant.
Eliminating multiples of $\vec X_1$ and $\vec X_2$ we have
\[
\vec X_3 = (1 - q) u \partial_u + v \partial_v + qw \partial_w .
\]
and the metric of $(V,h)$ is
\be
d \sigma^2 = \alpha u^{2p}(w du - dv)^2 + 2 \delta u^p dudw.
\ee

\subsubsection*{${\cal G}_3 VII_p$}
The metric functions are
\ba
A(u) & = & \alpha (u^2 - pu + 1)^{-1} \exp[-2pq^{-1} \tan^{-1} ((2u - p)/q)] ,
\nonumber\\
D(u) & = & \delta (u^2 - pu + 1)^{-1/2} \exp[-pq^{-1} \tan^{-1} ((2u - p)/q)] ,
\label{eq:G3VIIdconditions}
\ea
where $\alpha$ and $\delta$ are non-zero constants with $B(u)$, $C(u)$ arbitrary,
i.e., the metric of $(V,h)$ is given by (\ref{eq:G2Abeliancased}) with $A(u)$, $D(u)$ as
above. The ${\cal G}_3$ basis is
\ba
\vec X_1 & = & \partial_v, \qquad \vec X_2 = u\partial_v + \partial_w,
\nonumber\\
\vec X_3 & = & a(u) \partial_u + (uv + b(u)) \partial_v + (v + (p - u)w + c(u)) \partial_w
\nonumber
\ea
where $X^u = a(u) = (u^2 - pu + 1)$ and the functions $b(u)$, $c(u)$ are given by
\ba
& & a(u) B(u)_{,u} + [a(u)_{,u} + u] B(u) + (b(u)_{,u} - c(u)) A(u) + D(u) = 0 ,
\nonumber\\
& & a(u) C(u)_{,u} + 2 a(u)_{,u}  C(u) + 2 (b(u)_{,u} - c(u)) B(u) + 2 c(u)_{,u} D(u) = 0 .
\label{eq:G3VIIdconditions1}
\ea
The special case with $B(u) = C(u) = 0$ has $c(u) = c$ is constant and
\[
b(u) = c u - \alpha^{-1} \delta \int (u^2 - pu + 1)^{1/2} \exp[pq^{-1} \tan^{-1} ((2u - p)/q)]du.
\]
Eliminating multiples of $\vec X_1$ and $\vec X_2$ we have
\ba
\vec X_3 & = & (u^2 - pu + 1) \partial_u + (v + (p - u)w) \partial_w
\nonumber\\
& & + \biggl(uv - \alpha^{-1} \delta \int (u^2 - pu + 1)^{1/2}
\exp[pq^{-1} \tan^{-1} ((2u - p)/q)]du \biggr) \partial_v .
\nonumber
\ea

\subsubsection*{${\cal G}_3 VIII$}
In this case we use the metric (\ref{eq:G2nonAbeliancased}) of the 3-space admitting
a non-abelian ${\cal G}_2$ However, this leads to $B(u)=0$ which contradicts the
non-degeneracy condition, so this case cannot occur.

The canonical metric forms admitting transitive ${\cal G}_3 (d)$ algebras on $(V,h)$ are listed
in table \ref{tab:transitiveg3d}.
\begin{table}
\caption{Reducible 1+3 spacetimes admitting transitive ${\cal G}_3$ algebras on $(V,h)$,
with orbits of type (d). $\epsilon_0 = \pm 1$. Those metrics which are conformally reducible 2+2 spacetimes are omitted from the table.}
\footnotesize\rm
\begin{tabular*}{\textwidth}{@{}l*{15}{@{\extracolsep{0pt plus12pt}}l}}
\br
Algebra & $(V, h)$ &  Spacetime metric/${\cal G}_4$ basis & Condition \\
\mr

${\cal G}_3 III (d)$ & $(\ref{eq:G3IIId3spacemetric})$ &
$ds^2 = \epsilon_0 d\eta^2 + [\alpha u^{-2} w^2 - 2 B(u) w + C(u)] du^2$ & $\epsilon_0 \alpha \delta^2 > 0$ \\
& & $\hspace{10 pt} -2[\alpha u^{-2}w - B(u)] du dv + \alpha u^{-2} dv^2 + 2 \delta u^{-1} dudw$ & $u \ne 0$\\
& & $\vec X_0 = \partial_\eta$, $\vec X_1 = \partial_v$, $\vec X_2 = u\partial_v + \partial_w$, & \\
& & $\vec X_3 = u \partial_u + [v + b(u)] \partial_v + c(u) \partial_w$,
$b(u)$, $c(u)$ given by (\ref{eq:G3IIIdconditions}) & \\

${\cal G}_3 IV (d)$  & $(\ref{eq:G3IVd3spacemetric})$ &
$ds^2 = \epsilon_0 d\eta^2 + [\alpha e^{2u} w^2 - 2 B(u) w + C(u)] du^2$ & $\epsilon_0 \alpha \delta^2 < 0$ \\
& & $\hspace{10 pt} -2[\alpha e^{2u}w - B(u)] du dv + \alpha e^{2u} dv^2 + 2 \delta e^u dudw$ & \\
& & $\vec X_0 = \partial_\eta$, $\vec X_1 = \partial_v$, $\vec X_2 = u\partial_v + \partial_w$, & \\
& & $\vec X_3 = -\partial_u + (v + b(u)) \partial_v + (w + c(u)) \partial_w$, & \\
& & $b(u)$, $c(u)$ given by (\ref{eq:G3IVdconditions}) & \\

${\cal G}_3 VI_q (d)$ & $(\ref{eq:G3VId3spacemetric})$ &
$ds^2 = \epsilon_0 d\eta^2 + [\alpha u^{2p} w^2 - 2 B(u) w + C(u)] du^2 $ & $\epsilon_0 \alpha \delta^2 u^{4p} > 0$, \\
& & $\hspace{10 pt} - 2[\alpha u^{2p}w - B(u)] du dv + \alpha u^{2p} dv^2 + 2 \delta u^p dudw$ & $q \ne 0$, $q \ne 1$ \\
& & $\vec X_0 = \partial_\eta$, $\vec X_1 = \partial_v$, $\vec X_2 = u\partial_v + \partial_w$, & \\
& & $\vec X_3 = (1 - q) u \partial_u + (v + b(u)) \partial_v + (qw + c(u)) \partial_w$ & \\
& & $b(u)$, $c(u)$ given by (\ref{eq:G3VIdconditions}) & \\

${\cal G}_3 VII_p (d)$ & $(\ref{eq:G2Abeliancased})$ &
$ds^2 = \epsilon_0 d\eta^2 + [A(u) w^2 - 2 B(u) w + C(u)] du^2$ & $\epsilon_0 \alpha \delta^2 (u^2 - pu + 1)^{-2}$ \\
& & $\hspace{10 pt} - 2[A(u) w - B(u)] du dv + A(u) dv^2 + 2 D(u) dudw$ & $\hspace{10 pt}> 0$, \\
& & $\vec X_0 = \partial_\eta$, $\vec X_1 = \partial_v$, $\vec X_2 = u\partial_v + \partial_w$,
$\vec X_3 = (u^2 - pu + 1) \partial_u$ & $p^2 < 4$ \\
& & $\hspace{10 pt} + (uv + b(u)) \partial_v + (v + (p - u)w + c(u)) \partial_w$ & \\
& & $A(u)$, $D(u)$ given by (\ref{eq:G3VIIdconditions}),
$b(u)$, $c(u)$ given by (\ref{eq:G3VIIdconditions1}) &
\\

\br
\end{tabular*}
\label{tab:transitiveg3d}
\end{table}

\subsubsection{${\cal G}_3 IX$}
The 3-space and KVs are (from Petrov \cite{Petrov})
\ba
d \sigma^2  & = & \case{1}{2} \alpha [dx^2 + \sin^2 x dy^2]
+ \beta [-\cos(2z) dx^2 + \cos(2z) \sin^2 x dxdy]
\nonumber\\
& &+ \gamma [dz^2 + \cos^2 x dy^2 + 2 \cos xdydz] + \delta[-\sin(2z) dx^2
\nonumber\\
& &+ \sin (2z) \sin^2 xdy^2 + \cos (2z) \sin x dxdy] + \epsilon [2 \cos z \sin x \cos x dy^2
\nonumber\\
& &- 2 \sin z \cos x dx dy - 2 \sin z dxdz + 2 \cos z \sin x dydz]
\nonumber\\
& &+ \lambda [2 \sin z \sin x \cos x dy^2 + 2 \cos z \cos x dx dy + 2 \cos z dxdz \nonumber\\
& &+ 2 \sin z \sin x dydz].
\label{eq:petrovsg3ixmetric}
\ea
A basis for the ${\cal G}_3$ is
\ba
\vec X_1 & = & {\partial_y}, \nonumber\\
\vec X_2 & = & \cos y {\partial_x} - \cot x \sin y {\partial_y}
+ { \sin y  \csc x} {\partial_z}, \nonumber\\
\vec X_3 & = & - \sin y {\partial_x} - \cot x \cos y {\partial_y}
+ { \cos y \csc x} {\partial_z}. \nonumber
\ea

\subsubsection{${\cal G}_3$ (constant curvature 2-space admitting ${\cal G}_2$)}
The metric given by (\ref{eq:confflat2space}) represents a $(V,h)$ which admits at least
one hypersurface orthogonal Killing vector. This metric can be written
\be
d\sigma^2 = \epsilon_1 e^{2U(x^k)}du^2 + \Omega^2 (x^k) [\epsilon_2 (dx^1)^2 +
\epsilon_3 (dx^2)^2] ,
\label{eq:hometric}
\ee
where $k = 1,2$ and $\epsilon_1$, $\epsilon_2$ and $\epsilon_3$ are either all $+1$
or at most one of them is $-1$.
Due to this metric's properties, it is amenable to further investigation.
We consider the possible situations that arise when the 2-space is a space of constant curvature.
In this case the isometry groups of $(V,h)$ can be determined in a straightforward manner.
The three-dimensional metric is
\[
d\sigma^2 = \epsilon_1 e^{2U(y,z)}du^2 + \Omega^2(y,z) [\epsilon_2 dy^2 + \epsilon_3 dz^2] \, .
\]
Consider first the case $\epsilon_2 \epsilon_3 = -1$, i.e., the 2-space is of signature zero.
If the curvature of the 2-space is zero the metric of the 2-space can be written
\cite{CarotTupper}
\be
d \tau^2 = 2 \epsilon dv dw \, ,
\label{eq:dtau2zero}
\ee
where $\epsilon = \pm 1$, and if the 2-space is of non-zero curvature $\epsilon \lambda^2$
its metric can be written \cite{CarotTupper}
\be
d \tau^2 = \frac{\epsilon 4 dv dw}{\lambda^2 (v+w)^2} \, .
\label{eq:dtau2nonzero}
\ee

When $\epsilon_2 \epsilon_3 = +1$, i.e., the 2-space is of signature $+2$, the zero-curvature
2-space has the form
\be
d \tau^2 = 2  dz d {\bar z} \, ,
\label{eq:dtauzzbar0}
\ee
while, if the 2-space is of non-zero curvature $\epsilon \lambda^2$ its metric can be written
\be
d \tau^2 = \frac{4  dz d {\bar z}}{\lambda^2 (1 + \epsilon z {\bar z})^2} \, .
\label{eq:dtauzzbarnon0}
\ee
The 3-space metrics corresponding to (\ref{eq:dtau2zero}) to (\ref{eq:dtauzzbarnon0}), respectively,
can be written in the forms
\ba
d\sigma^2 & = & e^{2F(v,w)} du^2 + 2 \epsilon dv dw \, , \label{eq:dsigma21}\\
d\sigma^2 & = & \frac{2 e^{2F(v,w)}}{\lambda^2 (v+w)^2} (du^2 + 2 \epsilon e^{-2F(v,w)} dv dw) \, ,
\label{eq:dsigma22}\\
d\sigma^2 & = & \epsilon_1 e^{2F(z, {\bar z})} du^2 + 2 dz d {\bar z} \, , \label{eq:dsigma23}\\
d\sigma^2 & = & \frac{2 e^{2F(z, {\bar z})}}{\lambda^2 (1 + \epsilon z {\bar z})^2}
(\epsilon_1 du^2 + 2 e^{-2F(z, {\bar z})} dz d {\bar z}) \, , \label{eq:dsigma24}
\ea
where $z, {\bar z}$ are conjugate complex coordinates and $\epsilon_1 = \pm 1$. If the three KVs
of the 2-spaces (\ref{eq:dtau2zero}) to (\ref{eq:dtauzzbarnon0}) are labelled $Z_1$, $Z_2$ and $Z_3$
then, from Killing's equations the KV of the 3-spaces (\ref{eq:dsigma21}) to (\ref{eq:dsigma24})
are of the form
\be
\vec Y = Y^u \partial_u + a(u) Z_1 + b(u) Z_2 + c(u) Z_3 \, ,
\label{eq:generalKVof3space}
\ee
where $Y^u = Y^u(u,v,w)$ or $Y^u = Y^u(u,z, {\bar z})$, as appropriate,
$Z_i = Z_i(v,w)$ or $Z_i = Z_i(z, {\bar z})$.

Substituting the expression (\ref{eq:generalKVof3space}) into Killing's equations, we find that the
function $F$ satisfies the appropriate Liouville equation and this leads to 3-space solutions
admitting a number of KV. Some of these solutions admit only two KV and are special cases of the metrics
discussed in section \ref{eq:G2ofisometries}. Some admit four KV, but all of these are conformally
related to reducible 2+2 spacetimes which will not be considered further. However, several solutions
admitting three KV arise and we will now present one of these solutions for each Bianchi type
that arises.

The first example has a metric of the form
\[
d\sigma^2 = H(v, w) du^2 - \frac{4}{\lambda^2 (v+w)^2} dv dw \, .
\]

${\cal G}_3 \, VII_{q = 0}$
\[
H = \frac{k^2 (v^2 + 1) (w^2 + 1)}{\lambda^2 (v+w)^2} G^2(v, w) \, ,
\qquad
G = \tan^{-1} \left(  \frac{w-v}{1 + vw} \right) \, .
\]
The basis vectors for the Lie algebra are
\ba
\vec X_1 & = & \partial_u \, ,
\nonumber\\
\vec X_2 & = & 2k^{-1} G^{-1} \sin ku \, \partial_u
+ (v^2+1) \cos ku \, \partial_v - (w^2+1) \cos ku \, \partial_w \, ,
\nonumber\\
\vec X_3 & = & -2k^{-1} G^{-1} \cos ku \, \partial_u
+ (v^2+1) \sin ku \, \partial_v - (w^2+1) \sin ku \, \partial_w \, .
\nonumber
\ea

The following two examples have a metric of the form
\[
d\sigma^2 = H(v, w) du^2 + \frac{4}{\lambda^2 (v+w)^2} dv dw \, .
\]
${\cal G}_3 \, VIII$
\[
H = \frac{k^2 (v^2 + 1) (w^2 + 1)}{\mu^2 \lambda^2 (v+w)^2} \cosh^2 G(v, w) \, ,
\qquad
G = \mu \tan^{-1} \left( \frac{w-v}{1 + vw} \right) \, .
\]
The basis vectors for the Lie algebra are
\ba
\vec X_1 & = & \partial_u \, ,
\nonumber\\
\vec X_2 & = & 2k^{-1} \mu \tanh G \sin ku \, \partial_u
- (v^2+1) \cos ku \, \partial_v + (w^2+1) \cos ku \, \partial_w \, ,
\nonumber\\
\vec X_3 & = & -2k^{-1} \mu \tanh G \cos ku \, \partial_u
- (v^2+1) \sin ku \, \partial_v + (w^2+1) \sin ku \, \partial_w \, .
\nonumber
\ea

${\cal G}_3 \, VI_{q = -1}$
\[
H = \frac{k^2 (v^2 + 1) (w^2 + 1)}{\lambda^2 (v+w)^2} G^2(v, w) \, ,
\qquad
G = \tan^{-1} \left( \frac{w-v}{1 + vw} \right) \, .
\]
The basis vectors for the Lie algebra are
\ba
\vec X_1 & = & \partial_u \, ,
\nonumber\\
\vec X_2 & = & 2k^{-1} G^{-1} \sinh ku \, \partial_u
+ (v^2+1) \cosh ku \, \partial_v + (w^2+1) \cosh ku \, \partial_w \, ,
\nonumber\\
\vec X_3 & = & 2k^{-1} G^{-1} \cosh ku \, \partial_u
+ (v^2+1) \sinh ku \, \partial_v + (w^2+1) \sinh ku \, \partial_w \, .
\nonumber
\ea

${\cal G}_3 \, III$
\[
d\sigma^2 = P^{-1}(v) du^2 + 2 \epsilon dv dw \, .
\]
The basis vectors for the Lie algebra are
\[
\vec X_1 = \partial_u \, , \qquad
\vec X_2 = -\epsilon P(v) \partial_u + u \partial_w \, , \qquad
\vec X_3 = \partial_w \, .
\]

\subsection{${\mathcal G}_4$ of isometries acting multiply transitively on $(V,h)$.}
\label{G4transitive}
Each ${\cal G}_4$ algebra contains a ${\cal G}_3$ subalgebra. Thus each of the
3-spaces admitting a ${\cal G}_4$ are specializations of 3-spaces admitting ${\cal G}_3$.

However, two cases require special mention.

As discussed in sections \ref{eq:G2IonnullorbitsN2G1generatedbynullKV} and \ref{sec:g3fromg2typeb},
the ${\cal G}_3 \supset {\cal G}_2 I (b)$ algebras have 2-dimensional null orbits and there are
no metrics admitting ${\cal G}_3 \supset {\cal G}_2 II (b)$ algebras. Therefore, it is only necessary
to consider specialisations possessing ${\cal G}_2 I (b)$ subalgebras. These are pp-wave spacetimes.

Metrics admitting ${\cal G}_4$ algebras with orbit types $(a)$, $(c)$ and $(d)$ are
derived from the appropriate ${\cal G}_3$ metric for types $I- VII$. However, Petrov's
type $IX$ metric (\ref{eq:petrovsg3ixmetric}) does not possess a ${\cal G}_2$ subalgebra and
so cannot be used to classify the ${\cal G}_4$ metrics according to the ${\cal G}_2$
orbit types. Therefore, since the ${\cal G}_4 VIII$ algebra contains 2-dimensional abelian
subalgebras, we treat the metrics as specialisations of the appropriate ${\cal G}_2$ metrics.

The following diagram illustrates the analysis of the ${\cal G}_4$ structures.
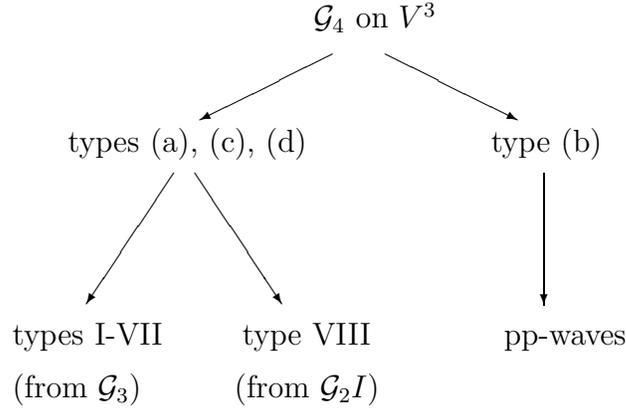
\begin{center}{
\begin{picture}(120,170)
\put(58,155){${\cal G}_4$ on $V^3$}

\put(65,145){\vector(-2,-1){50}}
\put(-35,108){types (a), (c), (d)}
\put(85,145){\vector(2,-1){50}}
\put(125,108){type (b)}

\put(5,100){\vector(-2,-3){33}}
\put(13,100){\vector(2,-3){33}}
\put(-56,35){types I-VII}
\put(-56,15){(from ${\cal G}_3$)}
\put(31,35){type VIII}
\put(28,15){(from ${\cal G}_2 I$)}

\put(145,100){\vector(0,-1){50}}
\put(130,35){pp-waves}

\end{picture}
}
\end{center}

\subsubsection{${\cal G}_4$ (from ${\cal G}_3 \supset {\cal G}_2 (a)$)}
In this section we derive all $(V,h)$ admitting a ${\cal G}_4$ with subalgebra
${\cal G}_3$ possessing a type $(a)$ ${\cal G}_2$ subalgebra. We find that the cases
${\cal G}_4 I_s$, ${\cal G}_4 II$, ${\cal G}_4 IV$, ${\cal G}_4 V$ and ${\cal G}_4 VI$
lead either to conformally reducible 2+2 spacetime, which may be flat, or have no solution
because the Killing equations for the fourth KV lead to a contradiction. The only cases
of interest are ${\cal G}_4 III_p$ and ${\cal G}_4 VII$.

\subsection*{${\cal G}_4 III_p$}
This case is a specialization of the 3-space of type ${\cal G}_3 II$ with metric (\ref{eq:metricg3IIa}).
The condition for the existence of the fourth KV is $p = 0$ and $\epsilon_1 = k = +1$. Thus only type
${\cal G}_4 III (p = 0)$ is possible.
The metric of the 3-space is
\be
d \sigma^2 = du^2 + dw^2 + (dv-udw)^2
\label{eq:g4IIIametric}
\ee
and the metric of $(M, {\hat g})$ is
\[
d \sigma^2 = -d \eta^2 + du^2 + dw^2 + (dv-udw)^2.
\]
The ${\cal G}_4 III (p = 0)$ basis is
\ba
\vec X_1 & = & \partial_v, \qquad \vec X_2 = \partial_w, \qquad \vec X_3 = \partial_u + w \partial_v,
\nonumber\\
\vec X_4 & = & w \partial_u + \case{1}{2} (w^2 - u^2) \partial_v - u \partial_w.
\nonumber
\ea

\subsection*{${\cal G}_4 VII$}
This case is a specialization of the 3-space of type ${\cal G}_3 VIII$ which has the metric
(\ref{eq:G2nonAbeliancasea}), i.e.,
\[
d\sigma^2 = \epsilon_1 du^2 + e^{-2w} A(u) dv^2 + 2 B(u) e^{-w} dv dw + C(u) dw^2
\]
governed by the equations (\ref{eq:g3viiicondition1}) - (\ref{eq:g3viiicondition5}).

Imposing the conditions $[X_1, X_4] = [X_2, X_4] = 0$ and the KV equations gives a fourth KV of the form
\[
\vec X_4 = h \partial_u + k e^w \partial_v + l \partial_w
\]
where $h$, $k$ and $l$ are constants. If $h = 0$ we find that either $k = l = 0$, i.e.,
$\vec X_4 = 0$, or $AC - B^2$ = 0 which implies that the metric is degenerate. Hence $h \ne 0$
and we shall put $h = 1$. The final condition $[X_3, X_4] = 0$ gives
\[
l c = 0, \qquad f(u)_{,u} + 2 l f(u) - kg(u) = 0, \qquad g(u)_{,u} + l g(u) + 2k = 0.
\]
From the first condition either $l = 0$ or $c = 0$. If $c = 0$ then the corresponding
spacetime $(M, {\hat g})$ is a conformally reducible 2+2 spacetime, which we discard. Hence
we have $l = 0$ and the KV equations give
\[
A(u) = A_0, \qquad B(u) = - k A_0 u + B_0, \qquad C(u) = k^2 A_0 u^2 - 2 k B_0 u + C_0
\]
such that
\be
2k (A_0 C_0 - B^2_0) = \epsilon_1 c A_0.
\label{eq:G4VIIcondition}
\ee
We note that $AC - B^2 = A_0 C_0 - B^2_0$.

Thus the 3-space is
\[
d\sigma^2 = \epsilon_1 du^2 + A_0 e^{-2w} dv^2
+ 2 (- k A_0 u + B_0) e^{-w} dv dw + (k^2 A_0 u^2 - 2 k B_0 u + C_0) dw^2
\]
such that (\ref{eq:G4VIIcondition}) holds. The coordinate transformations $u = a u' + k^{-1} A^{-1}_0 B_0$,
$v = ka (u' e^{-w} + v' / 2)$, $w = - w'$, where $a$ is a constant, transform $d\sigma^2$ into the
form (dropping primes)
\be
d\sigma^2 = (\epsilon_1 + A_0 k^2) a^2 du^2 + A_0 k^2 a^2 e^w du dv
+ \case{1}{4} A_0 k^2 a^2 e^{2w} dv^2 + m dw^2
\label{eq:g4VIIametric}
\ee
where $m = C_0 - B^2_0 A^{-1}_0$ is a non-zero constant. The ${\cal G}_4 VII$ basis is
\ba
\vec X_1 & = & \partial_v, \qquad \vec X_2 = v \partial_v - \partial_w,
\qquad \vec X_4 = \partial_u
\nonumber\\
\vec X_3 & = & 2 a^{-1} \epsilon_1 km e^{-w} \partial_u
+ (\case{1}{2} k a v^2 - 2 k^{-1} a^{-1} A^{-1}_0 m e^{-2w}) \partial_v - k a v \partial_w.
\label{eq:G4VIIbKVbasis}
\ea

The canonical metric forms admitting transitive ${\cal G}_4$ algebras on $(V,h)$ are listed
in table \ref{tab:transitiveg4}.
\begin{table}
\caption{Reducible 1+3 spacetimes admitting transitive ${\cal G}_4$ $(a)$, $(c)$ or $(d)$ algebras on $(V,h)$. Those metrics which are conformally reducible 2+2 spacetimes are omitted from the table.}
\footnotesize\rm
\begin{tabular*}{\textwidth}{@{}l*{15}{@{\extracolsep{0pt plus12pt}}l}}
\br
Algebra & $(V, h)$ &  Spacetime metric/${\cal G}_5$ basis & Condition \\
\mr

${\cal G}_4 III_p (a)$ & $(\ref{eq:g4IIIametric})$ &
$ds^2 = - d\eta^2 + du^2 + dw^2 + (dv-udw)^2$ & $-$ \\
& & $\vec X_0 = \partial_\eta$, $\vec X_1 = \partial_v$, $\vec X_2 = \partial_w$,
$\vec X_3 = \partial_u + w \partial_v$, & \\
& & $\vec X_4 = w \partial_u + \case{1}{2} (w^2 - u^2) \partial_v - u \partial_w$ & \\

${\cal G}_4 VII (a)$  & $(\ref{eq:g4VIIametric})$ &
$ds^2 = \epsilon_0 d\eta^2 + (\epsilon_1 + A_0 k^2) a^2 du^2 + A_0 k^2 a^2 e^w du dv$
& $a^2 \epsilon_0 \epsilon_1 < 0$\\
& & $\hspace{10 pt}+ \case{1}{4} A_0 k^2 a^2 e^{2w} dv^2 + m dw^2$ & \\
& & $\vec X_0 = \partial_\eta$, $\vec X_1 = \partial_v$, $\vec X_2 = v \partial_v - \partial_w$,
$\vec X_4 = \partial_u$ & \\
& &  $\vec X_3 = 2 a^{-1} \epsilon_1 km e^{-w} \partial_u
+ (\case{1}{2} k a v^2 - 2 k^{-1} a^{-1} A^{-1}_0 m e^{-2w}) \partial_v - k a v \partial_w$  & \\

${\cal G}_4 VIII (a)$ & $(\ref{eq:G4VIIIametric})$ &
$ds^2 = - d \eta^2 + du^2 + k^{-2} (\rho^2 \sin^2 (ku) + \cos^2 (ku)) dv^2$ & $\cos (ku) \ne 0$ \\
& & $\hspace{10 pt} + 2 \rho k^{-1} \sin (ku) dv dw + dw^2$ & \\
& & $\vec X_0 = \partial_\eta$, $\vec X_1 = \partial_v$, $\vec X_4 = \partial_w$, & \\
& & $\vec X_2 = k^{-1} \sin v \partial_u - \tan (ku) \cos v \partial_v
+ \rho k^{-1} \sec (ku) \cos v \partial_w$, & \\
& & $\vec X_3 = k^{-1} \cos v \partial_u + \tan (ku) \sin v \partial_v
- \rho k^{-1} \sec (ku) \sin v \partial_w$ & \\

${\cal G}_4 I_s (c)$ & $(\ref{eq:automaticG4I})$ &
$ds^2 = d\eta^2 + 2 du dw + k^2(dv + udw)^2$ & $q \ne 0$, $q \ne 1$, \\
& & $\vec X_0 = \partial_\eta$, $\vec X_1 = \partial_v$, $\vec X_2 =  \partial_w$,
$\vec X_3 = - \partial_u + w \partial_v$, $\vec X_4 = - u \partial_u + w \partial_w$ &
$k^2 > 0$\\

${\cal G}_4 I_s (d)$ & $(\ref{eq:g4Idmetric})$ &
$ds^2 = d\eta^2 + [k w^2 - 2 B(u) w + C(u)] du^2 -2[kw - B(u)] du dv$
& $p^2 < 4$ \\
& & $\hspace{10 pt} + k dv^2 + 2h dudw$ & $h^2 k^2 > 0$\\
& & $\vec X_0 = \partial_\eta$, $\vec X_1 = \partial_v$, $\vec X_2 = u\partial_v + \partial_w$,
$\vec X_3 = -\partial_u + f(u) \partial_v + c(u) \partial_w$, & \\
& & $\vec X_4 = -u\partial_u + u f(u) \partial_v + [w + u g(u) + \int g(u) du] \partial_w$, & \\

\br
\end{tabular*}
\label{tab:transitiveg4}
\end{table}

\subsubsection{${\cal G}_4$ (from ${\cal G}_3 \supset {\cal G}_2 (b)$)}
\label{sec:g4b}
As shown in section \ref{G2abelian}, $(V,h)$ of the form (\ref{eq:g2Ibmetric2})
will admit a ${\cal G}_4$ provided that either $A(u) = \alpha u^{-2}$ or
$A(u) = \alpha$, where $\alpha$ is a constant. In the first case $(V,h)$ admits
a ${\cal G}_4 I$ if $\alpha < 0$ or if $\alpha > 4$, a ${\cal G}_4 II$ if $\alpha = 4$,
or a ${\cal G}_4 III$ if $0 < \alpha <4$. In the second case $(V,h)$ admits a
${\cal G}_4 I$ or a ${\cal G}_4 III$ depending on whether $\alpha < 0$ or
$\alpha > 0$, respectively.

\subsubsection{${\cal G}_4$ (from ${\cal G}_3 \supset {\cal G}_2 (c)$)}
\label{sec:g4c}
In this section we derive all $(V,h)$ admitting a ${\cal G}_4$ with subalgebra
${\cal G}_3$ possessing a type $(c)$ ${\cal G}_2$ subalgebra.

\subsection*{${\cal G}_4 I_s$}
The 4-dimensional type $I$ algebra is a specialization of the 3-dimensional type $II$ algebra
(see table \ref{tab:liealgebras}). As mentioned in section \ref{sec:G3fromG2c} there is no
maximal ${\cal G}_3 II$. In fact, imposition of the ${\cal G}_3 II$ algebra leads
immediately to a ${\cal G}_4 I_0$.
The 3-space is
\be
d\sigma^2 = 2 du dw + k^2(dv + udw)^2
\label{eq:automaticG4I}
\ee
where $k$ is a non-zero constant and
\[
\vec X_1 =  \partial_v, \qquad \vec X_2 =  \partial_w, \qquad
\vec X_3 = - \partial_u + w \partial_v, \qquad
\vec X_4 = - u \partial_u + w \partial_w
\]

No further ${\cal G}_4$ 3-spaces of this class exist because they lead directly to the
${\cal G}_4 I_0$ above, or because the corresponding ${\cal G}_3$ space does not exist
or is flat, or the Killing equations lead to a contradiction.

\subsubsection{${\cal G}_4$ (from ${\cal G}_3 \supset {\cal G}_2 (d)$)}
\label{sec:G4fromG3fromG2a}
In this section we derive all $(V,h)$ admitting a ${\cal G}_4$ with subalgebra
${\cal G}_3$ possessing a type $(d)$ ${\cal G}_2$ subalgebra. As in section \ref{sec:g4c}
we find that only a ${\cal G}_4 I_0$ 3-space exists; all other ${\cal G}_4$ are excluded
for the same reasons as above.

\subsection*{${\cal G}_4 I_s$}
The 4-dimensional algebra type $I$ is a specialization of the 3-dimensional type $II$ algebra.
However, imposing the ${\cal G}_3 II$ conditions leads to a ${\cal G}_4 I (s = 0)$ solution,
that is, no maximal ${\cal G}_3 II$ solution is possible.
The 3-dimensional type $II$ algebra is a specialization of the 2-dimensional type $I$ algebra.
The type ${\cal G}_2 I (d)$ 3-space
metric is given by (\ref{eq:G2Abeliancased}). The ${\cal G}_3 II$ conditions lead to the metric
functions $A(u) = k$, $D(u) = h$, where $h$, $k$ are non-zero constants, with $B(u)$, $C(u)$
arbitrary, i.e., the metric of $(V, h)$ is
\ba
d \sigma^2 & = & [k w^2 - 2 B(u) w + C(u)] du^2 -2[kw - B(u)] du dv
\nonumber\\
& & + k dv^2 + 2h dudw
\label{eq:g4Idmetric}
\ea
and the basis for the ${\cal G}_3 II$ is
\ba
\vec X_1 & = & \partial_v, \qquad \vec X_2 = u\partial_v + \partial_w,
\qquad
\vec X_3 = -\partial_u + f(u) \partial_v + g(u) \partial_w
\label{eq:g4Ikvs}
\ea
with $f(u)$, $g(u)$ given by
\ba
f(u) & = & k^{-1} B(u) +  \int g(u) du
\label{eq:g4Idefnoff}\\
g(u) & = & \frac{1}{2hk} (k C(u) - B^2(u))
\label{eq:g4Idefnofg}
\ea
where we have eliminated multiples of $\vec X_1$ and $\vec X_2$. With no further restrictions
on $B(u)$ and $C(u)$, $(V, h)$ also admits a fourth KV given by
\be
\vec X_4 = -u\partial_u + u f(u) \partial_v + [w + u g(u) + \int g(u) du] \partial_w.
\label{eq:g4Ikv4}
\ee
No ${\cal G}_4 I (s \ne 0)$ solution is possible.

The special case with $B(u) = C(u) = 0$ has $f(u) = g(u) = 0$. The metric of $(V,h)$ is
\[
d \sigma^2 = k(w du - dv)^2+ 2h dudw
\]
and the basis is
\[
\vec X_1 = \partial_v, \qquad \vec X_2 = u\partial_v + \partial_w,
\qquad \vec X_3 = -\partial_u, \qquad \vec X_4 = -u\partial_u + w \partial_w.
\]

\subsubsection{${\cal G}_4 VIII$}
As discussed at the beginning of this section, metrics with ${\cal G}_4 VIII$ algebras
with orbit types $(a)$, $(c)$ and $(d)$ will be considered as specialisations of the
${\cal G}_2 I$ metrics. It can be shown that only type $(a)$ exists, with metric
\be
d \sigma^2 = du^2 + (k^{-1} \rho \sin (ku) dv + dw)^2
+ k^{-2} \cos^2 (ku) dv^2 .
\label{eq:G4VIIIametric}
\ee
The ${\cal G}_4$ basis is
\ba
\vec X_1 & = & \partial_v, \qquad \vec X_4 = \partial_w,
\nonumber\\
\vec X_2 & = & k^{-1} \sin v \partial_u - \tan (ku) \cos v \partial_v
+ \rho k^{-1} \sec (ku) \cos v \partial_w,
\nonumber\\
\vec X_3 & = & k^{-1} \cos v \partial_u + \tan (ku) \sin v \partial_v
- \rho k^{-1} \sec (ku) \sin v \partial_w .
\nonumber
\ea
The corresponding $(M, {\hat g})$ is
\[
d s^2 = - d \eta^2 + du^2 + k^{-2} (\rho^2 \sin^2 (ku) + \cos^2 (ku)) dv^2
+ 2 \rho k^{-1} \sin (ku) dv dw + dw^2.
\]

\section{Homothety Groups on $(V,h)$}
\label{sec:homothetygroups}
We wish to enumerate all relevant homothety algebras on three-dimensional manifolds $(V,h)$.
Homothetic algebras will be denoted ${\cal H}_r$, where $r$ is the
dimension the algebra. Note that an ${\cal H}_r$ on $(V,h)$ will
lead to a ${\cal H}_{r+1}$ on $(M, {\hat g})$ on account of the existence of the KV $\partial_u$.

As it is well known, if a proper HV $\vec X$ exists it is unique in
the sense that any other proper HV say $\vec X'$ will be a linear
combination of $\vec X$ and KVs.  Further, the Lie bracket of a HV
and a KV is always a KV  (it can also be zero, as zero is a KV).
From now on, we will always refer to a proper HV simply as a HV,
unless confusion may arise.

The above facts imply that an r-dimensional Lie algebra of
Homotheties, say $\mathcal{H}_r$, always contains an
(r-1)-dimensional Lie algebra of isometries $\mathcal{G}_{r-1}$.

In $(V,h)$, the maximal Killing algebra is 6-dimensional and the
metric $h$ is then of constant curvature, i.e., the Ricci scalar, $R$, is a constant.
If a HV $\vec X$  exists and if $\psi \ne 0$ is the homothetic constant, then
$ \mathcal{L}_{\vec X} R = -2 \psi R = 0$ which implies that $R = 0$ and $(V,h)$ is locally flat.

The case in which ${\cal G}_5$ acts on a $(V,h)$ is forbidden by Fubini's theorem.

We first consider separately the cases with an ${\cal H}_1$
with null orbits and with non-null orbits. The ${\cal H}_2$ metrics are obtained in a
similar way to the ${\cal G}_2$ metrics.
An ${\cal H}_3$ acting multiply transitively on a 2-space is impossible \cite{HallSteele90}.
To determine the relevant homothety algebras ${\cal H}_3$ acting on $(V,h)$ we take the ${\cal G}_2$ metrics
$(V,h)$ and demand that in addition each case admits an HV. Note that the ${\cal H}_2$ metrics
were not taken as the starting point since an ${\cal H}_3$ does not necessarily admit a
${\cal H}_2$ subalgebra.
There is only one ${\cal H}_4$ algebra relevant to this work, as shall be shown presently.
Hall and Steele \cite{HallSteele90} have shown that an ${\cal H}_4$ acting transitively on the $(V,h)$
must possess a ${\cal G}_3$ subalgebra with 2-dimensional orbits. However, such a ${\cal G}_3$ acting
on non-null 2-dimensional orbits leads to a $(M,\hat{g})$ which is a conformally reducible 2+2 spacetime,
and need not be considered any further. The metric (\ref{eq:g2Ibmetric2}) represents a $(V,h)$ with ${\cal G}_3$
acting on null 2-dimensional orbits and, in fact, the corresponding $(M,\hat{g})$ is given by metric
(\ref{eq:isometryclass10pp-wave}) and admits an ${\cal H}_6 \supset {\cal G}_5$.

Any conformally reducible 2+2 spacetimes obtained in this analysis can be discarded since they
are treated in \cite{CarotTupper}.

\subsection{$(V,h)$ admits a group ${\cal H}_1$ of homotheties.}
\label{eq:H1ofhomotheties}

We will treat separately the two cases: null orbits and non-null orbits.

\subsubsection{${\cal H}_1$ on null orbits}
Let $\vec l$ be a null HV and choose a null triad, say $\{\vec
l,\vec n,\vec x\}$ such that: $l^Al_A = n^An_A = 0$, $l^An_A=-1$,
$x^Ax_A=1$ and the remaining products are zero. From
$$l_{A/B} = \psi g_{AB} + F_{AB}$$ it follows upon contraction with
$l^A$ that $F_{AB}l^B = \psi l_A$, hence:
\begin{equation}\label{lAB}
    l_{A/B} = -2\psi l_A n_B + \psi x_A x_B
\end{equation}
and one then has for $\vec n$ and $\vec x$ in the triad
\begin{eqnarray}\label{nAB}
    n_{A/B} = 2\psi n_A n_B + \mu x_A l_B + \nu x_A n_B + \rho x_A
    x_B\\
    x_{A/B} = \mu l_A l_B + \nu l_A n_B + \rho l_A
    x_B + \psi n_A x_B \label{xAB}
\end{eqnarray}
and then the various Lie brackets can be evaluated to get:
\begin{equation}\label{brackets1}
    [\vec l, \vec n] = -2\psi \vec n -\nu \vec x, \quad [\vec l, \vec x] = - \psi \vec x -\nu \vec l, \quad
    [\vec n, \vec x] = -2\mu \vec l -\rho \vec x
\end{equation}

Now, a null rotation can be used in order to simplify the above
expressions, thus putting
\begin{equation}\label{nullrotation}
    \vec l' = \vec l, \quad \vec n' = \case{1}{2} P^2 \vec l + \vec
    n + P \vec x, \quad \vec x' = P\vec l + \vec x
\end{equation}
and dropping primes for convenience, one gets after some
straightforward calculations:
\begin{equation}\label{brackets2}
    [\vec l, \vec n] = -2\psi \vec n , \quad [\vec l, \vec x] = - \psi \vec x, \quad
    [\vec n, \vec x] = \alpha \vec l + \beta \vec n + \gamma \vec x
\end{equation}
Notice that $\vec l$ and $\vec n$ are surface-forming, and so are
$\vec l$ and $\vec x$, thus, we can choose coordinates, say $x^A =
v,u,x$ such that
\begin{equation}\label{triadvectors}
    \vec l = \partial_v, \quad \vec n = A(u,v,x)\partial_v +
    B(u,v,x)\partial_u, \quad \vec x = C(u,v,x)\partial_v +
    D(u,v,x)\partial_x
\end{equation}
and from (\ref{brackets2}) we get
$$\vec n = \exp{(-2\psi v)} \left(A_0(u,x) \partial_v  + B_0(u,x) \partial_x\right),
\qquad\vec x = \exp{(- \psi v)}\left(C_0(u,x)\partial_v + D_0(u,x)\partial_x\right).$$
Further, from the orthogonality relations $l^Al_A = n^An_A = 0$, etc. it follows that
\be
h_{AB} = \exp{(2\psi v)}
\left[ \begin{array}{ccc}
0 & -B_0^{-1} & 0 \\
-B_0^{-1} & 2A_0B_0^{-2} & C_0B_0^{-1} D_0^{-1}\\
0 & C_0B_0^{-1} D_0^{-1} & D_0^{-2}
\end{array} \right]
\label{matrix1}
\ee
and a coordinate transformation $v \rightarrow v$, $u \rightarrow u$
and $x \rightarrow x(u,x)$ exists such that $h_{ux}=0$ in the new
coordinates, that is:
\be
h_{AB} = \exp{(2 \psi v)}K(u,x)
\left[ \begin{array}{ccc}
0 & -1 & 0 \\
-1 & M(u,x) & 0\\
0 & 0 & N(u,x)
\end{array} \right]
, \qquad \vec l = \partial_v.
\label{matrix2}
\ee

\subsubsection{${\cal H}_1$ on non-null orbits}
Let now $\vec X$ be a non-null proper HV, that is
\begin{equation}
    X_{A/B} = \psi h_{AB} + F_{AB}, \quad F_{AB} = -F_{BA} \;\; \mathrm{Homothetic \; Bivector}
\end{equation}
There are now two possibilities:
\begin{enumerate}
  \item $\vec X$ is hypersurface orthogonal (h.o. for short); i.e.:
  $X_{[A}X_{B/C]} = 0$ or equivalently $F_{AB} = X_{[A}V_{B]}$ for
  some  covector $V_B$, that is: $\vec X$ is contained in the blade
  of its own bivector (note this  includes the case of $\vec X$ being a gradient, i.e.: $F_{AB} =0$.
  Assuming it has no fixed points, we can set
  up a coordinate system $x^A = u, x^\alpha$ adapted to $\vec X$ so that $\vec X =
  \partial_u$ and the metric in these coordinates is such that
  $h_{u \alpha}= 0$, imposing next that $\vec X$ is a HV we easily
  get
  \begin{equation}\label{metricnonnullHV1}
    d\sigma^2 = \exp{(2ku)}\left[ \epsilon \exp{V(x^\gamma)} du^2 +
    \bar{h}_{\alpha\beta}(x^\gamma) dx^\alpha dx^\beta\right]
  \end{equation}
  where $\bar{h}_{\alpha\beta}(x^\gamma) $ is a 2-metric and can therefore be
  diagonalized.

  \item $\vec X$ is not h.o. Assuming one is not at a fixed point,
  one can choose an adapted coordinate system, say $u,v,w$ such that
  $\vec X = \partial_u$, and the metric reads then:
  \begin{equation}
    d\sigma^2 = \exp{(2 \psi u)}\hat{h}_{AB}(v,w) dx^A dx^B.
  \end{equation}
  Now, $\hat{h}_{AB}$ depends just on $v$ and $w$ and coordinate
  changes can be performed so as to bring it to a simpler form, thus
  for instance, a change
  $$ u' = u + \alpha(v,w) $$
  allows one to set $\hat{h}_{uw} = 0$ without altering the form of
  the HV (i.e.: $\vec X = \partial_{u'}$). Next, a change of the
  form
  $$ v' = \beta(v,w), \quad \mathrm{and}\quad w' = \gamma(v,w) $$
  can be used to render the 2-metric in the $v,w$ plane in an
  explicitly conformally flat form, thus the whole line element can
  be written as:
  \ba
    d\sigma^2 = \exp{(2 \psi u)} [ \epsilon_1 A^2(v,w) du^2 + 2B(v,w) du dv
    \nonumber\\
    + C^2(v,w) (\epsilon_2 dv^2 + dw^2 ) ]
  \label{metricnonnullHV2nonho}
  \ea
\end{enumerate}
Table \ref{tab:h1} lists the corresponding 1+3 reducible spacetimes $(M, {\hat g})$.
\begin{table}
\caption{Canonical metric forms admitting ${\cal H}_1$ on $(V,h)$. The term hypersurface orthogonal is abbreviated to $h.o.$. $\epsilon_0 = \pm 1$ and $\epsilon_1 = \pm 1$ are not both negative.}
\footnotesize\rm
\begin{tabular*}{\textwidth}{@{}l*{15}{@{\extracolsep{0pt plus12pt}}l}}
\br
Algebra & $(V, h)$ & Spacetime metric/${\cal H}_2$ basis & Condition\\
\mr

${\cal H}_1$ on $N_1$ & $(\ref{matrix2})$ & $ds^2 = \epsilon_0 d\eta^2 + \exp(2 \psi v) K(u,x) [-2 du dv$
& $\epsilon_0 K(u,x) N(u,x) > 0$ \\
& & $\hspace{10 pt} + M(u,x) du^2 + N(u,x) dx^2]$ & \\
& & $\vec X_0 = \partial_\eta$, $\vec l = \partial_v$, & \\

${\cal H}_1$ on $S_1$ or $T_1$ & $(\ref{metricnonnullHV1})$ & $ds^2 = \epsilon_0 d\eta^2
+ \exp{(2 \psi u)} [ \epsilon_1 \exp{V(x^\gamma)} du^2$ &
$\epsilon_0 \epsilon_1 \det \bar{h} < 0$ \\
($\vec X$ is $h.o.$) & & $\hspace{10 pt} + \bar{h}_{\alpha\beta}(x^\gamma) dx^\alpha dx^\beta ]$ & \\
& & $\vec X_0 = \partial_\eta$, $\vec X = \partial_u$, & \\

${\cal H}_1$ on $S_1$ or $T_1$ & $(\ref{metricnonnullHV2nonho})$ & $ds^2 = \epsilon_0 d\eta^2
+ \exp{(2 \psi u)} [ \epsilon_1 A^2(v,w) du^2$ &
$\epsilon_0 C^2(v,w)$ \\
($\vec X$ is not $h.o.$) & & $\hspace{10 pt} + 2B(v,w) du dv + C^2(v,w) (\epsilon_2 dv^2 + dw^2 )$ &
$(\epsilon_1 \epsilon_2 A^2(v,w) C^2(v,w) - B^2(v,w))$ \\
& & $\vec X_0 = \partial_\eta$, $\vec X = \partial_u$, & $\hspace{10 pt}< 0$ \\

\br
\end{tabular*}
\label{tab:h1}
\end{table}

\subsection{$(V,h)$ admits a group ${\cal H}_2$ of homotheties.}
\label{eq:H2ofhomotheties}
The Lie brackets will have the form
\[
[H, X] = \lambda X
\]
where $\lambda = 0, 1$.

The metrics admitting ${\cal H}_2$ algebras are derived in a similar way to the metrics
admitting ${\cal G}_2$ algebras: The abelian and non-abelian cases are dealt with separately,
considering each sub-case $(a)$ - $(d)$ as follows:
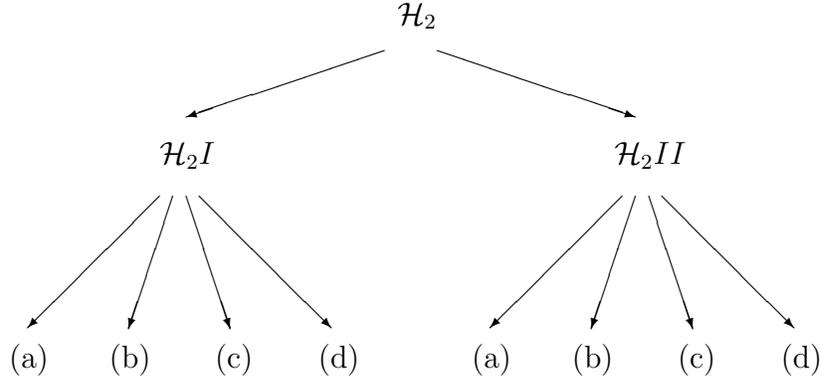
\begin{center}{
\begin{picture}(120,145)
\put(70,140){${\cal H}_2$}

\put(65,130){\vector(-3,-1){75}}
\put(-20,87){${\cal H}_2 I$}
\put(85,130){\vector(3,-1){75}}
\put(152,87){${\cal H}_2 II$}

\put(-20,75){\vector(-1,-1){50}}
\put(-15,75){\vector(-1,-3){16.75}}
\put(-10,75){\vector(1,-3){16.75}}
\put(-5,75){\vector(1,-1){50}}
\put(-77,10){(a)}
\put(-39,10){(b)}
\put(1,10){(c)}
\put(40,10){(d)}

\put(155,75){\vector(-1,-1){50}}
\put(160,75){\vector(-1,-3){16.75}}
\put(165,75){\vector(1,-3){16.75}}
\put(170,75){\vector(1,-1){50}}
\put(98,10){(a)}
\put(136,10){(b)}
\put(176,10){(c)}
\put(215,10){(d)}

\end{picture}
}
\end{center}
\subsubsection{The abelian case, ${\cal H}_2 I$.}
Starting with the general $(V,h)$ metric, in cases $(a)$, $(b)$ and $(c)$ we choose
coordinates such that the KV $\vec X = \partial_v$ and HV $\vec H = \partial_w$.
Similarly, starting with the $(V,h)$ metric, in case $(d)$ we choose
coordinates such that the KV $\vec X = \partial_v$ and HV $\vec H = u \partial_v + \partial_w$.
The corresponding metrics are all of the form
\[
d \Sigma^2 = e^{2 \psi w} d \sigma^2
\]
where $d \sigma^2$ is the 3-space metric of the corresponding ${\cal G}_2$ spacetime and
$\psi$ is the homothetic scalar. The $(V, h)$ metrics admitting an abelian ${\cal H}_2$ algebra are
as follows
\ba
(a) \qquad d \Sigma^2 & = & e^{2 \psi w} (\epsilon_1 du^2 + A(u) dv^2 + 2B(u) dv dw + C(u) dw^2)
\label{eq:H2metrica}\\
(b) \qquad d \Sigma^2 & = & e^{2 \psi w} (P^{-2}(u) dw^2 - 2 du dv - 2H(u) du^2)
\label{eq:H2metricb}\\
(c) \qquad d \Sigma^2 & = & e^{2 \psi w} (2 du dw + (A(u) dv + B(u) dw)^2)
\label{eq:H2metricc}\\
(d) \qquad d \Sigma^2 & = & e^{2 \psi w} [(A(u) w^2 - 2 B(u) w + C(u)) du^2 + A(u) dv^2
\nonumber\\
& & - 2(A(u)w - B(u)) du dv + 2 D(u) dudw].
\label{eq:H2metricd}
\ea
corresponding to metrics (\ref{eq:G2Abeliancasea}), (\ref{eq:g2Ibmetric2}),
(\ref{eq:G2Abeliancasec}), and (\ref{eq:G2Abeliancased}) respectively.
We note that there is some freedom in the choice of the KV and the HV. However, it can be shown that
the above choice, and corresponding metrics, encapsulate all possibilities.
Note that since (\ref{eq:g2Ibmetric2}) admits a ${\cal G}_4$, the metric (\ref{eq:H2metricb})
admits a ${\cal C}_4 \supset {\cal H}_2$.
The reducible 1+3 spacetimes $(M, {\hat g})$ with $(V, h)$ metrics admitting an abelian
${\cal H}_2$ algebra are given in table \ref{tab:H2abelian}.
\begin{table}
\caption{The reducible 1+3 spacetimes $(M, {\hat g})$ with $(V, h)$ metrics admitting
an abelian ${\cal H}_2$. $\epsilon_0 = \pm 1$ and $\epsilon_1 = \pm 1$ are not both negative.}
\footnotesize\rm
\begin{tabular*}{\textwidth}{@{}l*{15}{@{\extracolsep{0pt plus12pt}}l}}
\br
Algebra & $(V, h)$ & Spacetime metric/${\cal H}_3$ basis & Condition\\
\mr

${\cal H}_2 I (a)$ & $(\ref{eq:H2metrica})$ & $ds^2 = \epsilon_0 d\eta^2
+ e^{2 \psi w} (\epsilon_1 du^2 + A(u) dv^2 + 2B(u) dv dw + C(u) dw^2)$ &
$\epsilon_0 \epsilon_1 (A(u) C(u) - B^2(u))$ \\
& & $\vec X_0 = \partial_\eta$, $\vec X = \partial_v$, $\vec H = \partial_w$ & $\hspace{10 pt}< 0$ \\

${\cal H}_2 I (b)$ & $(\ref{eq:H2metricb})$ & $ds^2 = \epsilon_0 d\eta^2
+ e^{2 \psi w} (P^{-2}(u) dw^2 - 2 du dv - 2H(u) du^2)$ & $\epsilon_0 P^{-2}(u) > 0$ \\
& &  $\vec X_0 = \partial_\eta$, $\vec X = \partial_v$, $\vec H = \partial_w$ & \\

${\cal H}_2 I (c)$ & $(\ref{eq:H2metricc})$ & $ds^2 = \epsilon_0 d\eta^2
+ e^{2 \psi w} (2 du dw + (A(u) dv + B(u) dw)^2)$ & $\epsilon_0 A^2(u) > 0$ \\
& & $\vec X_0 = \partial_\eta$, $\vec X = \partial_v$, $\vec H = \partial_w$ & \\

${\cal H}_2 I (d)$ & $(\ref{eq:H2metricd})$ & $ds^2 = \epsilon_0 d\eta^2
+ e^{2 \psi w} [(A(u) w^2 - 2 B(u) w + C(u)) du^2 + A(u) dv^2$ & $\epsilon_0 A(u) D^2(u) > 0$ \\
& & $\hspace{10 pt} - 2(A(u)w - B(u)) du dv + 2 D(u) dudw]$ & \\
& & $\vec X_0 = \partial_\eta$, $\vec X = \partial_v$, $\vec H = u \partial_v + \partial_w$ & \\

\br
\end{tabular*}
\label{tab:H2abelian}
\end{table}
\subsubsection{The non-abelian case, ${\cal H}_2 II$.}
Since the Lie bracket of a KV and an HV is a KV then in cases $(a)$, $(b)$ and $(c)$
we set $\vec X = \partial_v$ and $\vec H = v \partial_v + \partial_w$, and in
case $(d)$ we set $\vec X = \partial_w$ and $\vec H = \partial_v + w \partial_w$.
The $(V, h)$ metrics admitting a non-abelian ${\cal H}_2$ algebra are all conformally
related to the corresponding non-abelian ${\cal G}_2$ metrics, and are as follows
\ba
(a) \qquad d \Sigma^2 & = & e^{2 \psi w}
(\epsilon_1 du^2 + A(u) e^{-2w} dv^2 + 2 B(u) e^{-w} dvdw
\nonumber\\
& & + C(u) dw^2)
\label{eq:H2metricnonabeliana}\\
(b) \qquad d \Sigma^2 & = & e^{2 \psi w} (P^{-2}(u) dw^2 - 2 e^{-w} du dv - 2H(u) du^2)
\label{eq:H2metricnonabelianb}\\
(c) \qquad d \Sigma^2 & = & e^{2 \psi w} [2 du dw + (A(u) e^{-w} dv + B(u) dw)^2]
\label{eq:H2metricnonabelianc}\\
(d) \qquad d \Sigma^2 & = & e^{2 \psi v} [2 dudv + 2 A(u) e^{-v} du dw + B(u) e^{-2v} dw^2]
\label{eq:H2metricnonabeliand}
\ea
corresponding to metrics (\ref{eq:G2nonAbeliancasea}), (\ref{eq:G2nonAbeliancaseb}),
(\ref{eq:G2nonAbeliancasec}), and (\ref{eq:G2nonAbeliancased}) respectively.
The reducible 1+3 spacetimes $(M, {\hat g})$ with $(V, h)$ metrics admitting a non-abelian
${\cal H}_2$ algebra are given in table \ref{tab:H2nonabelian}.
\begin{table}
\caption{The reducible 1+3 spacetimes $(M, {\hat g})$ with $(V, h)$ metrics admitting
a non-abelian ${\cal H}_2$. $\epsilon_0 = \pm 1$ and $\epsilon_1 = \pm 1$ are not both negative.}
\footnotesize\rm
\begin{tabular*}{\textwidth}{@{}l*{15}{@{\extracolsep{0pt plus12pt}}l}}
\br
algebra & $(V, h)$ & spacetime metric / ${\cal H}_3$ basis & condition\\
\mr

${\cal H}_2 II (a)$ & $(\ref{eq:H2metricnonabeliana})$ & $ds^2 = \epsilon_0 d\eta^2
+ e^{2 \psi w} (\epsilon_1 du^2 + A(u) e^{-2w} dv^2 + 2 B(u) e^{-w} dvdw$ &
$\epsilon_0 \epsilon_1 (A(u) C(u) - B^2(u))$ \\
& & $\hspace{10 pt} + C(u) dw^2)$ & $\hspace{10 pt}< 0$ \\
& & $\vec X_0 = \partial_\eta$, $\vec X = \partial_v$, $\vec H = v \partial_v + \partial_w$ &  \\

${\cal H}_2 II (b)$ & $(\ref{eq:H2metricnonabelianb})$ & $ds^2 = \epsilon_0 d\eta^2
+ e^{2 \psi w} (P^{-2}(u) dw^2 - 2 e^{-w} du dv - 2H(u) du^2)$ &
$\epsilon_0 > 0$ \\
& & $\vec X_0 = \partial_\eta$, $\vec X = \partial_v$, $\vec H = v \partial_v + \partial_w$ & \\

${\cal H}_2 II (c)$ & $(\ref{eq:H2metricnonabelianc})$ & $ds^2 = \epsilon_0 d\eta^2
+ e^{2 \psi w} (2 du dw + (A(u) e^{-w} dv + B(u) dw)^2)$ &
$\epsilon_0 A^2(u) > 0$ \\
& & $\vec X_0 = \partial_\eta$, $\vec X = \partial_v$, $\vec H = v \partial_v + \partial_w$ & \\

${\cal H}_2 II (d)$ & $(\ref{eq:H2metricnonabeliand})$ & $ds^2 = \epsilon_0 d\eta^2
+ e^{2 \psi v} [2 dudv + 2 A(u) e^{-v} du dw + B(u) e^{-2v} dw^2]$ & $\epsilon_0 B(u) > 0$ \\
& & $\vec X_0 = \partial_\eta$, $\vec X = \partial_w$, $\vec H = \partial_v + w \partial_w$ & \\

\br
\end{tabular*}
\label{tab:H2nonabelian}
\end{table}

\subsection{$(V,h)$ admits a group ${\cal H}_3$ of homotheties.}
\label{eq:H3ofhomotheties}
A homothety algebra ${\cal H}_r$, $r > 1$ will admit an isometry subalgebra
${\cal G}_{r-1}$. Therefore ${\cal H}_3$ metrics will be obtained from the
corresponding ${\cal G}_2$, rather than the ${\cal H}_2$ metrics.
Type ${\cal H}_3 IX$ cannot occur since Lie algebra type $IX$ admits no
2-dimensional subalgebra.
Lie algebra type $VIII$ is not permitted as a homothety algebra because it does not satisfy the
requirement that the Lie bracket of a KV and an HV must be a KV. Within the 3-dimensional Lie algebras
admitting a 2-dimensional subalgebra, only the type $VIII$ has a non-abelian 2-dimensional subalgebra.
Therefore, only the those homothety algebras ${\cal H}_3 \supset {\cal G}_2 I$ need be considered,
that is, types $I-VII$. Further, orbit type $(b)$ need not be considered since it was established in
section \ref{eq:G2ofisometries} that ${\cal G}_2 I (b)$ cannot occur since it leads directly to an
${\cal H}_4 \supset {\cal G}_3$, and ${\cal G}_2 II (b)$ has been ruled out from the discussion above.
Therefore, only orbit types $(a)$, $(c)$ and $(d)$ need be considered as follows:
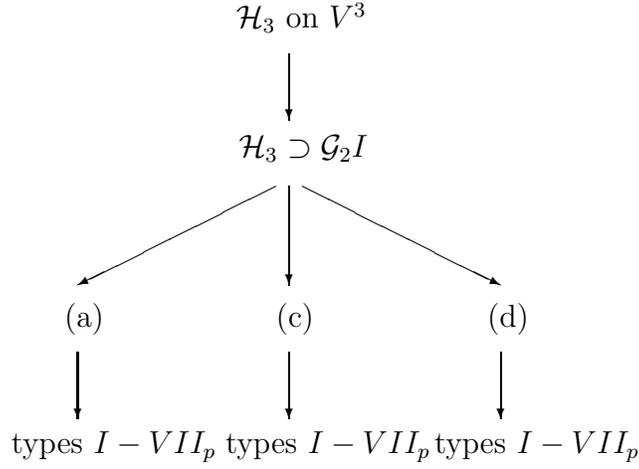
\begin{center}{
\begin{picture}(120,185)
\put(40,170){${\cal H}_3$ on $V^3$}

\put(60,160){\vector(0,-1){25}}
\put(41,121){${\cal H}_3 \supset {\cal G}_2 I$}

\put(60,110){\vector(0,-1){37.5}}
\put(55,110){\vector(-2,-1){75}}
\put(65,110){\vector(2,-1){75}}

\put(-25,57){(a)}
\put(55,57){(c)}
\put(135,57){(d)}
\put(-20,47){\vector(0,-1){25}}
\put(60,47){\vector(0,-1){25}}
\put(140,47){\vector(0,-1){25}}

\put(-45,9){types $I-VII_p$}
\put(36,9){types $I-VII_p$}
\put(115,9){types $I-VII_p$}

\end{picture}
}
\end{center}

\subsubsection{${\cal H}_3$ (from ${\cal G}_2$ (a))}
The 3-space with abelian ${\cal G}_2$ is given by equation (\ref{eq:G2Abeliancasea})
with basis $\vec X_1 = \partial_v$, $\vec X_2 = \partial_w$.

\subsubsection*{${\cal H}_3 I$}
There is no $(V, h)$ with a maximal ${\cal H}_3 I$. The 3-space metric can be written
\be
d \sigma^2 = \epsilon_1 du^2 + u^2 (dv^2 + dw^2)
\ee
and admits at least an ${\cal H}_4$.
The corresponding spacetime $(M, {\hat g})$ is conformally flat and can be rejected.

\subsubsection*{${\cal H}_3 II$}
The 3-space is
\ba
d \sigma^2 = \epsilon_1 du^2 + u^2 [\alpha dv^2 + 2(-\alpha \ln |u| + \beta) dvdw
\nonumber\\
+ (\alpha (\ln|u|)^2 -2 \beta \ln |u| + \gamma)dw^2]
\label{eq:metrich3IIa}
\ea
where $\alpha$, $\beta$ and $\gamma$ are constants such that $\beta^2 - \alpha \gamma \ne 0$.
The ${\cal H}_3$ basis is
\[
\vec X_1 = \partial_v, \qquad \vec X_2 = \partial_w,
\qquad
\vec H = u \partial_u + w \partial_v.
\]

\subsubsection*{${\cal H}_3 III$}
The 3-space is
\be
d \sigma^2 = \epsilon_1 du^2 + u^2[\alpha u^{-2 / \psi} dv^2
+ 2 \beta u^{-1/ \psi} dv dw + \gamma dw^2]
\label{eq:metrich3IIIa}
\ee
where $\alpha$, $\beta$ and $\gamma$ are constants such that $\beta^2 - \alpha \gamma \ne 0$.
The ${\cal H}_3$ basis is
\[
\vec X_1 = \partial_v, \qquad \vec X_2 = \partial_w,
\qquad
\vec H = \psi u \partial_u + v \partial_v.
\]

\subsubsection*{${\cal H}_3 IV$}
The 3-space is
\ba
d \sigma^2 = \epsilon_1 du^2 + e^{2(\psi - 1)/ \psi}[\alpha dv^2 + 2 \psi^{-1}(-\alpha \ln|u| + \beta)dvdw
\nonumber\\
+ \psi^{-2}(\alpha (\ln|u|)^2 - 2 \beta \ln|u| + \gamma) dw^2]
\label{eq:metrich3IVa}
\ea
where $\alpha$, $\beta$ and $\gamma$ are constants such that $\beta^2 - \alpha \gamma \ne 0$.
The ${\cal H}_3$ basis is
\[
\vec X_1 = \partial_v, \qquad \vec X_2 = \partial_w,
\qquad
\vec H = \psi u \partial_u + (v+w) \partial_v + w \partial_w.
\]

\subsubsection*{${\cal H}_3 V$}
The 3-space is
\be
d \sigma^2 = \epsilon_1 du^2 + u^{2(\psi - 1)/ \psi}( \alpha dv^2 + 2 \beta dvdw + \gamma dw^2)
\ee
where $\alpha$, $\beta$ and $\gamma$ are constants such that $\beta^2 - \alpha \gamma \ne 0$.
The corresponding spacetime $(M, {\hat g})$ is a conformally reducible 2+2 spacetime and can be rejected.

\subsubsection*{${\cal H}_3 VI_q$}
The 3-space is
\be
d \sigma^2 = \epsilon_1 du^2
+ u^2 ( \alpha u^{-2/ \psi} dv^2 + 2 \beta u^{-(1+q)/ \psi} dvdw + \gamma u^{-2q/ \psi} dw^2)
\label{eq:metrich3VIa}
\ee
where $q \ne 0$, $q \ne 1$ and $\alpha$, $\beta$ and $\gamma$ are constants such that
$\beta^2 - \alpha \gamma \ne 0$. The ${\cal H}_3$ basis is
\[
\vec X_1 = \partial_v, \qquad \vec X_2 = \partial_w,
\qquad
\vec H = \psi u \partial_u + v \partial_v + q w \partial_w.
\]

\subsubsection*{${\cal H}_3 VII_p$}
The 3-space metric is
\be
d \sigma^2 = \epsilon_1 du^2 + A(u) dv^2 + 2 B(u) dvdw + C(u) dw^2
\label{eq:metrich3VIIa}
\ee
with functions $A(u)$, $B(u)$ and $C(u)$ restricted by
\ba
\psi u A(u)_{,u} + 2 B(u) - 2 \psi A(u) = 0,
\nonumber\\
\psi u B(u)_{,u} - A(u) + (p - 2 \psi) B(u) + C(u) = 0,
\nonumber\\
\psi u C(u)_{,u} - 2 B(u) + 2(p-\psi) C(u) = 0.
\nonumber
\ea
The general solution for this system of equations is
\ba
A(u) & = & u^{2 - p / \psi}
\bigr[\alpha_1 + \alpha_2 \{ (1 - \case{1}{2} s^2) \cos (\frac{s}{\psi} \ln |u|)
- \case{1}{2} ps \sin (\frac{s}{\psi} \ln |u|) \}
\nonumber\\
& & + \alpha_3 \{ (1 - \case{1}{2} s^2) \sin (\frac{s}{\psi} \ln |u|)
+ \case{1}{2} ps \cos (\frac{s}{\psi} \ln |u|) \} \bigr],
\label{eq:metrich3VIIaconditiona}
\\
B(u) & = & \case{1}{2} u^{2 - p / \psi}
\bigr[\alpha_1 p + \alpha_2 \{ p \cos (\frac{s}{\psi} \ln |u|) - s \sin (\frac{s}{\psi} \ln |u|) \}
\nonumber\\
& & + \alpha_3 \{ p \sin (\frac{s}{\psi} \ln |u|) + s \cos (\frac{s}{\psi} \ln |u|) \} \bigr],
\label{eq:metrich3VIIaconditionb}
\\
C(u) & = & u^{2 - p/ \psi} \bigr[\alpha_1 + \alpha_2 \cos (\frac{s}{\psi} \ln |u|)
+ \alpha_3 \sin (\frac{s}{\psi} \ln |u|) \bigr]
\label{eq:metrich3VIIaconditionc}
\ea
where $\alpha_1$, $\alpha_2$, and $\alpha_3$ are arbitrary constants and $s = \sqrt{4 - p^2}$.
The ${\cal H}_3$ basis is
\[
\vec X_1 = \partial_v, \qquad \vec X_2 = \partial_w,
\qquad
\vec H = \psi u \partial_u - w \partial_v + (v + pw) \partial_w.
\]

The canonical metric forms admitting transitive ${\cal H}_3$ type $(a)$ algebras on $(V,h)$ are listed
in table \ref{tab:transitiveh3a}.

\begin{table}
\caption{Reducible 1+3 spacetimes admitting transitive ${\cal H}_3$ type $(a)$ algebras on $(V,h)$.
Those metrics which are conformally reducible 2+2 spacetimes are omitted from the table.}
\footnotesize\rm
\begin{tabular*}{\textwidth}{@{}l*{15}{@{\extracolsep{0pt plus12pt}}l}}
\br
${\cal H}_3$ algebra & $(V, h)$ & Spacetime metric and ${\cal H}_4$ basis & Condition \\
\mr

${\cal H}_3 II (a)$ & $(\ref{eq:metrich3IIa})$ &
$ds^2 = \epsilon_0 d\eta^2 + \epsilon_1 du^2 + u^2 [\alpha dv^2 + 2(-\alpha \ln |u| + \beta) dvdw$ &
$\epsilon_0 \epsilon_1 (\alpha \gamma - \beta^2) u^2 < 0$\\
& & $ \phantom{ds^2 =} + (\alpha (\ln|u|)^2 -2 \beta \ln |u| + \gamma)dw^2]$ & \\
& & $\vec X_0 = \partial_\eta$, $\vec X_1 = \partial_v$, $\vec X_2 = \partial_w$,
$\vec H = u \partial_u + w \partial_v$  & \\

${\cal H}_3 III (a)$ & $(\ref{eq:metrich3IIIa})$ &
$ds^2 = \epsilon_0 d\eta^2 + \epsilon_1 du^2 + u^2[\alpha u^{-2 / \psi} dv^2$ &
$\epsilon_0 \epsilon_1 (\alpha \gamma - \beta^2) u^{2(1 - 1 / \psi)} $\\
& & $ \phantom{ds^2 =} + 2 \beta u^{-1/ \psi} dv dw + \gamma dw^2]$ & $\hspace{10 pt}< 0$ \\
& & $\vec X_0 = \partial_\eta$, $\vec X_1 = \partial_v$, $\vec X_2 = \partial_w$,
$\vec H = \psi u \partial_u + v \partial_v$ & \\

${\cal H}_3 IV (a)$ & $(\ref{eq:metrich3IVa})$ & $ds^2 = \epsilon_0 d\eta^2 +
\epsilon_1 du^2 + e^{2(\psi - 1)/ \psi}[\alpha dv^2 $ &
$\epsilon_0 \epsilon_1 (\alpha \gamma - \beta^2) < 0$ \\
& & $\hspace{10 pt} + 2 \psi^{-1}(-\alpha \ln|u| + \beta)dvdw $ & \\
& & $\hspace{10 pt} + \psi^{-2}(\alpha (\ln|u|)^2 - 2 \beta \ln|u| + \gamma) dw^2]$ & \\
& & $\vec X_0 = \partial_\eta$, $\vec X_1 = \partial_v$, $\vec X_2 = \partial_w$, & \\
& & $\vec H = \psi u \partial_u + (v+w) \partial_v + w \partial_w$ & \\

${\cal H}_3 VI_q (a)$ & $(\ref{eq:metrich3VIa})$ & $ds^2 = \epsilon_0 d\eta^2 + \epsilon_1 du^2 $ &
$\epsilon_0 \epsilon_1 (\alpha \gamma - \beta^2) u^{2-2(1+q) / \psi}$ \\
& & $\hspace{10 pt} + u^2 ( \alpha u^{-2/ \psi} dv^2 + 2 \beta u^{-(1+q)/ \psi} dvdw + \gamma u^{-2q/ \psi} dw^2)$ & $\hspace{10 pt}< 0$ \\
& & $\vec X_0 = \partial_\eta$, $\vec X_1 = \partial_v$, $\vec X_2 = \partial_w$,
$\vec H = \psi u \partial_u + v \partial_v + q w \partial_w$ & \\

${\cal H}_3 VII_p (a)$ & $(\ref{eq:metrich3VIIa})$ &
$ds^2 = \epsilon_0 d\eta^2 + \epsilon_1 du^2 + A(u) dv^2 + 2 B(u) dvdw + C(u) dw^2$ &
$\epsilon_0 \epsilon_1 (p^2 \alpha_1 - s^2 \alpha_2 - s^2 \alpha_3)$ \\
& & Functions $A(u)$, $B(u)$, $C(u)$ given by
(\ref{eq:metrich3VIIaconditiona}) - (\ref{eq:metrich3VIIaconditionc})& $\hspace{10 pt}< 0$ \\
& & $\vec X_0 = \partial_\eta$, $\vec X_1 = \partial_v$, $\vec X_2 = \partial_w$, &  \\
& &  $\vec H = \psi u \partial_u - w \partial_v + (v + pw) \partial_w$ & \\

\br
\end{tabular*}
\label{tab:transitiveh3a}
\end{table}

\subsubsection{${\cal H}_3$ (from ${\cal G}_2$ (c))}
The 3-space with abelian ${\cal G}_2$ is given by equation (\ref{eq:G2Abeliancasec})
with basis $\vec X_1 = \partial_v$, $\vec X_2 = \partial_w$. As noted in sections
\ref{sec:g2IonN2nonullsubgroup}(c) and \ref{sec:g4b}, if $B(u)$ is proportional to $A(u)$,
or if $B(u) = 0$, the corresponding $(M, {\hat g})$ is an isometry class 10 (or class 11 or 13)
pp-wave spacetime. Where such cases arise in the following analysis they will not be considered
in further detail.

\subsubsection*{${\cal H}_3 I$}
There is no $(V, h)$ with a maximal ${\cal H}_3 I$. The 3-space can be written
\[
d \sigma^2 = 2 du dw + u (\alpha dv + \beta dw)^2
\]
where $\alpha \ne 0$ and $\beta$ are constants. Since $A(u)$ and $B(u)$ are proportional,
the corresponding $(M, {\hat g})$ is a
class 11 pp-wave spacetime admitting a ${\cal H}_5 \supset {\cal G}_4$ as considered in
section \ref{sec:g4b}.

\subsubsection*{${\cal H}_3 II$}
The 3-space is
\be
d \sigma^2 = 2 du dw + u [\alpha dv + (- \alpha \ln |u| + \beta) dw]^2
\label{eq:metrich3IIc}
\ee
where $\alpha \ne 0$ and $\beta$ are constants. The ${\cal H}_3$ basis is
\[
\vec X_1 = \partial_v, \qquad \vec X_2 = \partial_w,
\qquad
\vec H = u \partial_u + w \partial_v .
\]

\subsubsection*{${\cal H}_3 III$}
The 3-space is
\be
d \sigma^2 = 2 du dw + u [\alpha u^{-1 / 2 \psi} dv + \beta dw]^2
\label{eq:metrich3IIIc}
\ee
where $\alpha \ne 0$ and $\beta$ are constants. The ${\cal H}_3$ basis is
\[
\vec X_1 = \partial_v, \qquad \vec X_2 = \partial_w,
\qquad
\vec H = 2 \psi u \partial_u + v \partial_v .
\]

\subsubsection*{${\cal H}_3 IV$}
There are two cases. For $\psi \ne 1/2$ the 3-space is
\be
d \sigma^2 = 2 du dw + u^{2(\psi - 1)/(2 \psi -1)}
[\alpha dv + (- \alpha (2 \psi -1)^{-1} \ln |u| + \beta) dw]^2
\label{eq:metrich3IVc}
\ee
where $\alpha \ne 0$ and $\beta$ are constants. The ${\cal H}_3$ basis is
\[
\vec X_1 = \partial_v, \qquad \vec X_2 = \partial_w,
\qquad
\vec H = (2 \psi - 1) u \partial_u + (v + w) \partial_v + w \partial_w.
\]
For $\psi = 1/2$ the 3-space is
\be
d \sigma^2 = 2 du dw + e^{-u}
[\alpha dv + (- \alpha u + \beta) dw]^2
\label{eq:metrich3IVcii}
\ee
where $\alpha \ne 0$ and $\beta$ are constants. The ${\cal H}_3$ basis is
\[
\vec X_1 = \partial_v, \qquad \vec X_2 = \partial_w,
\qquad
\vec H = \partial_u + (v + w) \partial_v + w \partial_w.
\]

\subsubsection*{${\cal H}_3 V$}
There are two cases. For $\psi \ne 1/2$ the 3-space is
\[
d \sigma^2 = 2 du dw + u^{2(\psi - 1)/(2 \psi -1)}
[\alpha dv + \beta dw]^2
\]
where $\alpha \ne 0$ and $\beta$ are constants. The corresponding $(M, {\hat g})$ is a
type 11 pp-wave spacetime admitting a ${\cal H}_5 \supset {\cal G}_4$. When $\psi = 1$ the 3-space is flat.
When $\psi = 1/2$ the 3-space metric is
\[
d \sigma^2 = 2 du dw + e^{-u} dv^2.
\]
The corresponding $(M, {\hat g})$ is a type 13 pp-wave spacetime admitting a
${\cal H}_5 \supset {\cal G}_4$.

\subsubsection*{${\cal H}_3 VI_q$}
We note that $q \ne 0, 1$. For $\psi \ne q/2$ the 3-space metric is
\be
d \sigma^2 = 2 du dw +
u^{2(\psi - 1)/ (2 \psi - q)} [\alpha dv + \beta u^{(1 - q)/(2 \psi - q)} dw]^2
\label{eq:metrich3VIc}
\ee
where $\alpha \ne 0$ and $\beta$ are constants.
The ${\cal H}_3$ basis is
\[
\vec X_1 = \partial_v, \qquad \vec X_2 = \partial_w,
\qquad
\vec H = (2 \psi - q) u \partial_u + v \partial_v + q w \partial_w .
\]
For $\psi = q/2$ the 3-space metric is
\be
d \sigma^2 = 2 du dw +
[\alpha e^{(q - 2)u / (2c)} dv + \beta e^{-qu/(2c)} dw]^2
\label{eq:metrich3VIcii}
\ee
where $\alpha \ne 0$, $\beta$ and $c \ne 0$ are constants.
The ${\cal H}_3$ basis is
\[
\vec X_1 = \partial_v, \qquad \vec X_2 = \partial_w,
\qquad
\vec H = c \partial_u + v \partial_v + q w \partial_w .
\]

\subsubsection*{${\cal H}_3 VII_p$}
The 3-space is
\be
d \sigma^2 = 2 dudw + [A(u) dv +  B(u) dw]^2
\label{eq:metrich3VIIc}
\ee
where the functions $A(u)$ and $B(u)$ are restricted by
\ba
A(u)_{,u} a(u) + B(u) - \psi A(u) = 0,
\label{eq:H3cVIImetricfunctioncondition1}\\
B(u) [ B(u)_{,u} a(u) - A(u) + (p - \psi) B(u)] = 0,
\label{eq:H3cVIImetricfunctioncondition2}\\
f(u)_{,u} + A^{-2}(u) = 0,
\label{eq:H3cVIImetricfunctioncondition3}\\
a(u)_{,u} + A(u) B(u) f(u)_{,u} + p - 2 \psi = 0.
\label{eq:H3cVIImetricfunctioncondition4}
\ea
The ${\cal H}_3$ basis is
\[
\vec X_1 = \partial_v, \qquad \vec X_2 = \partial_w,
\qquad
\vec H = a(u) \partial_u + (-w + f(u)) \partial_v + (p w + v) \partial_w.
\]
From equations (\ref{eq:H3cVIImetricfunctioncondition1}) -
(\ref{eq:H3cVIImetricfunctioncondition4}) we find that $a(u)$ satisfies the equation
\be
a(u) a(u)_{,uu} - (a(u)_{,u})^2 = (p - 2 \psi) a(u)_{,u} + 1 - 2 p \psi + 4 \psi^2
\label{eq:H3cVIImetricfunctioncondition5}
\ee
corresponding to equation (\ref{eq:G3VIIIconditionsDefora}) in the ${\cal G}_3 VII_p$ case.
One solution of equation (\ref{eq:H3cVIImetricfunctioncondition5}) is $a(u) = ku$, where
$k$ is a constant given in terms of $p$ and $\psi$, but this leads to
$B(u) A^{-1}(u) = constant$, i.e., to a pp-wave solution. No general solution of equation
(\ref{eq:H3cVIImetricfunctioncondition5}) has been found. Note that $a(u) \ne 0$ for
3-dimensional orbits.

The canonical metric forms admitting transitive ${\cal H}_3$ type $(c)$ algebras on $(V,h)$ are listed
in table \ref{tab:transitiveh3c}.

\begin{table}
\caption{Reducible 1+3 spacetimes admitting transitive ${\cal H}_3$ type $(c)$ algebras on $(V,h)$.
Those metrics which are conformally reducible 2+2 spacetimes are omitted from the table.}
\footnotesize\rm
\begin{tabular*}{\textwidth}{@{}l*{15}{@{\extracolsep{0pt plus12pt}}l}}
\br
${\cal H}_3$ algebra & $(V, h)$ & Spacetime metric and ${\cal H}_4$ basis & Condition \\
\mr

${\cal H}_3 II (c)$ & $(\ref{eq:metrich3IIc})$ &
$ds^2 = \epsilon_0 d\eta^2 + 2 du dw + u [\alpha dv + (- \alpha \ln |u| + \beta) dw]^2$ &
$\epsilon_0 \alpha^2 u > 0$\\
& & $\vec X_0 = \partial_\eta$, $\vec X_1 = \partial_v$, $\vec X_2 = \partial_w$,
$\vec H = u \partial_u + w \partial_v$,  & \\

${\cal H}_3 III (c)$ & $(\ref{eq:metrich3IIIc})$ &
$ds^2 = \epsilon_0 d\eta^2 + 2 du dw + u [\alpha u^{-1 / 2 \psi} dv + \beta dw]^2$ &
$\epsilon_0  \alpha^2 u^{1 - 1 / \psi} > 0$\\
& & $\vec X_0 = \partial_\eta$, $\vec X_1 = \partial_v$, $\vec X_2 = \partial_w$,
$\vec H = 2 \psi u \partial_u + v \partial_v$ & \\

${\cal H}_3 IV (c)$ & $(\ref{eq:metrich3IVc})$ &
$ds^2 = \epsilon_0 d\eta^2 + 2 du dw $ &
$\epsilon_0 \alpha^2 u^{2(\psi - 1)/(2 \psi - 1)}$\\
$(\psi \ne 1/2)$ & & $\hspace{10 pt} + u^{2(\psi - 1)/(2 \psi -1)} [\alpha dv + (- \alpha (2 \psi -1)^{-1} \ln |u| + \beta) dw]^2$ & $\hspace{10 pt}> 0$ \\
& & $\vec X_0 = \partial_\eta$, $\vec X_1 = \partial_v$, $\vec X_2 = \partial_w$, & \\
& & $\vec H = (2 \psi - 1) u \partial_u + (v + w) \partial_v + w \partial_w$ & \\

${\cal H}_3 IV (c)$ & $(\ref{eq:metrich3IVcii})$ &
$ds^2 = \epsilon_0 d\eta^2 + 2 du dw + e^{-u} [\alpha dv + (- \alpha u + \beta) dw]^2$ &
$\epsilon_0 \alpha^2 > 0$\\
$(\psi = 1/2)$ & & $\vec X_0 = \partial_\eta$, $\vec X_1 = \partial_v$, $\vec X_2 = \partial_w$,
$\vec H = \partial_u + (v + w) \partial_v + w \partial_w$ & \\

${\cal H}_3 VI_q (c)$ & $(\ref{eq:metrich3VIc})$ &
$ds^2 = \epsilon_0 d\eta^2 + 2 du dw +
u^{2(\psi - 1)/ (2 \psi - q)} [\alpha dv $ &
$\epsilon_0 \alpha^2 u^{2(\psi - 1)/(2 \psi - q)}$\\
& & $\hspace{10 pt} + \beta u^{(1 - q)/(2 \psi - q)} dw]^2$ & $\hspace{10 pt}> 0$ \\
$(\psi \ne q/2)$ & & $\vec X_0 = \partial_\eta$, $\vec X_1 = \partial_v$, $\vec X_2 = \partial_w$,
$\vec H = (2 \psi - q) u \partial_u + v \partial_v + q w \partial_w$ & \\

${\cal H}_3 VI_q (c)$ & $(\ref{eq:metrich3VIcii})$ &
$ds^2 = \epsilon_0 d\eta^2 + 2 du dw +
[\alpha e^{(q - 2)u / (2c)} dv + \beta e^{-qu/(2c)} dw]^2$ &
$\epsilon_0 \alpha^2 > 0$\\
$(\psi = q/2)$ & & $\vec X_0 = \partial_\eta$, $\vec X_1 = \partial_v$, $\vec X_2 = \partial_w$,
$\vec H = c \partial_u + v \partial_v + q w \partial_w$ & \\

${\cal H}_3 VII_p (c)$ & $(\ref{eq:metrich3VIIc})$ &
$ds^2 = \epsilon_0 d\eta^2 + 2 dudw + [A(u) dv +  B(u) dw]^2$ &
$\epsilon_0 A^2(u) > 0$ \\
& & Functions $A(u)$, $B(u)$, $C(u)$ given by
(\ref{eq:H3cVIImetricfunctioncondition1}) - (\ref{eq:H3cVIImetricfunctioncondition4}) & $a(u) \ne 0$ \\
& & $\vec X_0 = \partial_\eta$, $\vec X_1 = \partial_v$, $\vec X_2 = \partial_w$, & \\
& & $\vec H = a(u) \partial_u + (-w + f(u)) \partial_v + (p w + v) \partial_w$ & \\

\br
\end{tabular*}
\label{tab:transitiveh3c}
\end{table}

\subsubsection{${\cal H}_3$ (from ${\cal G}_2$ (d))}
The 3-space with abelian ${\cal G}_2$ is given by metric (\ref{eq:G2Abeliancased}) with basis
$\vec X_1 = \partial_v$, $\vec X_2 = u \partial_v + \partial_w$.
Note that both metric functions $A(u)$ and $D(u)$ are non-zero for non-degeneracy.

\subsubsection*{${\cal H}_3 I$}
This type is not possible.

\subsubsection*{${\cal H}_3 II$}
The 3-space metric is
\be
d \Sigma^2 = e^{-2 \psi u} d \sigma^2
\label{eq:metrich3IId}
\ee
where $d \sigma^2$ is the metric (\ref{eq:g4Idmetric}), i.e., the 3-space metric of the
corresponding ${\cal G}_4 I (d)$. (Recall that the corresponding ${\cal G}_3 II(d)$
does not exist, see section \ref{sec:G4fromG3fromG2a}.)
The KV $\vec X_4$ of $d \sigma^2$, given by equation (\ref{eq:g4Ikv4}), becomes a proper CKV of
$d \Sigma^2$ but does not give rise to a symmetry of $(M, {\hat g})$. The KVs
$\vec X_1$, $\vec X_2$ of $d \sigma^2$ remain as KVs of $d \Sigma^2$
and thus of $(M, {\hat g})$ while the KV $\vec X_3$, given by equation (\ref{eq:g4Ikvs})
becomes a HV of $d \Sigma^2$ and leads to the HV of $(M, {\hat g})$ given by
\[
\vec H = \psi  \eta \partial_\eta - \partial_u + f(u) \partial_v + g(u) \partial_w
\]
where $f(u)$ and $g(u)$ are defined by (\ref{eq:g4Idefnoff}), (\ref{eq:g4Idefnofg}).

\subsubsection*{${\cal H}_3 III$}
The 3-space metric is
\be
d \Sigma^2 = u^{2 \psi} d \sigma^2
\label{eq:metrich3IIId}
\ee
where $d \sigma^2$ is metric (\ref{eq:G3IIId3spacemetric}), i.e., the 3-space metric of the
corresponding ${\cal G}_3 III (d)$. The ${\cal H}_3$ basis is given by the corresponding
${\cal G}_3 III (d)$ basis, with homothety $X_3$.

\subsubsection*{${\cal H}_3 IV$}
The 3-space metric is
\be
d \Sigma^2 = u^{- 2 \psi} d \sigma^2
\label{eq:metrich3IVd}
\ee
where $d \sigma^2$ is metric (\ref{eq:G3IVd3spacemetric}), i.e., the 3-space metric of the
corresponding ${\cal G}_3 IV (d)$. The ${\cal H}_3$ basis is given by the corresponding
${\cal G}_3 IV (d)$ basis, with homothety $X_3$.

\subsubsection*{${\cal H}_3 V$}
This type is not possible.

\subsubsection*{${\cal H}_3 VI_q$}
The 3-space metric is
\be
d \Sigma^2 = u^{2 \psi / (1 - q)} d \sigma^2
\label{eq:metrich3VId}
\ee
where $d \sigma^2$ is metric (\ref{eq:G3VId3spacemetric}), i.e., the 3-space metric of the
corresponding ${\cal G}_3 VI (d)$. The ${\cal H}_3$ basis is given by the corresponding
${\cal G}_3 VI (d)$ basis, with homothety $X_3$.

\subsubsection*{${\cal H}_3 VII_p$}
The 3-space metric is
\be
d \Sigma^2 = e^{4 \psi f(u) / q } d \sigma^2, \qquad f(u) = \tan^{-1}[(2u - p)/q]
\label{eq:metrich3VIId}
\ee
where $q = \sqrt{4 - p^2}$ and $d \sigma^2$ is metric (\ref{eq:G2Abeliancased})
with functions $A(u)$, $D(u)$ given by (\ref{eq:G3VIIdconditions}), i.e., the 3-space metric of the
corresponding ${\cal G}_3 VII (d)$. The ${\cal H}_3$ basis is given by the corresponding
${\cal G}_3 VII (d)$ basis, with homothety $X_3$.

The canonical metric forms admitting transitive ${\cal H}_3$ type $(d)$ algebras on $(V,h)$ are listed
in table \ref{tab:transitiveh3d}.

\begin{table}
\caption{Reducible 1+3 spacetimes admitting transitive ${\cal H}_3$ type $(d)$ algebras on $(V,h)$.
Those metrics which are conformally reducible 2+2 spacetimes are omitted from the table.}
\footnotesize\rm
\begin{tabular*}{\textwidth}{@{}l*{15}{@{\extracolsep{0pt plus12pt}}l}}
\br
${\cal H}_3$ algebra & $(V, h)$ & Spacetime metric and ${\cal H}_4$ basis & Condition \\
\mr

${\cal H}_3 II (d)$ & $(\ref{eq:metrich3IId})$ &
$ds^2 = \epsilon_0 d\eta^2 + e^{-2 \psi u} \{ [k w^2 - 2 B(u) w + C(u)] du^2 $ &
$\epsilon_0 h^2 k^2 > 0$\\
& & \phantom{$ds^2 = $} $- 2[kw - B(u)] du dv + k dv^2 + 2h dudw \}$ & \\
& & $\vec X_0 = \partial_\eta$, $\vec X_1 = \partial_v$, $\vec X_2 = u\partial_v + \partial_w$, & \\
& & $\vec H = -\partial_u + f(u) \partial_v + g(u) \partial_w$ & \\

${\cal H}_3 III (d)$ & $(\ref{eq:metrich3IIId})$ &
$ds^2 = \epsilon_0 d\eta^2 + u^{2 \psi} \{ [\alpha u^{-2} w^2 - 2 B(u) w + C(u)] du^2$ &
$\epsilon_0 \alpha \delta^2 > 0$ \\
& & $\hspace{10 pt} -2[\alpha u^{-2}w - B(u)] du dv + \alpha u^{-2} dv^2 + 2 \delta u^{-1} dudw \}$ & $u \ne 0$\\
& & $\vec X_0 = \partial_\eta$, $\vec X_1 = \partial_v$, $\vec X_2 = u\partial_v + \partial_w$, & \\
& & $\vec H = u \partial_u + [v + b(u)] \partial_v + c(u) \partial_w$,
$b(u)$, $c(u)$ given by (\ref{eq:G3IIIdconditions}) & \\

${\cal H}_3 IV (d)$  & $(\ref{eq:metrich3IVd})$ &
$ds^2 = \epsilon_0 d\eta^2 + u^{2 \psi} \{ [\alpha e^{2u} w^2 - 2 B(u) w + C(u)] du^2$ & $\epsilon_0 \alpha \delta^2 > 0$ \\
& & $\hspace{10 pt} -2[\alpha e^{2u}w - B(u)] du dv + \alpha e^{2u} dv^2 + 2 \delta e^u dudw \} $ & \\
& & $\vec X_0 = \partial_\eta$, $\vec X_1 = \partial_v$, $\vec X_2 = u\partial_v + \partial_w$, & \\
& & $\vec H = -\partial_u + (v + b(u)) \partial_v + (w + c(u)) \partial_w$, & \\
& & $b(u)$, $c(u)$ given by (\ref{eq:G3IVdconditions}) & \\

${\cal H}_3 VI_q (d)$ & $(\ref{eq:metrich3VId})$ &
$ds^2 = \epsilon_0 d\eta^2 + u^{2 \psi / (1 - q)} \{ [\alpha u^{2p} w^2 - 2 B(u) w + C(u)] du^2 $ & $\epsilon_0 \alpha \delta^2 u^{4p} > 0$, \\
& & $\hspace{10 pt} - 2[\alpha u^{2p}w - B(u)] du dv + \alpha u^{2p} dv^2 + 2 \delta u^p dudw \}$ & $q \ne 0$, $q \ne 1$ \\
& & $\vec X_0 = \partial_\eta$, $\vec X_1 = \partial_v$, $\vec X_2 = u\partial_v + \partial_w$, & \\
& & $\vec H = (1 - q) u \partial_u + (v + b(u)) \partial_v + (qw + c(u)) \partial_w$ & \\
& & $b(u)$, $c(u)$ given by (\ref{eq:G3VIdconditions}) & \\

${\cal H}_3 VII_p (d)$ & $(\ref{eq:metrich3VIId})$ &
$ds^2 = \epsilon_0 d\eta^2 + e^{4 \psi f(u) / q } \{ [A(u) w^2 - 2 B(u) w + C(u)] du^2$ & $\epsilon_0 \alpha \delta^2 (u^2 - pu + 1)^{-2}$ \\
& & $\hspace{10 pt} - 2[A(u) w - B(u)] du dv + A(u) dv^2 + 2 D(u) dudw \}$ & $\hspace{10 pt}> 0$, \\
& & $\vec X_0 = \partial_\eta$, $\vec X_1 = \partial_v$, $\vec X_2 = u\partial_v + \partial_w$, & $p^2 < 4$ \\
& & $\vec H = (u^2 - pu + 1) \partial_u + (uv + b(u)) \partial_v$ & \\
& & $\hspace{10 pt} + (v + (p - u)w + c(u)) \partial_w$ & \\
& & $A(u)$, $D(u)$ given by (\ref{eq:G3VIIdconditions}),
$b(u)$, $c(u)$ given by (\ref{eq:G3VIIdconditions1}) & \\

\br
\end{tabular*}
\label{tab:transitiveh3d}
\end{table}

\section{Discussion and Examples}
\label{sec:discussion}
We have shown that the following possibilities exist for 1+3 reducible spacetimes:

$(1)$ If $(V,h)$ is of constant curvature, then it admits two or more GCKV and $(M, {\hat g})$
is conformally flat and thus is a conformally reducible 2+2 spacetime.

$(2)$ If $(V,h)$ admits only one GCKV, which is not a null GKV, then $(M, {\hat g})$ is not
conformally flat, but is a conformally reducible 2+2 spacetime.

$(3)$ If $(V,h)$ admits only one GCKV, which is a null GKV, then $(M, {\hat g})$ is a pp-wave
spacetime and the dimension, $r$, of the symmetry group
(${\cal S}_r$, ${\cal H}_r$, or ${\cal G}_r$) of $(V,h)$ is increased to $r+2$ in $(M, {\hat g})$.

$(4)$ If $(V,h)$ admits no GCKV, then $(M, {\hat g})$ admits only HV and KV and the dimension,
$r$, of the symmetry group (${\cal H}_r$ or ${\cal G}_r$) of $(V,h)$ is
increased to $r+1$ in $(M, {\hat g})$.

The first two possibilities were dealt with in \cite{CarotTupper}. The second two
possibilities are discussed in sections \ref{sec:ih} and \ref{sec:homothetygroups},
where we enumerated all possible isometry and homothety Lie algebras on the
3-dimensional spacelike and timelike hypersurfaces of the 1+3 spacetimes.

Throughout this work we have pointed out instances where the spacetimes in question are
conformally reducible 2+2 spacetimes. There may be conformally reducible 1+3 spacetimes
present which are also conformally reducible 2+2 spacetimes although this may not be apparent
from their particular coordinate representation. The condition for a spacetime to be reducible 2+2
is that it admits two independent null recurrent vector fields. Further, only Petrov type
$D$ or $O$ are possible.
Oktem \cite{oktem76} shows that, for a non-zero Weyl tensor, the two independent recurrent vector
fields are the principal null directions of the Weyl tensor. It follows that the recurrent vector fields
will also be the principal null directions of the Weyl tensor for any conformally related spacetime,
i.e., a conformally reducible 2+2 spacetime. If the spacetime is type $O$
then it is automatically a conformally reducible 2+2 spacetime. If the spacetime is a type $D$
conformally reducible 2+2 spacetime, then the principal null directions will satisfy
the conditions of Theorem 1 in \cite{CarotTupper}, which we shall quote here

\begin{thm}\label{thm:CarotTupper}
Let $(M,g)$ be a spacetime. If there exists a function $\mu: M \rightarrow \mathbb{R}$ and null vectors
$l^a$ and $k^a$ ($l^a k_a = -1$) satisfying
\[
l_{a;b} = \alpha e^{-\mu} l_a l_b - \mu_{,a} l_b + (\mu_{,c} l^c) g_{ab} \, ,
\qquad
k_{a;b} = -\alpha e^{-\mu} k_a l_b - \mu_{,a} k_b + (\mu_{,c} k^c) g_{ab} \, ,
\]
then $(M,g)$ is conformally related to a reducible 2+2 spacetime with conformal factor $e^{2\mu}$.
Here $\alpha$ is some real function of the coordinates associated with the integral distribution
spanned by ${\hat l}^a = e^\mu l^a$ and ${\hat k}^a = e^\mu k^a$.
\end{thm}

We now present some examples of reducible and conformally reducible 1+3 spacetimes which, in general,
are well known and which illustrate some of the results presented here. A number of examples
of 1+3 pp-wave spacetimes are given in the text so no further examples of such spacetimes will
be given here.

\subsubsection*{Example 1}
Consider the ${\cal G}_4 VII$ metric of $(V,h)$ given by (\ref{eq:g4VIIametric}) with KVs given by
(\ref{eq:G4VIIbKVbasis}). The special case given by the choice $\epsilon_1 = 1$, $A_0 k^2 = -2$,
$m = a^2$ results in the spacetime $(M, {\hat g})$ with metric
\be
ds^2 = a^2 [d \eta^2 - (du + e^w dv)^2 + \case{1}{2} e^{2w} dv^2 + dw^2]
\label{eq:example1godel}
\ee
which is the G\"{o}del metric admitting the five KVs
\ba
\vec X_1 & = & \partial_v, \qquad \vec X_2 = v \partial_v - \partial_w,
\qquad \vec X_4 = \partial_u, \qquad \vec X_5 = \partial_\eta,
\nonumber\\
\vec X_3 & = & -2e^{-w} \partial_u - (e^{-2w} + v^2 / 2) \partial_v + v \partial_w.
\ea
The spacetime with metric
\[
d \Sigma^2 = \eta^{-2} ds^2,
\]
where $ds^2$ is the G\"{o}del metric (\ref{eq:example1godel}), also satisfies the
field equations for a comoving perfect fluid with a non-zero cosmological constant
$\Lambda$. The density and pressure are given by
\[
\mu = \case{1}{2} a^{-2} \eta^2 - 3 a^{-2} - \Lambda,
\qquad
p = \case{1}{2} a^{-2} \eta^2 + 3 a^{-2} + \Lambda,
\]
and the appropriate energy conditions are satisfied if $\Lambda \le -3a^{-2}$.
The KV remain as KV except for $\vec X_5$ which becomes a proper CKV with
$\psi = - \eta^{-1}$.

\subsubsection*{Example 2}
The perfect fluid spacetime (Senovilla \cite{senovilla}) with metric
\be
ds^2 = G^2(y) d\Sigma^2
\ee
where $G(y) = a^{-1} k \sin (ay)$ and the metric of $(M, {\hat g})$ is given by
\be
d\Sigma^2 = G^{-2}(y) dy^2 + F^{-1}(x) dx^2 + F(x) d \phi^2 - x^2(dt + bx^{-2} d\phi)^2
\ee
is a conformally reducible 1+3 spacetime. Here $F(x) = m \ln (x/c) + b^2 x^{-2} - k^2 x^2$
and $a$, $b$, $c$, $k$ and $m$ are constants. The 3-space $(V, h)$ admits only the two
KV $\vec X_1 = \partial_t$, $\vec X_2 = \partial_\phi$, which is an example of the ${\cal G}_2 I$
structure discussed in section \ref{G2onV2}. The spacetime $(M, {\hat g})$ admits the additional
KV $\vec X_3 = G^{-1}(y) \partial_y$ and the conformally related spacetime $(M, g)$ retains the
KVs $\vec X_1$, $\vec X_2$ but $\vec X_3$ becomes a proper CKV with $\psi = - G^{-2} G_{,y}$.

\subsubsection*{Example 3}
Consider the conformally reducible 1+3 perfect fluid spacetime of Allnutt \cite{allnutt}, viz.
\be
ds^2 = e^{-2ax} d\Sigma^2
\ee
where the metric of $(M, {\hat g})$ is given by
\be
d\Sigma^2 = e^{2ax} dx^2 + t^{1+n} dy^2 + t^{1-n} dz^2 - dt^2 \, .
\ee
$(M, g)$ is a Bianchi $VI_h$ solution and so admits three KVs.

As in the previous example, the 3-space $(V,h)$ admits two KVs, $\vec X_1 = \partial_y$,
$\vec X_2 = \partial_z$, but also admits a HV,
$\vec X_3 = t \partial_t + \case{1}{2} (1-n) y \partial_y + \case{1}{2} (1+n) z \partial_z$,
and thus is an example of a ${\cal H}_3 VI$ space given by equation (\ref{eq:metrich3VIc})
after a suitable coordinate transformation.
We find that $(M, {\hat g})$ admits the
KV $\vec X_4 = e^{-ax} \partial_x$ in addition to $\vec X_1$ and $\vec X_2$ \cite{Ramos}. Restoring
the conformal factor $e^{-2ax}$ we find that for $(M, g)$ $\vec X_1$ and $\vec X_2$ remain
as KV, $\vec X_3$ becomes the third KV while $\vec X_4$ becomes a proper CKV with
$\psi = a e^{-ax}$.

\subsubsection*{Example 4}
The null electromagnetic field solution found by Datta and Raychaudhuri \cite{DattaRaychaudhuri}
has metric
\[
ds^2 = \rho^{-1/2} [d \rho^2 + dz^2 + f(\rho)^{-1} \rho^{5/2} d\phi^2
- f(\rho) \rho^{1/2} (dt^2 - \rho f(\rho)^{-1} d\phi^2)^2]
\]
where $f(\rho) = 4b^2 \rho^2 + c \rho \ln |\rho|$ and $b$, $c$ are constants. The corresponding
$(V,h)$, i.e.,
\[
d \sigma^2 = d \rho^2 - f(\rho) \rho^{1/2} dt^2 + 2 \rho^{3/2} dt d\phi
\]
admits a ${\cal G}_2 I$ with basis, $\vec X_1 = \partial_\phi$, $\vec X_2 = \partial_t$ where $\vec X_1$
is null and $\vec X_1$, $\vec X_2$ are not orthogonal. The spacetime $(M, g)$ thus admits
three KV and no other symmetries.

If the constant $c = 0$ the metric of $(V,h)$ becomes
\[
ds^2 = d \rho^2 - 4b^2 \rho^{5/2} dt^2 + 2 \rho^{3/2} dt d\phi .
\]
Thus also admits an HV,
$\vec H = \rho \partial_\rho + \case{3}{4} \phi \partial_\phi - \case{1}{4} t \partial_t$,
and so $(V,h)$ is of type $H_3 V$. The corresponding spacetime $(M, g)$, i.e.,
\[
\rho^{-1/2} (d \rho^2 + dz^2) - 4b^2 \rho^2 dt^2 + 2 \rho dt d \phi
\]
admits the three KVs $\vec X_1$, $\vec X_2$, $\vec X_3 = \partial_z$ and the HV
\[
\vec H = 4 z \partial_z + 4 \rho \partial_\rho + 3 \phi \partial_\phi - t \partial_t,
\qquad \psi = 3.
\]

\subsubsection*{Example 5}
We now illustrate Theorem 4, part 4(a), by considering a special type of plane wave spacetime $(M,g)$
with metric (the conformal symmetries of which are given in \cite{keanetupper})
\be
ds^2 = dy^2 + dx^2 - 2 du dv - 2u^{-4} x^2 du^2 \, .
\ee
The energy-momentum tensor for this spacetime is that of a null fluid. The 3-space $(V,h)$
given by
\be
d\sigma^2 = dx^2 - 2 du dv - 2u^{-4} x^2 du^2
\ee
admits a null GKV, $\vec \xi = \partial_v$, and a SCKV
$\vec X_1 = u^2 \partial_u + \case{1}{2} x^2 \partial_v +ux \partial_x$. Thus the requirements
of the theorem are satisfied and $(M, g)$ admits a proper SCKV given by
\[
\vec Y = u^2 \partial_u + \case{1}{2} (x^2 + y^2) \partial_v + ux \partial_x + uy \partial_y \, .
\]
The 3-space $(V, h)$ also admits a HV which leads to a HV of $(M, g)$ given by
$\vec H = 2 v \partial_v + x \partial_x + y \partial_y$, and two KVs, which are also
KVs of $(M, g)$ given by $\vec X_2 = p(u)_{,u} \partial_v + p(u) \partial_x$ and
$\vec X_3 = q(u)_{,u} \partial_v + q(u) \partial_x$ where $p(u) = u \cos (\sqrt{2} / u)$
and $q(u) = u \sin (\sqrt{2} / u)$. Apart from the obvious KV $\vec X_4 = \partial_y$,
$(M, g)$ also admits a fifth KV $\vec X_5 = y \partial_v + u \partial_y$. Thus $(M, g)$ admits
one proper SCKV, one HV and five KV, including a null GKV.

\subsubsection*{Example 6}
Siklos \cite{Siklos} found a family of pure radiation solutions with non-zero cosmological
constant $\Lambda$; the metric of $(M, g)$ is
\[
ds^2 = \frac{3}{|\Lambda| x^2} d \Sigma^2
\]
where $d \Sigma^2$ is the metric of $(M, {\hat g})$ given by
\[
d \Sigma^2 = dx^2 + dy^2 - 2 du dv - 2 x^{2k} du^2
\]
where $k(2k-3) > 0$ for positive energy. $(M, {\hat g})$ is a pp-wave spacetime of isometry
class 1(i) \cite{keanetupper} and is an example of the case $\rho = 0$ in Theorem 4, case (4a),
so no proper SCKV exists. The corresponding $(V,h)$ admits two KVs, $\vec X_1 = \partial_u$,
$\vec X_2 = \partial_v$ and a HV,
$\vec X_3 = (k - 1) u \partial_u - (k + 1) v \partial_v - x \partial_x$,
while $(M, {\hat g})$ admits the additional KVs, $\vec X_4 = \partial_y$,
$\vec X_5 = y \partial_v + u \partial_y$ and the HV
$\vec X_3 = (k - 1) u \partial_u - (k + 1) v \partial_v - x \partial_x - y \partial_y$.
Restoring the conformal factor $3 / |\Lambda| x^2$, we find that $\vec X_1$, $\vec X_2$,
$\vec X_4$, $\vec X_5$ remain as KVs while $\vec X_3$ becomes a fifth KV so that
$(M, {\hat g})$ admits a ${\cal G}_5$.

\section*{Acknowledgments}
JC acknowledges financial support from the Spanish Ministry of Education through grant no.
FPA 2004-0366 and also the `Govern de les Illes Balears' through its programme `Suport als
Grups d'Investigaci\'{o} Competitius'.

\end{document}